\documentclass[cite,12pt]{article}
\makeatletter
\@addtoreset{equation}{section}

\makeatother
\usepackage{graphicx}
\usepackage{dcolumn}
\usepackage{bm}
\usepackage{cite}

\begin{document}

\begin{titlepage}
\null
\begin{flushright}

\end{flushright}

\vskip 1.8cm
\begin{center}

  {\LARGE \bf 
Lattice QCD with fixed topology
}

\vskip 2.3cm
\normalsize

  {\Large Hidenori Fukaya
\footnote{
Ph.D thesis submitted to Department of Physics, Kyoto University
on January 5th 2006.
\\~~~~~~E-mail: fukaya@yukawa.kyoto-u.ac.jp
}
}

\vskip 0.5cm

  {\large \it Yukawa Institute for Theoretical Physics,\\ 
Kyoto University, Kyoto 606-8502, Japan
              }

\vskip 2cm

A Dissertation in candidacy for \\
the degree of Doctor of Philosophy

\vskip 1.5cm

\begin{figure}[hbtp]
  \centering
  \includegraphics[width=6cm]{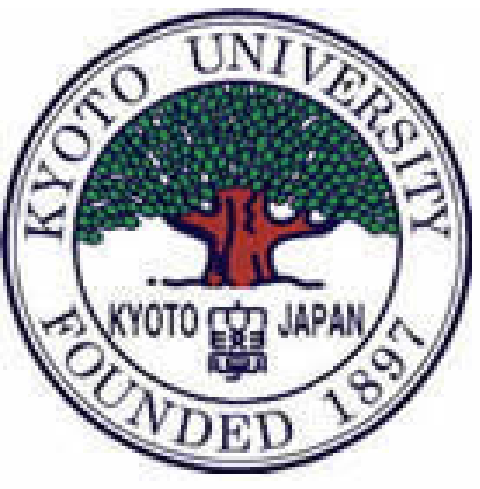}
\end{figure}
\end{center}

\vspace{0.5cm}

\end{titlepage}

\vspace*{4cm}
\begin{center}
\begin{abstract}
The overlap Dirac operator, which satisfies the Ginsparg-Wilson
relation, realizes exact chiral symmetry on the lattice.
It also avoids fermion doubling but its locality and smoothness
are subtle. In fact, the index theorem on the lattice 
implies that there are certain points where the overlap Dirac operator
has discontinuities.
Aside from the theoretical subtleties, this non-smoothness also 
raises practical problems in numerical simulations,
especially in $N_f\neq 0$ full QCD case.
One must carefully calculate 
the lowest eigenvalue of the Dirac operator at each
hybrid Monte Carlo step, in order to catch sudden jumps of
the fermion determinant on the topology boundaries
(reflection/refraction). 
The approximation of the sign function, which is a crucial point
in implementing the overlap Dirac operator, 
gets worse near the discontinuities. 

A solution may be 
to concentrate on a fixed topological sector in the full theory.
It is known that an ``admissibility'' condition, which suppress
small plaquette values, 
preserves topology of the gauge fields and
improves the locality of the overlap Dirac operator at the same time.
In this thesis, we test a gauge action which automatically generates
``admissible'' configurations, as well as (large) negative mass Wilson
fermion action which would also keep the topology.
The quark potential and the topology stability are investigated
with different lattice sizes and different couplings.
Then we discuss the effects of these new approaches on 
the numerical cost of the overlap fermions.

The results of quenched QCD 
in the $\epsilon$-regime are also presented
as an example of the lattice studies with fixed topology.
Remarkable quark mass and topology dependences 
of meson correlators allow us to
determine the fundamental parameters of the effective theory,
in which the exact chiral symmetry with the Ginsparg Wilson relation
plays a crucial role.
\end{abstract}
\end{center}

\baselineskip 7mm

\clearpage

\tableofcontents

\baselineskip 6.22mm

\section{Introduction}

Lattice QCD has played an important role in elementary 
particle physics, in particular, in studying 
the low energy dynamics of hadrons.
One can nonperturbatively 
calculate hadron masses or decay constants 
with Monte Carlo simulations. 
As compensation, however, 
the lattice discretization of space-time
spoils a lot of symmetries of the gauge theory.
Violation of the 
translational symmetry would be an easiest example. 

It is well known that chiral symmetry is not compatible
with the absence of fermion doubling, 
due to the periodic properties of
the lattice Dirac operator in 
momentum space
\cite{Nielsen:1980rz,Nielsen:1981xu}.
A popular 
prescription to avoid the appearance of unphysical modes,
or doublers, is
adding a so-called Wilson term \cite{Wilson:1975id}
to the naive subtraction operator;
\begin{eqnarray}
\label{eq:Wilsonfermion}
D_W &=& \frac{1}{2}\gamma_{\mu}
(\nabla_{\mu}+\nabla_{\mu}^*) -\frac{ra}{2}\nabla_{\mu}^*
\nabla_{\mu}.
\end{eqnarray}
where $r$ is Wilson parameter (we set $r=1$.).
\footnote{The other
notations used here and in the following of this paper are
summarized in appendix \ref{sec:notations}.}  
This term gives a large mass to the doublers which are
decoupled from the theory.
The Wilson term, however, violates chiral symmetry.
A well known difficulty with Wilson fermion is renormalization.
The complicated operator mixing, additive 
quark mass corrections, have to be calculated.
Actually these are obstacles to obtain reliable 
numerical data, especially in the chiral limit.

With the overlap Dirac operator
\cite{Neuberger:1997fp,Neuberger:1998wv}, as well as
the other Dirac operators 
\cite{Kaplan:1992bt,
Shamir:1993zy,Furman:1994ky,Wiese:1993cb,
Hasenfratz:1998ri},
which satisfies the Ginsparg-Wilson relation 
\cite{Ginsparg:1981bj},
one can construct lattice gauge theories 
which have the exact chiral symmetry \cite{Luscher:1998pq}.
Although the good chiral behaviors in applying the overlap operator
to QCD are reported in both theoretical and numerical studies,
its locality properties and smoothness with respect to
the gauge fields are not so obvious.
Since it has a term proportional to 
$1/\sqrt{a^2H_W^2}\equiv 1/\sqrt{(\gamma_5(aD_W-1-s))^2}$, where $s$ is
a fixed parameter in the region $|s|<1$, 
the near zero modes of $H_W$ can contaminate the locality or
smoothness properties.  
In fact, it is not difficult to see that there exist some points
where the overlap Dirac operator, $aD$,
is not smooth by noting the fact
that the index of $aD$ can take integer values only.
Also practically, near zero modes of $H_W$ 
causes some problems in the numerical simulations.
Small eigen-modes of $|H_W|$ lower the convergence of 
polynomial or rational expansion of $1/\sqrt{a^2H_W^2}$.
For example, to keep a certain accuracy,
the order of the Chebyshev polynomial
has to be proportional to $1/\lambda_{\mathrm{min}}$,
where
$\lambda_{\mathrm{min}}$ is the minimum eigenvalue of $|H_W|$.
In full QCD with the dynamical overlap fermion
\cite{Narayanan:1995sv,
Neuberger:1999re,
Bode:1999dd,
Fodor:2003bh,
Arnold:2003sx,
Cundy:2004pz,
Cundy:2005pi,
DeGrand:2004nq,
DeGrand:2005vb,
Schaefer:2005qg,
Egri:2005cx,
Cundy:2005mr}, 
one would have to carefully perform the 
hybrid Monte Carlo \cite{Duane:1987de} 
updating near $H_W\sim 0$ points 
since sudden changes of the trajectories,
reflection or refraction,
should occur due to sudden jumps of the fermion determinant.
Thus, at least, the smallest eigenvalue of $H_W$ always 
needs to be monitored in conventional methods of lattice QCD 
simulations, which is very 
time consuming and it is known that reflection/refraction itself 
has systematic errors when one employs the pseudo-fermion method
\cite{Egri:2005cx}.
Because of these difficulties, no lattice QCD study with the
$N_f\neq 0$ overlap Dirac fermions has been done
except for the cases with a very small lattice size. \\

An interesting solution might be prohibiting 
the topology change along the simulations.
It is known that under a smoothness condition on the plaquette 
variables \cite{Luscher:1998kn, Luscher:1998du, Luscher:1999un,
Hernandez:1998et},
\begin{equation}
  \label{eq:admi}
  ||1-P_{\mu\nu}(x)|| < \epsilon \;\;\;\;
  \mbox{for all ($x$, $\mu,\nu)$},
\end{equation}
which is called the ``admissibility'' bound,
any eigenvalues of $H_W$ are non-zero (we denote $|H_W|>0$)
and the topological charge can be conserved if  
$\epsilon$, which is a fixed number, is sufficiently small 
\cite{Hernandez:1998et, Neuberger:1999pz}.
Furthermore, when $|H_W|>0$, the locality is
also guaranteed.
The ``admissibility'' condition, Eq.(\ref{eq:admi}), is
automatically satisfied if one takes a type of gauge action
which diverges when $||1-P_{\mu\nu}(x)|| \to \epsilon$
\cite{Fukaya:2003ph, Fukaya:2004kp,
Shcheredin:2004xa,Bietenholz:2004mq,Shcheredin:2005ma,
Nagai:2005fz,Fukaya:2005cw,Bietenholz:2005rd}.
   
Topology transitions can also be suppressed by including 
the factor, $\det H_W^2$,
in the functional integral \cite{private}. 
The inclusion of this factor was previously considered
in a study of domain-wall fermions 
\cite{Vranas:1999rz, Hernandez:2000iw, Izubuchi:2002pq}, 
where the aim was to reduce the effects of
the finiteness of the lattice in the 5th dimension.
If any eigenvalue approaches near $H_W=0$ along the
simulation,
the determinant, $\det H_W^2$, would give a very small
Boltzmann weight, and such a trajectory would 
be rejected.
Since $\det H_W^2 = \det (D_W-(1+s)/a)^2$ is equivalent to 
$N_f=2$ Wilson fermion determinant 
with a negative cutoff-scale mass, 
it would not have any effects on the low energy physics.
Moreover, the numerical cost of this determinant is expected 
to be much smaller than that of the dynamical overlap fermions.

What can we do with the configurations 
in a fixed topological sector ?
A straightforward application would be QCD 
in the so-called $\epsilon$-regime
\cite{Gasser:1983yg, Gasser:1987ah,Gasser:1987zq,Hansen:1990un,Hansen:1990yg, Leutwyler:1992yt},
where the linear extent of the space-time is smaller
than the pion Compton wave length $L \ll 1/m_{\pi}$.
In this regime 
(though it is an unphysical small-volume situation), 
it is believed that one would be able to evaluate the
pion decay constant and the chiral condensates,
which are
the fundamental parameters of the 
chiral perturbation theory (ChPT)\cite{Gattringer:2005ij,
Gattringer:2005qy}.
They should be evaluated in lattice QCD studies
without taking the large volume limit, 
since the finite volume effects are already 
involved on the ChPT side
\cite{
Hernandez:1999cu, Prelovsek:2002qs,
Giusti:2002sm,
Giusti:2003gf,
Bietenholz:2003bj,
Giusti:2003iq,
Giusti:2004yp,
Ogawa:2005jn,
Fukaya:2005yg,
Mehen:2005fw,
Bietenholz:2005ip,
Giusti:2005sx,
Damgaard:2005ys,
Bietenholz:2005rc,
Shcheredin:2005ew,
Damgaard:2006pu
}.\\

In this thesis, we study
\begin{enumerate}
\item The practical feasibility of the
topology conserving gauge action 
which keeps the ``admissibility'' bound, Eq.(\ref{eq:admi}), 
as well as the Wilson fermion action with a large negative mass.
A careful analysis on the gluonic quantity and comparison with
that with the standard plaquette action have to be done.
\item How much stable the topological charge can be, with
these topology conserving actions. 
\item Their effect on the numerical cost of 
the overlap Dirac operator.
\item The determination of the low energy constants of quenched 
chiral perturbation theory in the $\epsilon$-regime 
in a fixed topological sector. This study would be helpful 
when $N_f\neq 0$ simulations with the dynamical overlap fermion
are done in the future works.
\end{enumerate}

We start with
the theoretical details on  
the overlap Dirac operator and topology of the lattice 
gauge fields in Sec.~\ref{sec:GWtop}.
The technical issues of our numerical study 
is presented in Sec.~\ref{sec:simulation}.
To test their practical application, 
the static quark potential with different couplings and $\epsilon$
and with/without the negative mass Wilson fermions
is investigated
(Sec.~\ref{sec:qpotential}).
We study the parameter dependence of the topological charge stability 
in Sec.~\ref{sec:top}.
Then the effects on 
the overlap Dirac operator are discussed in Sec.~\ref{sec:overlap}.
The numerical result of quenched lattice QCD in the $\epsilon$-regime
is presented in Sec. \ref{sec:eregime}.
Conclusions and discussions are given in Sec.~\ref{sec:conclusion}.

The main papers contributed to this thesis are
\begin{itemize}
\item H.~Fukaya, S.~Hashimoto, T.~Hirohashi, K.~Ogawa and T.~Onogi,
{\it``Topology conserving gauge action and the overlap-Dirac operator,''}
 Phys.\ Rev.\ D {\bf 73}, 014503 (2006)  [arXiv:hep-lat/0510116] 
\cite{Fukaya:2005cw},
\item H.~Fukaya, S.~Hashimoto and K.~Ogawa,
{\it``Low-lying mode contribution to the quenched meson correlators in the
epsilon-regime,''}
Prog.\ Theor.\ Phys.\  {\bf 114} (2005) 451
[arXiv:hep-lat/0504018]\cite{Fukaya:2005yg}.
\end{itemize}
Refer also \cite{Fukaya:2003ph, Fukaya:2004kp} which are similar studies
in 2-dimensions as a good test ground.

\section{The overlap Dirac operator and topology}
\label{sec:GWtop}

The overlap Dirac operator \cite{Neuberger:1997fp,Neuberger:1998wv}
is defined by
\begin{equation}
  \label{eq:ov}
  D = \frac{1}{\bar{a}} 
  \left( 1 + \gamma_5 \frac{aH_W}{\sqrt{a^2H_W^2}} \right),
  \;\;\;
  \bar{a} = \frac{a}{1+s},\;\;\;aH_W=\gamma_5(aD_W-1-s),
\end{equation}
which satisfies the Ginsparg-Wilson relation \cite{Ginsparg:1981bj}
\begin{equation}
\label{eq:GWrelation}
  \gamma_5D+D\gamma_5=\bar{a}D\gamma_5D,
\end{equation}
and $\gamma_5$-hermiticity $D^{\dagger}=\gamma_5 D\gamma_5$.
Here $s$ is a real parameter which satisfies $|s|<1$.
The Dirac operator Eq.(\ref{eq:ov}) is gauge covariant and
has no fermion doubling, as one can see in the Fourier
transform of $aD$ in the free case
\footnote{Here we show $s=0$ case for simplicity.}; 
\begin{eqnarray}
a\tilde{D}(p)&=&1-\left(1-\frac{1}{2}a^2\hat{p}^2
-ia\gamma_{\mu}\tilde{p_{\mu}}\right)
\left(1+\frac{1}{2}a^4\sum_{\mu < \nu}
\hat{p_{\nu}}\hat{p_{\nu}}\right)^{-1/2}
\sim ia\gamma_{\mu}p_{\mu}+O((ap_{\mu})^2)
,\nonumber\\
\hat{p_{\mu}}&=& (2/a)\sin (ap_{\mu}/2),\;\;\;\;
\tilde{p_{\mu}}\;\;=\;\; (1/a)\sin (ap_{\mu}),
\end{eqnarray}
where the term $a^2\hat{p}^2/2$ at $a p_{\mu}\sim \pi$ 
in any direction $\mu$ gives large mass, 
so that the doublers are decoupled and 
the continuum limit can be properly taken
\cite{Luscher:1998pq}.
The Ginsparg-Wilson relation guarantees that 
the fermion action 
\begin{equation}
S_F = \sum_x \bar{\psi}(x)D\psi(x),
\end{equation}
is exactly invariant under the chiral rotation,
even for finite lattice spacings;
\begin{eqnarray}
\psi \to e^{i\alpha\gamma_5(1-\bar{a}D)}\psi,\;\;\;\;\;
\bar{\psi}\to\bar{\psi}e^{i\alpha\gamma_5}.
\end{eqnarray}
Thus the chiral symmetry at classical level is realized on
the lattice.\\

In order to see quantum level properties of chiral symmetry, 
let us consider the eigenstates of the operator $\bar{a}\gamma_5 D$
\cite{Fujikawa:2000my,Fujikawa:2000qw}.
Note that if an eigen-mode of $\bar{a}\gamma_5 D$ has the eigenvalue
$\pm 2$, it is also the eigenstate of $\bar{a}D$ 
with the eigenvalue $2$,
which means that this mode has $\pm$ chirality.
It is easier to see that zero modes of  $\bar{a}\gamma_5 D$ can be
taken as eigen-modes of $\gamma_5$. 
Every other mode of $\bar{a}\gamma_5 D$ with 
eigenvalue $\lambda_n$ in the range $0<|\lambda_n|<2$ 
is not chiral but has its pair 
with eigenvalue $-\lambda_n$ through the Ginsparg-Wilson relation;
\begin{eqnarray}
(\bar{a}\gamma_5 D )| \lambda_n \rangle &=& \lambda_n | \lambda_n \rangle,
\nonumber\\
(\bar{a}\gamma_5 D )\gamma_5(1-\bar{a}D/2)
| \lambda_n \rangle &=& - \lambda_n 
\gamma_5(1-\bar{a}D/2) | \lambda_n  \rangle.
\end{eqnarray}
These modes have another interesting property,
\begin{eqnarray}
\langle \lambda_n | \gamma_5 | \lambda_n \rangle 
= \lambda_n/2,
\end{eqnarray}
which is again obtained from the Ginsparg-Wilson relation;
\begin{eqnarray}
0&=&
\langle \lambda_n | 
\{\gamma_5 (\gamma_5 \bar{a}D)+ (\gamma_5 \bar{a}D) \gamma_5 
-(\gamma_5 \bar{a}D)(\gamma_5 \bar{a}D) \}| \lambda_n \rangle
\nonumber\\
&=& 2\lambda_n \langle \lambda_n | \gamma_5 | \lambda_n \rangle
- \lambda_n^2 \langle \lambda_n | \lambda_n \rangle.
\end{eqnarray}
Now it is obvious to see 
\begin{eqnarray}
\label{eq:ASindex1}
0=\mbox{Tr} \gamma_5 &=&
\sum_{\lambda_n=0}\langle \lambda_n | \gamma_5 | \lambda_n \rangle 
+ \sum_{0 < |\lambda_n|< 2 }\langle \lambda_n | \gamma_5 | \lambda_n \rangle 
+\sum_{|\lambda_n|=2}\langle \lambda_n | \gamma_5 | \lambda_n \rangle 
\nonumber\\
&=&n_+-n_- +\sum_{0 < |\lambda_n|< 2 }\frac{\lambda_n}{2}  +N_+-N_-
\nonumber\\
&=&
n_+-n_- +N_+-N_-,
\end{eqnarray}
and 
\begin{eqnarray}
\mbox{Tr} \;\;\;\bar{a}\gamma_5 D&=&
\sum_{0 < |\lambda_n|< 2 } \lambda_n
+\sum_{|\lambda_n|=2} 2
\nonumber\\
&=&
2(N_+-N_-) \;\;\;=\;\;\; 2(n_- -n_+) \;\;\;\equiv \;\;\; -2Q,
\end{eqnarray}
where $N_{\pm}$ denotes the number of $\lambda_n=\pm 2$ modes and
$n_{\pm}$ is that of the zero modes with $\pm$ chirality.
It is known that the equation
\begin{eqnarray}
\label{eq:ASindex2}
-\frac{1}{2 a^4}\mbox{tr}_c\mbox{tr}_s
\gamma_5 \bar{a}D(x,x) &=& \frac{1}{32\pi^2} \mbox{tr}_c
\epsilon_{\mu\nu\rho\sigma}F_{\mu\nu}(x)F_{\rho\sigma}(x) + O(a^2),
\end{eqnarray}
can be obtained with the perturbative expansion  
$P_{\mu\nu}(x)=e^{ia^2(F_{\mu\nu}(x)+O(a))}$.
Eq.(\ref{eq:ASindex1}) and Eq.(\ref{eq:ASindex2}) show
that the Atiyah-Singer index theorem is restored 
in the continuum limit.
In this way, the overlap Dirac operator properly establishes
the quantum aspects of chiral symmetry, 
the anomaly or the index theorem,
on the lattice. \\

It is, however, not difficult to see that 
the overlap Dirac operator is not a smooth function
of link variables.
Consider two gauge configurations, one of which 
$U^0_{\mu}(x)=e^{iA^0_{\mu}(x)}$ gives the index $Q=0$ 
and another is $U^1_{\mu}(x)=e^{iA^1_{\mu}(x)}$ 
with $Q=1$, where $A^i_{\mu}(x)$ denotes Lie algebra of $SU(3)$.
One obtains a smooth path which connects 
$U^0_{\mu}(x)$ and $U^1_{\mu}(x)$, for example, 
$U^t_{\mu}(x)=e^{i(tA^1_{\mu}(x)+(1-t)A^0_{\mu}(x))}$, while
$Q$ can take integers only.
There must be at least one jump 
from $Q=0$ to $Q=1$,
at some $t$ along the path, $(0,1)$.
This jump of $Q$ indicates
non-smoothness of the overlap Dirac operator.
Hern\'andez {\it et al.} \cite{Hernandez:1998et}
showed that this discontinuity
occurs exactly when $H_W$ has zero mode.
They also proved that if the configuration space is 
limited such that $|H_W|>u$ holds for a certain positive number $u$,   
then the locality is also 
guaranteed\footnote{Recently it is discussed that the mobility edge of $H_W$ plays
a more important role for the locality property. See refs.
\cite{Svetitsky:2005qa,Golterman:2005xa,Golterman:2005fe,
Golterman:2004cy,Golterman:2003qe}.}.
To see what happens when an eigenvalue crosses $H_W=0$,
let us rewrite the overlap Dirac operator,
\begin{equation}
  \label{eq:ov2}
  D = \frac{1}{\bar{a}} 
  \left( 1 + \frac{(aD_W-1-s)}
{\sqrt{(aD_W-1-s)^{\dagger}(aD_W-1-s)}} \right).
  \;\;\;
\end{equation}
Since every complex eigenvalue $\lambda^{D_W}$ 
of $(aD_W-1-s)$ makes 
a pair with its complex conjugate $\lambda^{D_W*}$, 
only real modes can pass through $(aD_W-1-s)=0$ or $H_W=0$.
Note that any real modes are chiral;
$[D_W, \gamma_5]|\lambda^{D_W}\rangle=0$, 
and these real modes are the eigenstates of $D$ which eigenvalues 
take 0 or $2/\bar{a}$ only, 
which means that crossing $H_W=0$ is always accompanied by 
topology changes.

In the numerical studies, 
there is a very crucial advantage of suppressing 
the small eigenvalues of $|H_W|$ .
In order to implement the overlap Dirac operator, 
the sign function in Eq.(\ref{eq:ov}) has to be 
expressed in the polynomial or rational expansion;
\begin{eqnarray}
\mbox{sgn}(x)=\frac{x}{\sqrt{x^2}}&=& a_0 + a_1 x + a_2 x^2  + \cdots 
\;\;\;\;\;\mbox{(polynomial)} \nonumber\\
&& \mbox{or}\nonumber\\
&=& b_0 \frac{(x-b_1)(x-b_2)\cdots}{(x-c_1)(x-c_2)\cdots}
\;\;\;\;\;\mbox{(rational)},
\end{eqnarray}
whose errors are controlled to some desired accuracy.
The order of polynomial or rational function one needs
is known to be a monotonous increasing function 
of the condition number $\kappa$:
\begin{equation}
\kappa \equiv \frac{\lambda_{\mathrm{max}}}{\lambda_{\mathrm{min}}},
\end{equation}
where $\lambda_{\mathrm{min}}$ and $\lambda_{\mathrm{max}}$ denote
the lowest and the highest eigenvalues respectively.
For example, the accuracy of the Chebyshev polynomial approximation
of sign function, $\mbox{sgn}_{\mathrm{Cheb}}(aH_W)$, with 
degree $N_{\mathrm{poly}}$ is empirically known as 
$\sim A\exp (-BN_{\mathrm{poly}}/\kappa)$,
where $A$ and $B$ are constants \cite{Giusti:2002sm}.
Apparently the approximation of finite order $N_{\mathrm{poly}}$ would
break down when $\lambda_{\mathrm{min}}=1/\kappa=0$.
In the hybrid Monte Carlo simulation with the dynamical 
overlap fermion (See appendix \ref{sec:HMC}.)\cite{Duane:1987de}, 
the problem would be worse, since one has to
consider the discontinuity of the fermion determinant
\cite{Fodor:2003bh,
Arnold:2003sx,
Cundy:2004pz,
Cundy:2005pi,
DeGrand:2004nq,
DeGrand:2005vb,Schaefer:2005qg}.
Every molecular dynamics step in the simulation trajectory, 
one would have to monitor the nearest zero-mode of $H_W$, 
and judge if the topology change occurs or not. 
Then careful recalculation of the link updates has to be done to
determine the simulation trajectory to enter the other
topological sector (refraction) or go back to the previous sector
(reflection). This procedure would require an enormous numerical 
cost when one tries the Monte Carlo simulation in a large volume.
If one omits this step it would make the acceptance rate very low
due to overlooking sudden jumps of the determinant. 
Keeping topology and assuring smoothness of the
determinant allows us to avoid this procedure.
Thus, to exclude configurations which gives 
$H_W\sim 0$ is useful not only theoretically to construct   
a sound quantum lattice theory with a smooth fermion determinant
(in particular, it is essential for
the chiral gauge theory
\cite{
Luscher:1998kn,
Luscher:1998du,
Luscher:1999un,
Luscher:2000zd,
Fujiwara:1999fi,
Fujiwara:1999fj,
Suzuki:2000ii,
Adams:2000yi,
Kikukawa:2000kd,
Igarashi:2000zi,
Kikukawa:2001mw,
Igarashi:2002zz,
Kadoh:2003ii,
Kadoh:2004uu,
Matsui:2004dc,
Kadoh:2005fa} 
and also applied to the supersymmetric theory
\cite{
Sugino:2004qd,
Sugino:2004uv,
Suzuki:2005dx,
Sugino:2006uf}
or non-commutative spaces
\cite{
Nagao:2005st,
Aoki:2006sb
}
.),
but also in practically, to reduce the numerical cost of 
calculating the overlap Dirac operator both in quenched and 
unquenched simulations.
Here we would like to make two proposals that would 
avoid the appearance of $H_W\sim 0$ modes or topology changes.
One is the modification of the gauge action and 
the another is the additional fermion action.\\

For the former solution,
one can construct a gauge action, which generates link variables
respecting the ``admissibility'' bound Eq.(\ref{eq:admi})
\cite{Luscher:1998du,
Shcheredin:2004xa,Bietenholz:2004mq,Shcheredin:2005ma,
Nagai:2005fz,Fukaya:2005cw,Bietenholz:2005rd}. 
A simplest example is
\begin{equation}
  \label{eq:admiaction}
  S_G = \left\{
    \begin{array}{ll}\displaystyle
      \beta\sum_{P}\frac{1-\mbox{ReTr}P_{\mu\nu}(x)/3}
      {1-(1-\mbox{ReTr}P_{\mu\nu}(x)/3)/\epsilon},
      & \;\;\; \mbox{when} \;\;
      1-\mbox{ReTr}P_{\mu\nu}(x)/3 < \epsilon,
      \\
      \infty & \;\;\; \mbox{otherwise}
    \end{array}
    \right..
\end{equation}
Every ``admissible'' gauge configuration with very small 
$\epsilon$ keeps $|H_W|>0$ and therefore,
keeps topological structure of gauge fields.
In order to explain
how ``admissible'' gauge fields can 
preserve the topological charge,
the best example would be the U(1) gauge theory 
in two-dimensions, 
for which we can define an exact geometrical definition of the
topological charge \cite{Luscher:1998kn,Luscher:1998du}
\begin{eqnarray}
  Q_{\mathrm{geo}} &=& \frac{1}{2\pi}
  a^2\sum_{x}\frac{1}{2}\epsilon_{\mu\nu}
  F^{\mathrm{lat}}_{\mu\nu}(x),\nonumber\\
  a^2F^{\mathrm{lat}}_{\mu\nu}(x)&=&
  -i\ln (P_{\mu\nu}(x)),\;\;\; -\pi <
  a^2F^{\mathrm{lat}}_{\mu\nu}(x)\leq \pi.
\end{eqnarray}
$P_{\mu\nu}(x)$ denotes the plaquette in the U(1) theory.
In two dimensions, $Q_{\mathrm{geo}}$ gives an integer on
the lattices with the periodic boundary condition.
Since the jump from 
$F^{\mathrm{lat}}_{\mu\nu}(x)=-\pi$ to 
$F^{\mathrm{lat}}_{\mu\nu}(x)=+\pi$ is allowed,
the topology can change easily.
It is the U(1) version of the L\"uscher's admissibility bound
\begin{equation}
  1-\mbox{Re} P_{\mu\nu}(x) < \epsilon,
\end{equation}
with $\epsilon < 2$, that can prevent these topology
changes because the point 
$F^{\mathrm{lat}}_{\mu\nu}(x) = \pm\pi$
is excluded under this condition.
Furthermore, it can be shown that $Q_{\mathrm{geo}}$ is
equivalent to the index of the overlap fermion (with $s=0$) 
if $\epsilon < 1/5$ is satisfied. 
For the non-abelian gauge theories in 
higher dimensions, we do not have the exact geometrical
definition of the topological charge.
But it is quite natural to assume 
that a similar mechanism concerning the compactness 
of the link variables allows us to preserve the index of 
the overlap-Dirac operator for very small $\epsilon$.
Note, however, that the value of $\epsilon\sim 1/20$ is too
tight for the simulation with the cutoff around 2.0GeV.
For practical purposes, we would like to try
much larger $\epsilon$, as we still expect that the topology 
is stabilized well in practical sampling of gauge
configurations, even though the topology change would not be
prevented completely. 
It is also notable that the difference between the gauge action 
Eq.(\ref{eq:admiaction}) and the Wilson plaquette action is 
only of $O(a^4)$ and one can safely take the continuum limit.
Also, the positivity which would be lost \cite{Creutz:2004ir}
at cut-off scale by the restriction on the configuration space,
is restored as $\epsilon/a^2\to \infty$.

For another solution to suppress small eigenvalues of $|H_W|$, 
the additional fermion determinant
\begin{equation}
\label{eq:nwf}
\det a^2H_W^2
\end{equation}
could be effective \cite{Vranas:1999rz, Izubuchi:2002pq, private}.
This determinant would prevent 
the appearance of near zero-modes of $H_W$, by
giving small Boltzmann weights to the configurations 
which have $H_W\sim 0$ modes.
Since it is equivalent to adding 2-flavor Wilson fermion
with a large negative mass at cut-off scale, $-(1+s)/a$, 
it would not affect the low-energy physics, 
and be decoupled like other fermion doublers.
This additional determinant may require more numerical cost 
than that of the gauge action, Eq.(\ref{eq:admiaction}),
but it would be negligible compared to the cost of 
the dynamical overlap fermions.

\newpage
\section{Lattice simulations}\label{sec:simulation}

In this section, we explain our setups for the numerical simulations.

\subsection{Quenched QCD with admissible gauge fields}\label{sec:quenchsim}

Although several types of the gauge action that generate
the ``admissible'' gauge fields satisfying the bound 
Eq.(\ref{eq:admi}) are proposed
\cite{Shcheredin:2004xa,Bietenholz:2004mq, Bietenholz:2005rd},
we take the simplest choice: Eq.(\ref{eq:admiaction}).
We use three values of $1/\epsilon$: 1, 2/3, and 0.
Note that $1/\epsilon=0$ corresponds to the conventional
Wilson plaquette gauge action.
The value $1/\epsilon = 2/3$ is the boundary,
below this value the gauge links can take any value in the 
gauge group $SU(3)$ (all configurations are admissible.) 
and the positivity is guaranteed
\cite{Creutz:2004ir}.

The link fields are generated with the standard hybrid
Monte Carlo algorithm \cite{Duane:1987de} (See appendix \ref{sec:HMC}).
The molecular dynamics step size $\Delta \tau$ is taken 
in the range 0.01--0.02 and the number of steps in an unit
trajectory, $N_{mds}$, is 20--40. 
Every molecular dynamics step, we check whether the
condition $1-\mbox{ReTr}P_{\mu\nu}(x)/3 < \epsilon$ is
satisfied or not.
In fact, 
no violation of this bound was observed 
in our simulations.
For thermalization, we discarded at least 2000 trajectories
before measuring observables.

To generate topologically non-trivial gauge configurations,
we use the initial link variables,
\begin{eqnarray}
U_1(x)=\left(
\begin{array}{l}
e^{2\pi ix_4Q_{\mathrm{init}}/L^2}\;\;\;\;\;\;\;\;\;\;\\
\;\;\;\;\;\;\;\;1\\
\;\;\;\;\;\;\;\;\;\;\;\;\;
e^{-2\pi ix_4Q_{\mathrm{init}}/L^2}
\end{array}
\right),&
U_2(x)=\left(
\begin{array}{l}
1\;\;\;\;\;\;\;\;\;\;\\
\;\;\;\;\;\;\;\;
e^{2\pi i x_3 \delta_{x_2,L-1}/L}\\
\;\;\;\;\;\;\;\;\;\;\;\;\;\;\;\;\;
e^{-2\pi ix_3 \delta_{x_2,L-1}/L}
\end{array}
\right),\nonumber\\
U_3(x)=\left(
\begin{array}{l}
1\;\;\;\;\;\;\;\;\;\;\\
\;\;\;\;\;\;e^{-2\pi ix_2/L^2}\\
\;\;\;\;\;\;\;\;\;\;\;\;\;\;\;\;\;
e^{2\pi ix_2/L^2}
\end{array}
\right),&
U_4(x)=\left(
\begin{array}{l}
e^{-2\pi i x_1 Q_{\mathrm{init}}\delta_{x_4,L-1}/L}\\
\;\;\;\;\;\;\;\;\;1\\
\;\;\;\;\;\;\;\;\;\;\;\;\;\;\;\;\;
e^{2\pi ix_1 Q_{\mathrm{init}}\delta_{x_4,L-1}/L}
\end{array}
\right),
\end{eqnarray}
which is a discretized version of a classical solution 
with the topological charge $Q=Q_{\mathrm{init}}$ on
a four-dimensional torus \cite{Gonzalez-Arroyo:1997uj}.
We confirmed that the topological charge assigned in this way
agrees with the index of the overlap operator with $s$ = 0.6.

The simulation parameters and the plaquette 
expectation values (for the run with the initial
configuration with $Q_{\mathrm{init}}=0$ or 1) are summarized 
in Table~\ref{tab:param}.
The length of unit trajectory is 0.07--0.4, and the step size
is chosen such that the acceptance rate is larger than
$\sim$~70\%. 

\begin{table}[ptb]
\begin{center}
\begin{tabular}{cccccccc}
\hline\hline
Lattice & $1/\epsilon$ & $\beta$ & $\Delta \tau$ & $N_{mds}$ &
acceptance & $Q_{\mathrm{init}}$ &plaquette \\
\hline
\\
$12^4$ & 1 & 1.0 & 0.01 & 40 & 89\% &0& 0.539127(9)\\
quenched& & 1.2 & 0.01 & 40 & 90\% &0& 0.566429(6)\\
& & 1.3 & 0.01 & 40 & 90\% &0& 0.578405(6)\\
& 2/3 & 2.25 & 0.01 & 40 & 93\% &0& 0.55102(1)\\
&  & 2.4 & 0.01 & 40 & 93\% &0& 0.56861(1)\\
&  & 2.55 & 0.01 & 40 & 93\% &0& 0.58435(1)\\
& 0 & 5.8 & 0.02 & 20 & 69\% &0& 0.56763(5)\\
&  & 5.9 & 0.02 & 20 & 69\% &0& 0.58190(3)\\
&  & 6.0 & 0.02 & 20 & 68\% &0& 0.59364(2)\\
\\
$16^4$ 
& 1 & 1.3 & 0.01 & 20 & 82\% &0& 0.57840(1)\\
quenched& & 1.42 & 0.01 & 20 & 82\% &0& 0.59167(1)\\
& 2/3 & 2.55 & 0.01 & 20 & 88\% &0& 0.58428(2)\\
& & 2.7 & 0.01 & 20 & 87\% &0& 0.59862(1)\\
&0  & 6.0 & 0.01 & 20 & 89\% &0& 0.59382(5)\\
&  & 6.13 & 0.01 & 40 & 88\% &0& 0.60711(4)\\
\\
$20^4$ 
& 1 & 1.3 & 0.01 & 20 & 72\% &0& 0.57847(9)\\
quenched &  & 1.42 & 0.01 & 20 & 74\% &0& 0.59165(1)\\
& 2/3 & 2.55 & 0.01 & 20 & 82\% &0& 0.58438(2)\\
&  & 2.7 & 0.01 & 20 & 82\% &0& 0.59865(1)\\
& 0 & 6.0 & 0.015 & 20 & 53\% &0& 0.59382(4) \\
&  & 6.13 & 0.01& 20& 83\% &0& 0.60716(3)\\
\\
$14^4$ 
& 1 & 0.75 & 0.01 & 15 & 72\% &0& 0.52260(2)\\
with $\det H_W^2$& 2/3 & 1.8 & 0.01 & 15 & 87\% &0& 0.52915(3)\\
& 0 & 5.0 & 0.01 & 15 & 88\% &0& 0.55374(6)\\
\\
$16^4$ 
& 1 & 0.8 & 0.007 & 60 & 79\% &+1& 0.53091(1)\\
with $\det H_W^2$& 2/3 & 1.75 & 0.008 & 50 & 89\% &+1& 0.52227(3)\\
& 0 & 5.2 & 0.008 & 50 & 93\% &+1& 0.57577(3)\\
\\
\hline
\end{tabular}
\end{center}
\caption{
  Simulation parameters and the plaquette expectation values.
}
\label{tab:param}
\end{table}

\subsection{Cooling method to measure the topological charge}

\begin{figure*}[tbp]
  \centering
  \includegraphics[width=8cm]{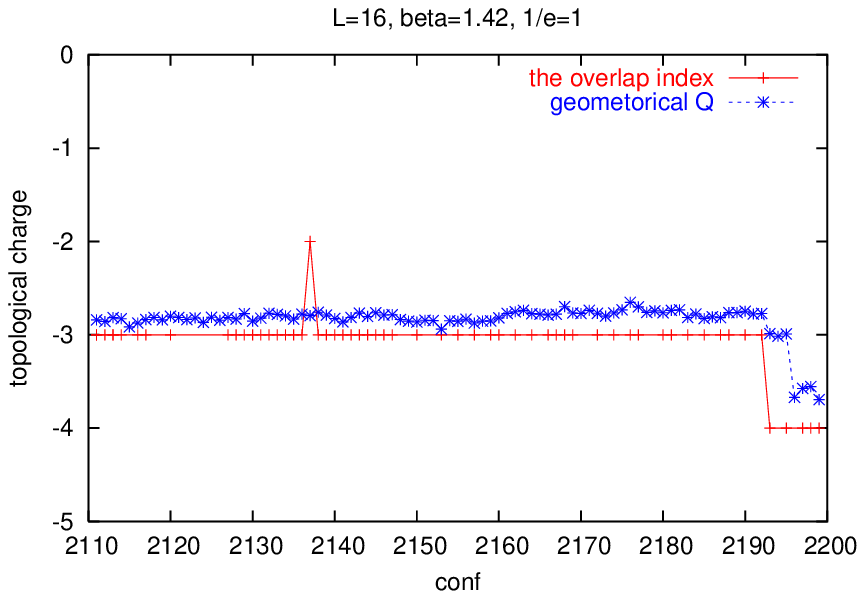}
  \includegraphics[width=8cm]{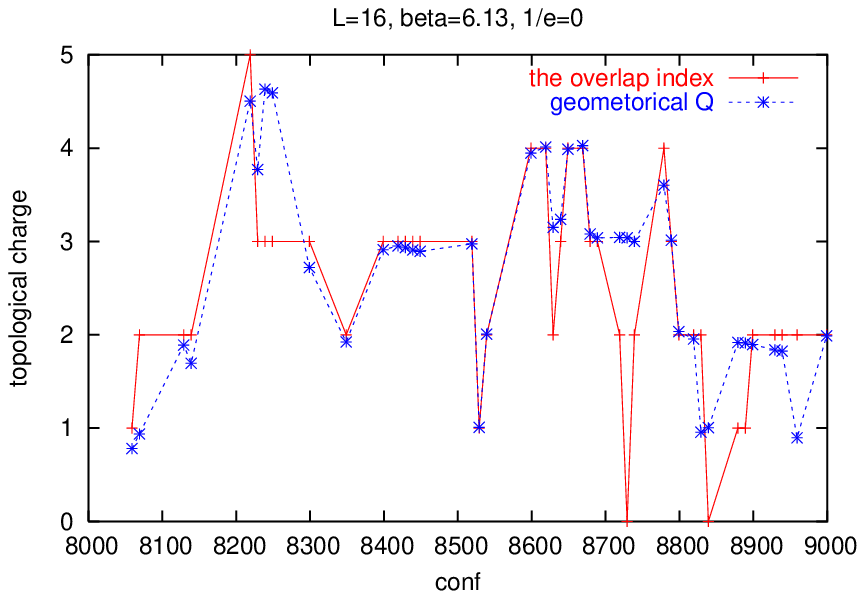}
  \caption{
    Comparison of the topological charge $Q$;
    the geometrical definition $Q=Z_{geo}Q_{geo}$ after 
    ``cooling'', and the index of the overlap-Dirac operator 
    with $s=0.6$. 
        The agreement is better for $1/\epsilon=1$ (left) than
    for $1/\epsilon=0$ (right).
  }
  \label{fig:Qhis}
\end{figure*}

In order to measure the topological charge, we develop a new
``cooling'' method.
It is achieved by the hybrid Monte Carlo steps with an
exponentially increasing coupling,
$\beta_{\mathrm{cool}}$, and decreasing step size,
$\Delta\tau_{\mathrm{cool}}$, as a function of trajectory,
$n_t$, {\it i.e.} 
\begin{eqnarray}
  \beta_{\mathrm{cool}}
  &=&\beta_{\mathrm{init}} \times(1.5)^{n_t},
  \nonumber\\
  \Delta\tau_{\mathrm{cool}} &=&
  \Delta\tau_{\mathrm{init}} \times(1.5)^{-n_t/2},
\end{eqnarray}
using the gauge action Eq.(\ref{eq:admiaction}) 
with fixed $1/\epsilon=1/\epsilon_{\mathrm{cool}}$.
Note that 
$\sqrt{\beta_{\mathrm{cool}}}\Delta\tau_{\mathrm{cool}}$
is fixed in order to keep each step small.
In this method, one can ``cool'' the configuration smoothly,
keeping the admissibility bound, Eq.(\ref{eq:admi}), with
$1/\epsilon=1/\epsilon_{\mathrm{cool}}$. 
The parameters are chosen as
$(\beta_{\mathrm{init}},\Delta\tau_{\mathrm{init}},1/\epsilon_{\mathrm{cool}})$
= (2.0, 0.01, 1) for the configurations generated with
$1/\epsilon=1$, 
and (3.5, 0.01, 2/3) for the configurations with
$1/\epsilon=2/3$ or $1/\epsilon=0$. 
Even for the gauge fields generated with the Wilson plaquette
gauge action ($1/\epsilon=0$), 
the condition $1/\epsilon_{\mathrm{cool}}=2/3$ can be used
because it allows all values of SU(3).
The link variables are cooled down after 50--200 steps
close to a classical solution in each topological sector. 
Then the geometrical definition of the topological charge
\cite{Hoek:1986nd},
\begin{equation}
  \label{eq:topgeo}
  Q_{geo}
  \equiv
  \frac{1}{32\pi^2}
  \sum_{x}\epsilon^{\mu\nu\rho\sigma} \mbox{ReTr}
  \left(P_{\mu\nu}(x)P_{\rho\sigma}(x)\right),
\end{equation}
for the ``cooled'' configuration gives a number close to an
integer times a universal factor $Z_{geo}^{-1}$, namely, 
$Q=Z_{geo}Q_{geo}$ is close to an integer.
The value $Z_{geo}= 0.923(4)$ is calculated 
through {\it would-be} $Q=1$ gauge configurations.
As Fig.~\ref{fig:Qhis} shows, the topological charge
$Q$ is consistent with the index of the overlap-Dirac
operator with $s$ = 0.6, 
which is measured as described
in Section~\ref{sec:setupov}.
The consistency seems better for $1/\epsilon=1$ than for the
standard Wilson plaquette action ($1/\epsilon=0$).

\subsection{Large negative mass Wilson fermion}

Now let us discuss the lattice QCD with the determinant of
Wilson fermion with negative mass at cut-off scale $-(1+s)/a$ 
as seen in Eq.(\ref{eq:nwf}).
This is a more direct way 
to suppress the small eigenvalue of $H_W$, 
or keep topological charge, since the determinant 
$\det a^2H_W^2$ would give a very 
small Boltzmann weight if $H_W$ touches zero.
For the gauge action, we use the topology conserving action
Eq.(\ref{eq:admiaction}) again with three values of 
$1/\epsilon$: 1, 2/3, and 0.
We set the hopping parameter $\kappa_{hop}=1/(8-(1+s))=0.208$, 
which is known to be an optimal choice for
the locality of the overlap Dirac operator \cite{Bietenholz:2003mi}, 
and the conventional pseudo-fermion method 
is performed in the hybrid Monte Carlo steps to calculate
the fermion determinant.
For thermalization, we performed more than 500 trajectories
before measuring observables.

The topological charge is evaluated in the same way as explained
in the previous subsection, namely, by the geometrical definition
Eq.(\ref{eq:topgeo})
after 100-200 of the quenched HMC steps with increasing $\beta$, 
switching off the Wilson fermion. 
Although the topological charge is expected to be more stable 
in this method, there may be a large scaling violation.
Careful comparison with quenched studies without the determinant
would be important.

The simulation parameters are summarized in Table.~\ref{tab:param}

\subsection{Numerical implementation of the overlap Dirac operator}
\label{sec:setupov}
\begin{figure}[tbp]
  \centering
  \includegraphics[width=10cm]{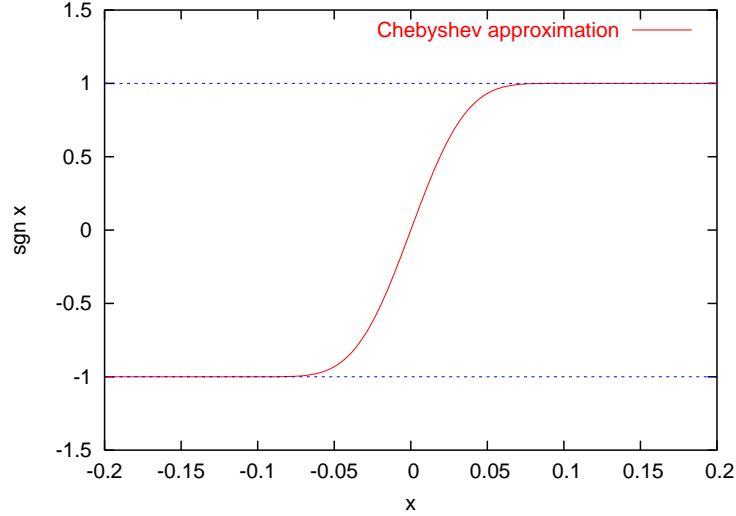}
  \caption{
 The Chebyshev approximation of sgn function.
 We set $N_{\mathrm{poly}}=60$ and the range $[a=0.1, b=1]$. 
  }
  \label{fig:Chevfig}
\end{figure}

For the overlap Dirac operator, Eq.(\ref{eq:ov}), we use 
the Wilson Dirac Hamiltonian $aH_W=\gamma_5(aD_W-1-s)$ with $s=0.6$,
which is empirically known as an optimal choice to suppress the 
small eigen-modes of $|aH_W|$ at $1/a\sim$2.0GeV 
in quenched QCD.
The sign function is approximated by 
the Chebyshev polynomial (See Fig.~\ref{fig:Chevfig});
\begin{eqnarray}
\label{eq:cheb}
\mbox{sgn}_{\mathrm{Cheb}} (aH_W)
= aH_W/ 
\mbox{sqrt}_{\mathrm{Cheb}}(a^2H_W^2), 
\end{eqnarray}
where the function $\mbox{sqrt}_{\mathrm{Cheb}}(a^2H_W^2)$ in the range
[a,b] is written
\begin{eqnarray}
1/\mbox{sqrt}_{\mathrm{Cheb}}(x)&\equiv &
\sum_{i=0}^{N_{\mathrm{poly}}} c_i T_i(x^{\prime}),\;\;\;\;\;
 x^{\prime}=\frac{b+a}{b-a}-\frac{2}{b-a}x\nonumber\\
c_i &=& \frac{2}{\pi}\int_0^{\pi} 
\frac{\sqrt{2}}{\sqrt{(b+a)-(b-a)\cos \phi}}\cos n\phi d\phi ,\nonumber\\
T_i(t)&=&\cos n\phi,\;\;\;\;\;t=\cos \phi.
\end{eqnarray}
Note that the Chebyshev basis $T_i(x)$ (of order $i$) satisfies
the orthogonal relation
\begin{equation}
\int_{-1}^1 \frac{dt}{\sqrt{1-t^2}} T_i(t)T_j(t) = \delta_{ij}. 
\end{equation}
We use the numerical package ARPACK \cite{ARPACK}, which
implements the implicitly restarted Arnoldi method,
 to measure the lowest eigenvalue $a = \lambda_{\mathrm{min}}^2$ and
the highest $b = \lambda_{\mathrm{max}}^2$.
In some cases, we calculate the 10 lowest modes explicitly,
project them out of $H_W$, and
use the approximation Eq.(\ref{eq:cheb}) after the projection,
in the range $[a_{11}, b]$ where we use the 11th eigenvalue
for the lower bound; $a_{11}=\lambda_{11}^2$.  

\begin{figure}[tbp]
  \centering
  \includegraphics[width=8cm]{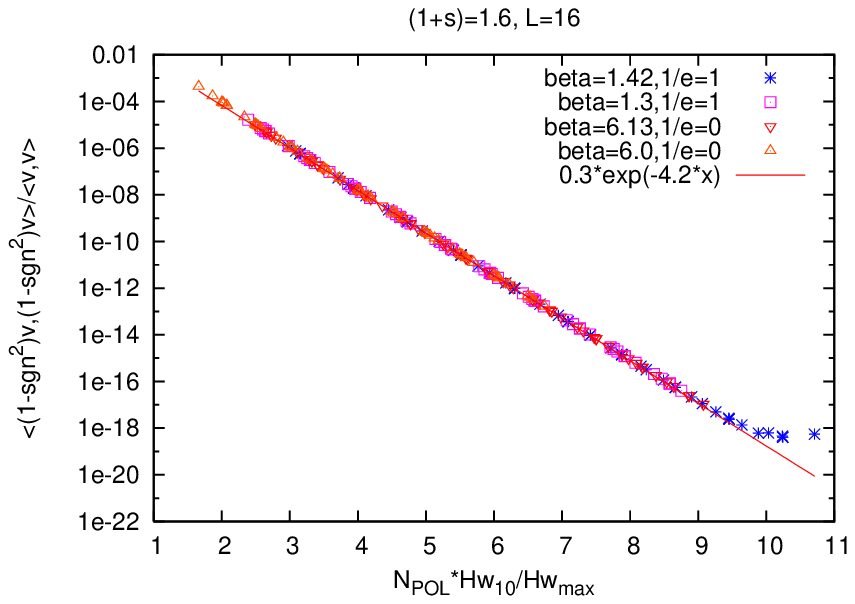}
  \includegraphics[width=8cm]{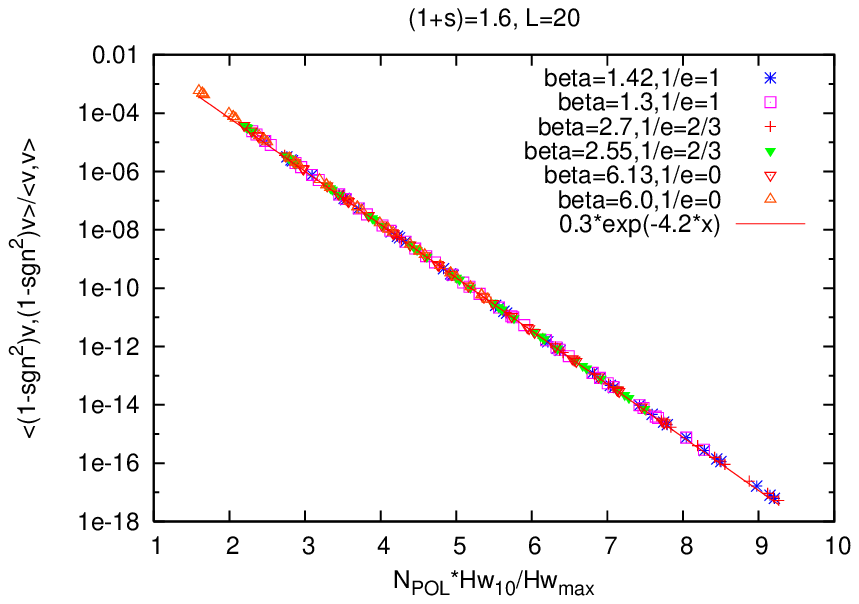}
  \caption{
    The accuracy Eq.(\ref{eq:Chebacc3}) as a function of
    $N_{\mathrm{pol}}/\kappa$ on a 16$^4$ lattice (left) 
    and a 20$^4$ lattice.
    We use 4 gauge configurations and 10 values of
    $N_{\mathrm{pol}}$ = 60--195 for each parameter set.
  }
  \label{fig:ABdeterm}
\end{figure}

In order to determine the order of the polynomial, $N_{\mathrm{poly}}$, 
it is important to note that the accuracy is expressed as a function of
$N_{\mathrm{poly}}$ and the condition number 
$\kappa=\lambda_{\mathrm{max}}/\lambda_{\mathrm{min}}=\sqrt{b/a}$,
\begin{equation}
  \label{eq:Chebacc3}
  \frac{\langle v |(1-\mbox{sgn}_{\mathrm{Cheb}}^2(aH_W))^2|v \rangle}
  {\langle v|v\rangle}  
  \sim A \exp (-BN_{\mathrm{poly}}/\kappa),
\end{equation} 
for a random vector $|v\rangle$, where the constants
$A\sim 0.3$ and $B\sim 4.2$ are
turned out to be independent of $\beta$, 
$\epsilon$ and the lattice size $L$, as seen in 
Fig.~\ref{fig:ABdeterm}.
Note that the lhs. of Eq.(\ref{eq:Chebacc3}) 
is zero when the sign function is exact and the approximation
breaks down when $1/\kappa=0$.

We also use ARPACK to calculate the eigenvalues
of $P_-DP_-$ and $P_+DP_+$, where $P_{\pm}=(1\pm\gamma_5)/2$.
The index can be obtained from the number of zero-modes of 
chirally projected Dirac operators.
To construct the non-zero mode, $|\lambda \rangle$, of $D$, 
the formula
\begin{equation}
|\lambda \rangle = 
\frac{D-\lambda^*}{\mbox{Im}\lambda}P_{+}|\lambda \rangle
=\frac{D-\lambda^*}{\mbox{Im}\lambda}P_{-}|\lambda \rangle,
\end{equation}
is useful.
 
\newpage
\section{Wilson loops and the static quark potential}\label{sec:qpotential}

In order to confirm the practical feasibility 
of the topology conserving actions,
Eq.(\ref{eq:admiaction}) and Eq.(\ref{eq:nwf}), 
a careful comparison with the Wilson plaquette action 
should be done.  
In this section, 
we study gluonic quantities or Wilson loops to determine 
the lattice spacing, to estimate the scaling violations 
due to the change in the actions, 
and to test the perturbation theory 
with tadpole improved coupling.
Then we can judge whether our naive expectation that
$1/\epsilon$ effect is small (of order $O(a^4)$) or the
effect of the determinant, $\det H_W^2$, is negligible,  
is true or not.

\subsection{The static quark potential}

In the following, we assume that the topology of the gauge
field does not affect the Wilson loops when the lattice size
is large, say $L>1$ fm, and choose the run with $Q=0$ or $Q=1$ 
initial configuration for the measurement.

Wilson loops, $W(\vec{r},t)$'s, are measured
using the smearing technique according to \cite{Bali:1992ab}, 
where the spatial separation $\vec{r}/a$ is taken to be an
integer multiples of vectors
$\vec{v}=(1,0,0)$, $(1,1,0)$,
$(2,1,0)$, $(1,1,1)$, $(2,1,1)$ and  $(2,2,1)$.
Assuming the Wilson loop is an exponential function 
\begin{equation}
\langle W(\vec{r},t)\rangle =\exp(- V(\vec{r}) t),
\end{equation}
for large $t/a$, 
we extract the static quark potential $a V(\vec{r})$.
The measurements are done every 20 trajectories and 
the errors are estimated by the jackknife method.

As a reference scale to determine the lattice spacing, 
we evaluate the Sommer scales
$r_0$ and $r_c$ \cite{Guagnelli:1998ud,Necco:2001xg} defined
by $r_0^2 F(r_0)=1.65$ and $r_c^2 F(r_c)=0.65$, respectively.
The force $F(r)$ on the lattice is given by a
differentiation of the potential 
in the direction of $\vec{u}/a=(1,0,0)$;
\begin{equation}
  a^2 F(r_I)
  =
  \frac{aV(\vec{r})-aV(\vec{r}-\vec{u})}{|\vec{u}/a|},
\end{equation}
for $\vec{r}/|\vec{r}|=(1,0,0)$.
$r_I$ is introduced to cancel $O(a)$ errors in
the short distances at tree level,
\begin{eqnarray}
  \label{eq:r_I}
  \frac{1}{4\pi (r_I/a)^2}&=&
  -\frac{aG(\vec{r})-aG(\vec{r}-\vec{u})}{|\vec{u}/a|},
  \nonumber\\
  aG(\vec{r})&=&\int^{\pi}_{-\pi}\frac{d^3k}{(2\pi)^3}
  \frac{\prod^3_{j=1}\cos(r_jk_j/a) }{4\sum_{j=1}^3 \sin^2(k_j/2)}.
\end{eqnarray}

In Table~\ref{tab:sommerscale} we present the values of the
Sommer scales $r_0/a$, $r_c/a$, their ratio $r_c/r_0$, and the
lattice spacing $a$ (We assume $r_0\sim 0.5$fm.).
The data of $aV(\vec{r})$ and $r_I^2F(r_I)$ in
the case that $\vec{r}/a$ is an integer multiples of
$\vec{u}/a$ are given in 
Tables~\ref{tab:VF1}, \ref{tab:VF2} and \ref{tab:VF3}.

\begin{table}[tbp]
\begin{center}
\begin{tabular}{cccccccc}
\hline\hline
Lattice size & $1/\epsilon$ & $\beta$ & statistics & $r_0/a$ &
$r_c/a$ & $r_c/r_0$ & $a$ \\
\hline\\
$12^4$ & 1 & 1.0 & 3800 & 3.257(30) & 1.7081(50)& 0.5244(52)& 0.15fm\\
quenched         & & 1.2 & 3800 &4.555(73)&2.319(10)&0.5091(81)&0.11fm\\
         & & 1.3 & 3800 &5.140(50)&2.710(14)&0.5272(53)&0.1fm\\
     & 2/3 & 2.25 & 3800 &3.498(24)&1.8304(60)&0.5233(41)&0.14fm\\
        &  & 2.4 & 3800 &4.386(53)&2.254(10)&0.5141(61)&0.11fm\\
        &  & 2.55 & 3800 & 5.433(72) & 2.809(18)&0.5170(67)&0.09fm \\\\
$16^4$ & 1 & 1.3 & 2300 & 5.240(96) & 2.686(13) & 0.5126(98)&0.1fm \\
quenched    & & 1.42 & 2247 & 6.240(89) & 3.270(26)    &0.5241(83)&0.08fm \\ 
     & 2/3 & 2.55 & 1950 & 5.290(69) & 2.738(15) & 0.5174(72)&0.09fm \\
         & & 2.7 & 2150 & 6.559(76) & 3.382(22) & 0.5156(65)&0.08fm \\
\\$14^4$ 
& 1 & 0.75 & 162 & 4.24(15) & 2.240(37) & 0.528(24)& 0.12fm \\
with $\det H_W^2$    
     & 2/3 & 1.8 & 261 & 4.94(19) & 2.361(26) & 0.478(19)&0.10fm \\
         & 0 & 5.0 & 162 & 4.904(90) & 2.691(42) & 0.549(13)&0.10fm \\
\\$16^4$ 
& 1 & 0.8 & 207 & 4.81(17) & 2.442(48) & 0.508(20)&0.10fm \\
with $\det H_W^2$    
     & 2/3 & 1.75& 189 & 4.71(19) & 2.279(48) & 0.484(22)&0.11fm \\
         & 0 & 5.2 & 225 & 7.09(17) & 3.462(55) & 0.489(13)&0.07fm \\\\
\multicolumn{3}{l}{continuum limit \cite{Necco:2001xg}}
&&&& 0.5133(24)\\
\hline
\end{tabular}
\end{center}
\caption{
  Sommer scales $r_0/a$, $r_c/a$ and their ratio.
  Here we assume $r_0\sim$0.5fm.
}\label{tab:sommerscale}
\end{table}

\begin{table}[tbp]
\renewcommand{\arraystretch}{0.9}
\begin{tabular}{ccc|cc|cc}
\hline\hline
& $1/\epsilon=1$& & \multicolumn{2}{c|}{12$^4$ quenched} 
& \multicolumn{2}{c}{16$^4$ quenched} \\
$\beta$ & $r/a$ & $r_I/a$ & $aV(\vec{r})$ & $r_I^2F(r_I)$ 
& $aV(\vec{r})$ & $r_I^2F(r_I)$ \\
\hline
1.0 & 1 &      & 0.50459(20) &            & & \\
    & 2 & 1.36 & 0.77828(61) & 0.5056(10) & & \\
    & 3 & 2.28 & 0.9629(15)  & 0.9520(69) & & \\
    & 4 & 3.31 & 1.1176(27)  &  1.691(26) & & \\
    & 5 & 4.36 & 1.2623(45)  &  2.751(80) & & \\
    & 6 & 5.39 & 1.4052(77)  &  4.33(22)  & & \\
\hline
1.2 & 1 &      & 0.44877(16) &            & & \\
    & 2 & 1.36 & 0.65982(39) & 0.38993(65)& & \\
    & 3 & 2.28 & 0.78291(80) & 0.6346(34) & & \\
    & 4 & 3.31 & 0.8775(13)  & 1.034(10)  & & \\
    & 5 & 4.36 & 0.9588(29)  & 1.545(45)  & & \\
    & 6 & 5.39 & 1.0322(47)  & 2.23(12)   & & \\
\hline
1.3 & 1 &      & 0.42730(10) &             & 0.42709(20) &            \\
    & 2 & 1.36 & 0.61711(34) & 0.35252(99) & 0.61710(66) & 0.35099(68)\\
    & 3 & 2.28 & 0.72140(69) & 0.53909(48) & 0.72130(92) & 0.5490(29) \\
    & 4 & 3.31 & 0.7977(12)  & 0.848(14)   & 0.7961(15)  & 0.8325(81) \\
    & 5 & 4.36 & 0.8608(21)  & 1.240(36)   & 0.8583(23)  & 1.180(32)  \\
    & 6 & 5.39 & 0.9230(25)  & 1.887(85)   & 0.9150(27)  & 1.809(79)  \\
    & 7 & 6.41 &             &             & 0.9636(51)  & 1.93(24)   \\
    & 8 & 7.43 &             &             & 1.0215(51)  & 3.09(37)   \\
\hline
1.42 & 1 &     &             &             & 0.40443(15) & \\
    & 2 & 1.36 &             &             & 0.57416(43) & 0.31444(58)\\
    & 3 & 2.28 &             &             & 0.66091(75) & 0.4567(22)\\
    & 4 & 3.31 &             &             & 0.7200(12) & 0.6583(61)\\
    & 5 & 4.36 &             &             & 0.7691(17) & 0.940(14)\\
    & 6 & 5.39 &             &             & 0.8076(24) & 1.189(48)\\
    & 7 & 6.41 &             &             & 0.8457(30) & 1.675(64)\\
    & 8 & 7.43 &             &             & 0.8832(37) & 1.91(14)\\
\hline
\end{tabular}
\caption{
  Potential and force values in the case that $\vec{r}/a$
  is an integer multiples of the unit vector
  $\vec{u}/a=(1,0,0)$. 
  Results with $1/\epsilon$ = 1 (quenched).
}\label{tab:VF1}
\end{table}

\begin{table}[tbp]
\renewcommand{\arraystretch}{0.9}
\begin{tabular}{ccc|cc|cc}
\hline\hline
& $1/\epsilon = 2/3$
& & \multicolumn{2}{c|}{12$^4$ quenched} & 
\multicolumn{2}{c}{16$^4$ quenched} \\
$\beta$ & $r/a$ & $r_I/a$ & $aV(\vec{r})$ & $r_I^2F(r_I)$ 
& $aV(\vec{r})$ & $r_I^2F(r_I)$ \\
\hline
2.25 & 1 &      & 0.48470(15) & \\
     & 2 & 1.36 & 0.74012(57) & 0.47189(97) \\
     & 3 & 2.28 & 0.9077(13)  & 0.8640(56) \\
     & 4 & 3.31 & 1.0463(22)  & 1.515(21) \\
     & 5 & 4.36 & 1.1701(38)  & 2.353(64) \\
     & 6 & 5.39 & 1.2901(58)  & 3.64(15) \\
\hline
2.4 & 1 &      &   0.44908(12) & \\
    & 2 & 1.36 & 0.66434(41)   & 0.39770(70) \\
    & 3 & 2.28 & 0.79152(84)   & 0.6557(37) \\
    & 4 & 3.31 & 0.8889(15)    & 1.065(12) \\
    & 5 & 4.36 & 0.9749(23)    & 1.635(32) \\
    & 6 & 5.39 & 1.0541(30)    & 2.401(74) \\
\hline
2.55 & 1 &      & 0.42013(11) &             & 0.42042(16) &            \\
     & 2 & 1.36 & 0.60682(36) & 0.34493(58) & 0.60786(51) & 0.34590(72)\\
     & 3 & 2.28 & 0.70826(72) & 0.5230(28)  & 0.71227(95) & 0.5337(32) \\
     & 4 & 3.31 & 0.7806(13)  & 0.7913(90)  & 0.7878(16)  & 0.8211(93) \\
     & 5 & 4.36 & 0.8430(18)  & 1.187(18)   & 0.8538(22)  & 1.210(21)  \\
     & 6 & 5.39 & 0.8986(23)  & 1.686(37)   & 0.9157(29)  & 1.765(47)  \\
     & 7 & 6.41 &             &             & 0.9710(43)  & 2.229(84)  \\
     & 8 & 7.43 &             &             & 1.0266(52)  & 2.94(15)   \\
\hline
2.7 & 1 &      &            &             & 0.39590(15) & \\
    & 2 & 1.36 &            &             & 0.56100(44) & 0.30650(53)\\
    & 3 & 2.28 &            &             & 0.64733(62) & 0.4456(22)\\
    & 4 & 3.31 &            &             & 0.70527(90) & 0.6329(56)\\
    & 5 & 4.36 &            &             & 0.7528(14) & 0.907(14)\\
    & 6 & 5.39 &            &             & 0.7937(19) & 1.309(28)\\
    & 7 & 6.41 &            &             & 0.8321(24) & 1.531(44)\\
    & 8 & 7.43 &            &             & 0.8703(29) & 2.035(80)\\
\hline
\end{tabular}
\caption{
  Same as Table~\ref{tab:VF1}, but 
  with $1/\epsilon$ = 2/3 (quenched).
}\label{tab:VF2}
\end{table}

\begin{table}[tbp]
\renewcommand{\arraystretch}{0.9}
\begin{center}
\begin{tabular}{cc|cc|cc}
\hline\hline
$1/\epsilon=1$ &
 & \multicolumn{2}{c|}{14$^4$ $\det H_W^2$ $\beta=0.75$} & 
\multicolumn{2}{c}{16$^4$ $\det H_W^2$ $\beta=0.8$} \\
$r/a$ & $r_I/a$ & $aV(\vec{r})$ & $r_I^2F(r_I)$ 
& $aV(\vec{r})$ & $r_I^2F(r_I)$ \\
\hline
 1 &      & 0.50199(51) &             & 0.48708(37) & \\
      2 & 1.36 & 0.7198(18) & 0.4025(31) & 0.6943(13) &0.3829(22)\\
      3 & 2.28 & 0.8480(32) & 0.661(14)  & 0.8137(25)&0.616(14)\\
      4 & 3.31 & 0.9457(50)  & 1.068(49)  & 0.8936(42)&0.873(45)\\
      5 & 4.36 & 1.037(55)  & 1.74(12)   & 0.9681(57)&1.416(87)\\
      6 & 5.39 & 1.097(85)  & 1.83(26)   & 1.0334(85)&1.98(28)\\
      7 & 6.41 &             &            &1.110 (11)&3.01(37)\\
 8 & 7.43 &             &             & 1.151(14)&2.26(80)\\
\hline
\\
\hline\hline
$1/\epsilon=2/3$ &
 & \multicolumn{2}{c|}{14$^4$ $\det H_W^2$ $\beta=1.8$} & 
\multicolumn{2}{c}{16$^4$ $\det H_W^2$ $\beta=1.75$} \\
$r/a$ & $r_I/a$ & $aV(\vec{r})$ & $r_I^2F(r_I)$ 
& $aV(\vec{r})$ & $r_I^2F(r_I)$ \\
\hline
 1 &      & 0.49349(42)&             & 0.50630(46)& \\
      2 & 1.36 & 0.7125(15) &0.4046(25)&0.7363(18)&0.4249(31)\\
      3 & 2.28 & 0.8323(25) &0.6178(91)&0.8605(35) &0.640(20)\\
      4 & 3.31 & 0.9294(48) &1.062(40)&0.9680(68) & 1.175(65)\\
      5 & 4.36 & 1.0000(68) &1.34(12)& 1.0423(88)  &1.41(16)\\
      6 & 5.39 & 1.0655(92)  &1.99(24)&1.118(15)   &2.30(39)\\
      7 & 6.41 & &                   & 1.185(17)   &2.64(55)\\
 8 & 7.43 &      &                   & 1.261(24)   &4.1(1.2)\\
\hline
\\
\hline\hline
$1/\epsilon=0$ &
 & \multicolumn{2}{c|}{14$^4$ $\det H_W^2$ $\beta=5.0$} & 
\multicolumn{2}{c}{16$^4$ $\det H_W^2$ $\beta=5.2$} \\
$r/a$ & $r_I/a$ & $aV(\vec{r})$ & $r_I^2F(r_I)$ 
& $aV(\vec{r})$ & $r_I^2F(r_I)$ \\
\hline
 1 &      & 0.45795(42)&             & 0.42080(26)&\\
      2 & 1.36 & 0.6547(12)&0.3634(19)&0.58846(72)&0.3098(13)\\
      3 & 2.28 & 0.7606(22)&0.5466(84)&0.6722(12) &0.4318(48)\\
      4 & 3.31 & 0.8369(32)&0.833(28)&0.7246(16) &0.617(14)\\
      5 & 4.36 & 0.9000(51)&1.200(70)&0.7745(25)  &0.872(43)\\
      6 & 5.39 & 0.9730(75)& 2.21(13)&0.8120(27)  &1.137(60)\\
      7 & 6.41 &          &         &0.8426(37)  &1.21(10)\\
      8 & 7.43 &          &         &0.8762(39)  &1.84(15)\\
\hline
\\
\end{tabular}
\end{center}
\caption{
  Same as Table~\ref{tab:VF1}, with $\det H_W^2$.
}\label{tab:VF3}
\end{table}

\begin{figure}[tbp]
  \centering
  \includegraphics[width=10cm]{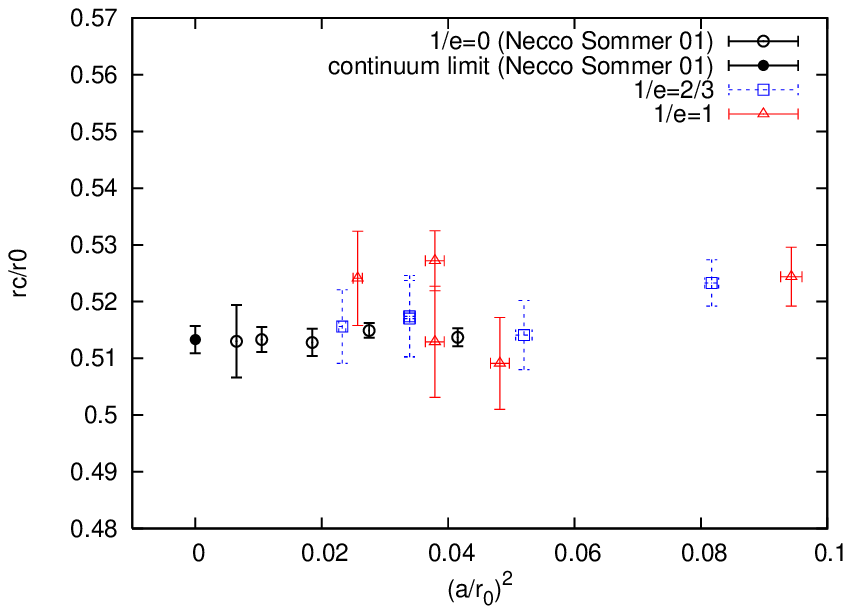}
  \caption{
    Quenched results (without $\det H_W^2$) of $r_c/r_0$.
    Squares and triangles are data for the topology
    conserving gauge action with $1/\epsilon$ = 2/3 and 1,
    respectively. 
    Open circles represent the Wilson plaquette
    gauge action (from \cite{Necco:2001xg}) and the filled
    circle is their continuum limit.
  }
  \label{fig:rcr0}
  \centering
  \includegraphics[width=10cm]{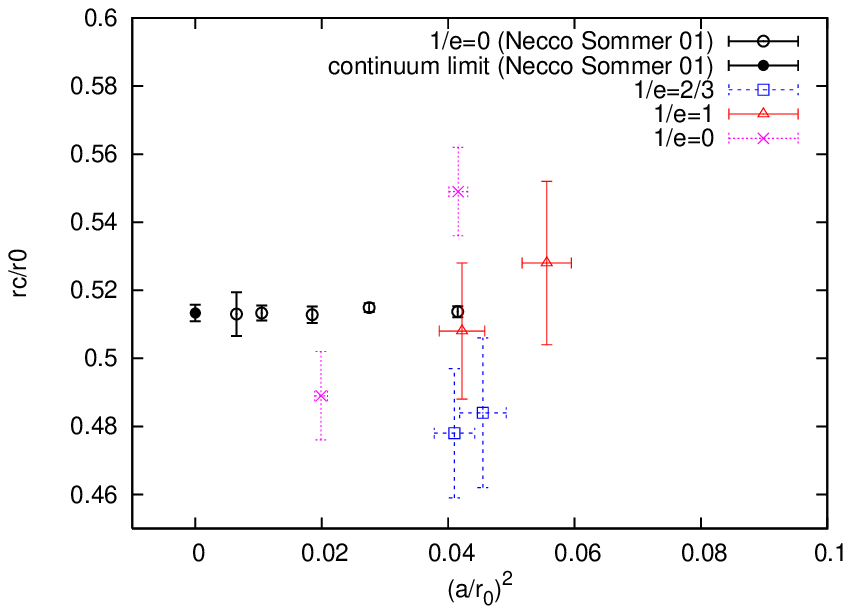}
  \caption{
 The same as Fig.\ref{fig:rcr0} with $\det H_W^2$.
  }
  \label{fig:rcr0det}
\end{figure}

$r_c/r_0$ is a good quantity to estimate the scaling violation, 
comparing with the value  $r_c/r_0=0.5133(24)$ 
in the continuum limit, which was obtained
with the plaquette action \cite{Necco:2001xg}.
Fig.~\ref{fig:rcr0} shows the $a^2$ dependence of 
this ratio with different values of $1/\epsilon$.
The quenched results with $1/\epsilon$ = 2/3 and 1 
agree very well with 
Ref \cite{Necco:2001xg} except for the coarsest lattice points
around $(a/r_0)^2\simeq$ 0.1.
Also in the case 
with $\det H_W^2$, as seen in Fig.~\ref{fig:rcr0det},
any large scaling violation has not been seen
although the statistics are much poorer.

\begin{figure*}[tbp]
  \centering
  \includegraphics[width=8cm]{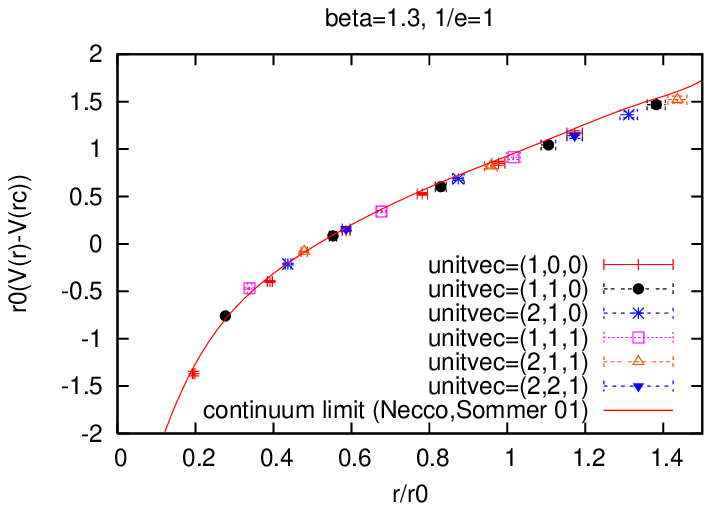}
  \includegraphics[width=8cm]{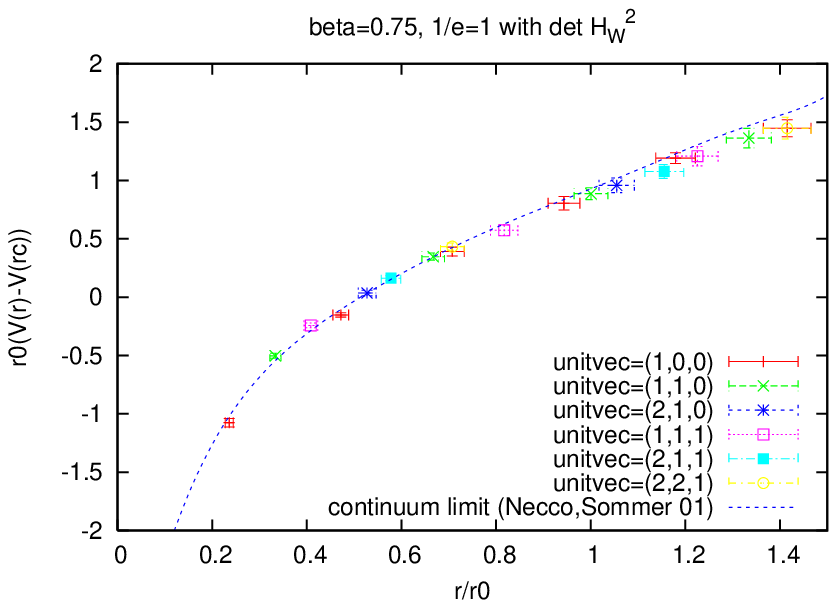}
\caption{
    Static quark potential at $\beta=1.3$, $1/\epsilon$ = 1
    (quenched) on a $12^4$ lattice (left) and $\beta=0.75$, $1/\epsilon$ = 1
    (with $\det H_W^2$) on a $14^4$ lattice (right) . 
    The curve is the continuum limit obtained by an
    interpolation of the data of \cite{Necco:2001xg}.
    Different symbols show $V(\vec{r})$'s with different
    orientations parallel to $\vec{v}$'s.
  }
  \label{fig:V(r)}
  \centering
  \includegraphics[width=8cm]{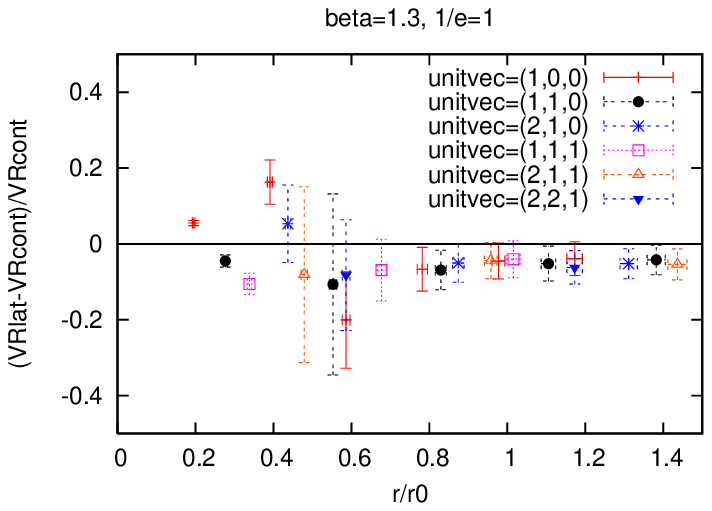}
  \includegraphics[width=8cm]{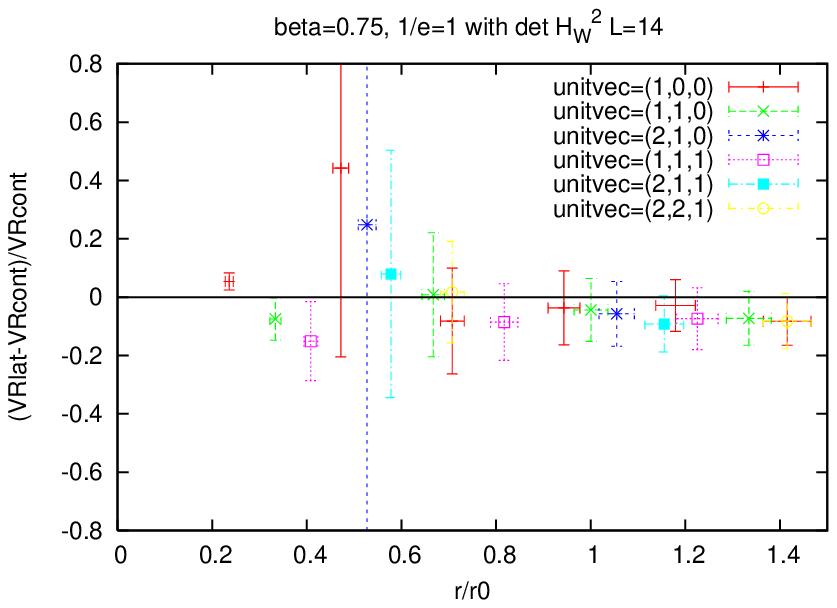}
  \caption{
    Violation of rotational symmetry 
    $(\hat{V}(\vec{r})-\hat{V}_{\mathrm{cont}}(|\vec{r}|))/
    \hat{V}_{\mathrm{cont}}(|\vec{r}|)$,
    where $\hat{V}_{\mathrm{cont}}(|\vec{r}|)$ denotes the
    continuum limit.
    Results for $\beta$ = 1.3, $1/\epsilon$ = 1 (quenched)
    and $\beta$ = 0.75, $1/\epsilon$ = 1 (with $\det H_W^2$) 
    are shown.
    The error of $\hat{V}_{\mathrm{cont}}(|\vec{r}|)$
    is not taken into account ($< 1 $\%).}
  \label{fig:V(r)_rot}
\end{figure*}

\begin{figure}[tbhp]
  \centering
  \includegraphics[width=8cm]{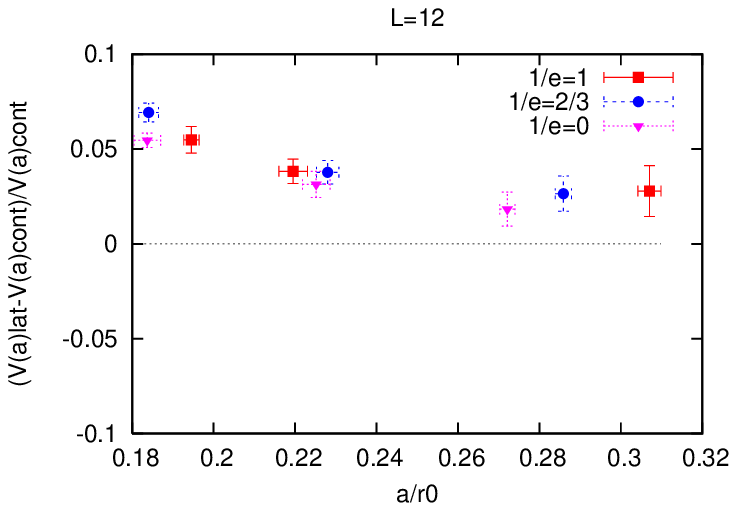}
  \includegraphics[width=8cm]{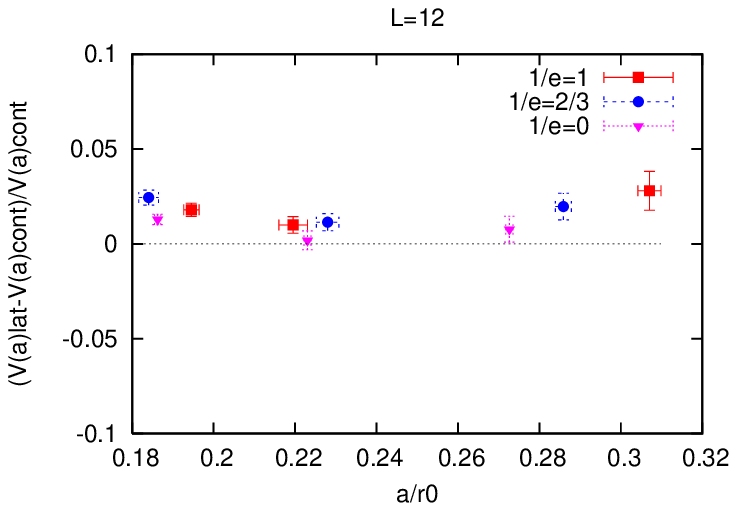}
  \caption{
    $(\hat{V}(\vec{r})-\hat{V}_{\mathrm{cont}}(|\vec{r}|))/
\hat{V}_{\mathrm{cont}}(|\vec{r}|)$ 
    for $\vec{r}=(1,0,0)$.
    Quenched results with different $\beta$ and $1/\epsilon$ values
    are plotted as a function of $a/r_0$.
    Left and right panels show the plot without and with the
    tree level improvement of the argument $\vec{r}$,
    respectively.
    See the text for details.
  }
  \label{fig:rotation}
\end{figure}

Fig.~\ref{fig:V(r)} presents the quark potential
itself in a dimensionless combination, {\it i.e.}
$\hat{V}(\vec{r})\equiv r_0(V(\vec{r})-V(r_c))$ 
as a function of $|\vec{r}|/r_0$.
$V(r_c)$ is evaluated by an interpolation of the data 
in the direction $\vec{r}/|\vec{r}|=(1,0,0)$.
The data at $\beta$ =1.3, $1/\epsilon=1$ in the quenched case and
at $\beta=0.75$, $1/\epsilon=1$ with $\det H_W^2$  
are plotted together with the curve representing the continuum limit
obtained in \cite{Necco:2001xg}.
The agreement is satisfactory (less than two sigma) for long
distances $|\vec{r}|/r_0 >$ 0.5. 
For short distances, on the other hand, 
one can see deviations of order 10\%, as shown in
Fig.~\ref{fig:V(r)_rot}, where the ratio
$(\hat{V}(\vec{r})-\hat{V}_{\mathrm{cont}}(|\vec{r}|))
/\hat{V}_{\mathrm{cont}}(|\vec{r}|)$ is plotted.
$\hat{V}_{\mathrm{cont}}(|\vec{r}|)$ represents the curve in
the continuum limit drawn in Fig.~\ref{fig:V(r)}.
The plots at $\vec{r}/a$ = $(1,0,0)$ and $(2,0,0)$ 
obviously deviate from zero in the upward direction, 
while the plots at $(1,1,0)$ and $(1,1,1)$ are lower than zero.
This fact indicates the rotational symmetry violation 
in a short distance.
Let us concentrate on the point $(1,0,0)$ as a function of the lattice
spacing, as seen in Fig.~\ref{fig:rotation} (left panel).
We find that the size of the violation is quite similar to that with
Wilson plaquette action and independent of $1/\epsilon$. 
One might think that the rotational symmetry goes worse 
in the continuum limit, but one should note the fact that
the relevant scale of the observable also diverges as $1/a$.
After a correction at the tree level by introducing
$d_I$ as $1/(4\pi d_I)=G(d)$, which is an analogue of $r_I$
in Eq.(\ref{eq:r_I}) but is defined for the potential, 
we obtain the plot on the right panel of Fig.~\ref{fig:rotation},
where one can see an improvement.
The remaining deviation should be of order $\alpha_s(1/a)$, 
which would vanish as $\sim 1/\ln(1/a)$ near the continuum limit.
The data at $\vec{r}=(1,0,0)$ with $\det H_W^2$ 
have larger statistical errors but show a very similar behavior.

These observations show that both of scaling violations 
and rotational symmetry violations are reasonable and the
topology conserving actions are feasible 
in the numerical studies.


Finally, we confirm our assumption that the topology
does not affect the quark potential by measuring $r_0$ for
two initial value of $Q$ (0 and $-3$).
Measurements are done on a $16^4$ lattice at $\beta$ = 1.42,
$1/\epsilon$ = 1, for which the probability of the topology
change is extremely small as discussed in Sec. \ref{sec:top}.
Our results are $r_0/a$ = 6.24(9) for the $Q=0$ initial
condition and 6.11(13) for $Q=-3$.
Details are seen in Table. \ref{tab:Qdep}.

\begin{table}[btp]
\begin{center}
\caption{
 The quark potential with different initial topological charge.
 The results are at $1/\epsilon=1.0$ and $\beta=1.42$ on the 
 $L=16$ lattice.
 Any large discrepancies are not seen.
}
\begin{tabular}{cccccccc}
\hline\hline
$Q_{\mathrm{init}}$ & $\Delta \tau$ & $N_{mds}$ & acceptance & Stab$_Q$
& plaquette & $r_0/a$ & $r_c/r_0$ \\
\hline
0 & 0.01 & 20 & 82\% & 961 & 0.59167(1) & 6.240(89) & 0.5241(83)\\
-3 & 0.01 & 20 & 83\% & 514 & 0.59162(1) & 6.11(13) & 0.513(12)\\
\hline
\end{tabular}\label{tab:Qdep}
\end{center}
\end{table}

\newpage
\subsection{Perturbative renormalization of the gauge coupling}
\label{sec:pert}
From $1\times 1$ Wilson loop, $W(1,1)$, the so-called 
``mean field improved'' bare gauge coupling can be defined.
In quenched study with the standard Wilson plaquette action, 
it is known that the perturbation theory 
with mean field improvement coupling converges very well.
Here we would like to define the mean field improved coupling 
with the topology conserving gauge action, Eq.(\ref{eq:admiaction}), 
evaluate the coupling renormalization
and see its convergence in 2-loop perturbation theory.

Two-loop corrections to the gauge coupling for general 
one-plaquette actions 
(constructed by the plaquette only and no rectangular
term is involved.) is calculated by Ellis {\it et al.}
\cite{Ellis:1983af}.
Using their formula, the renormalized gauge couping $g_{M}$
defined in the ``Manton'' scheme is given by
\begin{eqnarray}
\label{eq:manton1}
  \frac{1}{g_{M}^2}
  = \frac{1}{g^2} + A_1 + A_2 g^2,
\end{eqnarray}
where $g$ denotes the bare coupling and 
the coefficients $A_1$ and $A_2$ are calculated as
\begin{eqnarray}
  A_1 & = &
  s_4\frac{2N_c^3-3}{N_c}
  +t_4(N_c^2+1),
  \nonumber\\
  A_2 & = &
  a_R [ s_4 (2N_c^2-3) + t_4 N_c (N_c^2+1) ]
  +s_6 \frac{15(N_c^4-3N_c^2+3)}{8N_c^2}
  \nonumber\\
  &&
  +u_6\frac{3(2N_c^2-3)(N_c^2+3)}{8N_c}
  +t_6\frac{3}{8}(N_c^2+1)(N_c^2+3)
  \nonumber\\
  && 
  -s_4^2 \frac{9N_c^4-30N_c^2+36}{2N_c^2}
  -2s_4 t_4 \frac{(2N_c^2-3)(N_c^2+2)}{N_c}
  -t_4^2 (N_c^2+1)(N_c^2+2).
\end{eqnarray}
Here, the parameters are $N_c=3$, 
$s_4=-1/4!$, 
$s_6=1/6!$, 
$t_4=1/(4N_c\epsilon)$,
$t_6=1/(8N_c^2\epsilon^2)$, 
$u_6=-1/(4!N_c\epsilon)$, and 
$a_R=-0.0011(2)$.
The values of the next-to-leading and
next-to-next-to-leading order coefficients $A_1$ and $A_2$,
when $1/\epsilon=1,2/3$ and 0,
are given in Table~\ref{tab:A1A2}.

Let us define the mean field improved coupling 
$\bar{g}^2$ by
\begin{equation}
  \label{eq:boosted}
  \frac{1}{\bar{g}^2} = \frac{P}{g^2}
  \left(
    \frac{1}{1-(1-P)/\epsilon} +
    \frac{(1-P)/\epsilon}{(1-(1-P)/\epsilon)^2}
  \right),
\end{equation}
with the measured value of the plaquette expectation value
$P=\langle W(1,1)\rangle$ (see Table~\ref{tab:param}).
It is defined from a coefficient of $F_{\mu\nu}^2$ when
we rewrite $P_{\mu\nu}=P\exp(ia^2F_{\mu\nu})$ and expand the
action Eq.(\ref{eq:admiaction})\cite{Lepage:1992xa} .

A perturbative expectation value of the plaquette, or 
$1\times 1$ Wilson loop, is evaluated with 
the general one-plaquette action
by Heller {\it et al.} \cite{Heller:1995pu} as
\begin{eqnarray}
  \langle W(1,1)\rangle 
  &=&
  1 - g^2 \frac{(N_c^2-1)}{N_c}\bar{W}_2(1,1)
  - g^4 (N_c^2-1)X(1,1)\nonumber\\
  &&
  + g^4 \frac{(2N_c^2-3)(N_c^2-1)}{6N_c^2}\bar{W}_2(1,1)^2
  - g^4 \frac{(N_c^2-1)}{6N_c}C Z(1,1).
\end{eqnarray}
Here $\bar{W}_2(1,1)$ and $X(1,1)$ are from the
original calculation \cite{Heller:1984hx} for the
Wilson plaquette gauge action, and 
$Z(1,1)=(1-1/V)\bar{W}_2(1,1)/4$ 
(on a symmetric lattice $V=L^4$) is introduced for
generalization.
Their values are $\bar{W}_2(1,1)=1/8$, 
$X(1,1)=-1.01\times 10^{-4}$ and $Z(1,1)=1/32$ in the
infinite volume limit.
The constant $C$ is written
\begin{equation}
  C = \left[\sum_R 6g^2 \frac{s_R(\beta)T(R)C_2(R)}{d_R}-N_c\right],
\end{equation}
where $C_2(R)$ denotes the quadratic Casimir operator in a
representation $R$ of the group $SU(N_c)$.
$d_R$ is the dimension of the representation $R$, and
$T(R)$ is defined such that 
$\mbox{Tr}_R(t^a t^b) = T(R)\delta^{ab}$
for the group generator $t^a$.
The coupling $s_R(\beta)$ is defined as a coupling when
we rewrite the gauge action in terms of a general 
form of the one-plaquette action,
\begin{equation}
  S_G = \sum_{x,\mu,\nu}\sum_R s_R(\beta)
  \left[
    1-\frac{1}{d_R} \mbox{Re}\mbox{Tr}_R P^R_{\mu\nu}(x)
  \right],
\end{equation}
where $P_{\mu\nu}^R$ denotes the plaquette variable in the $R$
representation.
The values of these parameters for the topology conserving
gauge action Eq.(\ref{eq:admiaction}) are
\begin{equation}
  s_3(\beta) = \left(1+\frac{11}{6\epsilon}\right)\beta,\;\;\;
  s_6(\beta) = -\frac{1}{3\epsilon}\beta,\;\;\;
  s_8(\beta) = -\frac{4}{9\epsilon}\beta,
\end{equation}
The other parameters are 
$T(3)=1/2$, $T(6)=5/2$, $T(8)=3$,
$C_2(3)=4/3$, $C_2(6)=10/3$ and $C_2(8)=3$.
With these numbers, we obtain $C = 5-20/\epsilon$
and we finally get
\begin{equation}
  \langle W(1,1)\rangle
  = 1 - \frac{g^2}{3} +
  \left(\frac{5}{18\epsilon}-\frac{5}{144}\right)g^4.
\end{equation}

Since the perturbative relation between the bare coupling $g$
and the boosted coupling $\bar{g}$ Eq.(\ref{eq:boosted}) 
is given by 
\begin{equation}
  \frac{1}{\bar{g}^2}
  = \frac{1}{g^2}+B_1 + B_2 g^2,
\end{equation}
where
\begin{equation}
  B_1 = -\frac{1}{3}
  \left(1-\frac{2}{\epsilon}\right),
  \;\;\;
  B_2 =
  \left(1-\frac{2}{\epsilon}\right)\left(
    \frac{5}{18\epsilon}-\frac{5}{144}
  \right)-\frac{2}{9\epsilon}+\frac{1}{3\epsilon^2},
\end{equation}
one obtains the Manton coupling 
in terms of the boosted coupling as
\begin{equation}
  \label{eq:g2conv_tad}
  \frac{1}{g_M^2} = 
  \frac{1}{\bar{g}^2}
  + (A_1-B_1) + (A_2-B_2) \bar{g}^2.
\end{equation}
Numerical values of $B_i$ and $A_i-B_i$ are listed in
Table~\ref{tab:A1A2}.
One can see the effect of the mean field improvement;
the two-loop coefficient $A_2$ is significantly reduced by
reorganizing the perturbative expansion as in
Eq.(\ref{eq:g2conv_tad}).

\begin{table}[tbp]
  \centering
  \begin{tabular}{ccccccc}
    \hline\hline
    $1/\epsilon$ & $A_1$ & $A_2$ & $B_1$ & $B_2$ &
    $A_1-B_1$ & $A_2-B_2$\\
    \hline
    0   & $-$0.20833 & $-$0.03056 
        & $-$0.33333 & $-$0.03472
        &    0.12500 &    0.00416\\
    2/3 &    0.34722 & $-$0.04783
        &    0.11111 & $-$0.05015
        &    0.23611 &    0.00233\\
    1   &    0.62500 & $-$0.10276
        &    0.33333 & $-$0.13194
        &    0.29167 &    0.02919\\
    \hline
  \end{tabular}
  \caption{
    Next-to-leading and next-to-next-to-leading order
    coefficients of the coupling renormalization for
    various $\epsilon$. 
    Mean field improved coefficients $A_1-B_1$, $A_2-B_2$ are
    also presented. 
    See the text for details.
  }
  \label{tab:A1A2}
\end{table}
\begin{figure}[bthp]
  \centering
  \includegraphics[width=9.5cm]{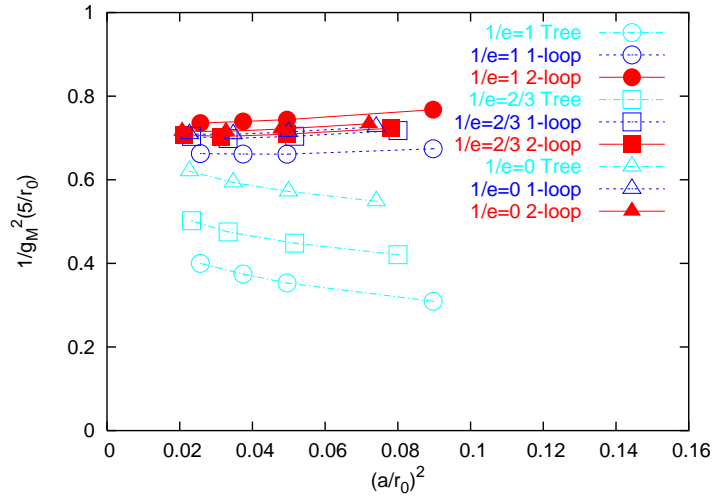}
  \caption{
    $1/g_M^2$ in the Manton scheme are plotted.
    The mean field improved expression Eq.(\ref{eq:g2conv_tad}) is
    used with the measured plaquette expectation value.
    Different symbols distinguish the value of $1/\epsilon$
    (1 for circles, 2/3 for squares, 0 for triangles).
    Open symbols with dot-dashed line represent tree-level
    results and open symbols with dashed lines are one-loop.
    The best results including two-loop corrections are
    shown by filled symbols with solid lines.
  }
  \label{fig:manton}
\end{figure}

With these results, the Manton coupling 
is obtained for each lattice parameters. 
In Fig.~\ref{fig:manton}, 
we plot the coupling evaluated at a reference scale $5/r_0$ 
as a function of $a^2/r_0^2$. 
We use the two-loop renormalization equation to run $g_M$
to the reference scale.
Although $g_M$ is very different at the tree level,
the one-loop results are already in good agreement among the
different values of $1/\epsilon$.
The two-loop corrections show that the
perturbative expansion converges very well and a good agreement
among different $1/\epsilon$.
Moreover, the scaling toward the continuum limit 
seems also good in the two-loop level plots.

\section{Stability of the topological charge}\label{sec:top}

\subsection{Admissibility condition and topology stability}


As discussed in Sec.\ref{sec:GWtop}, it is difficult to 
construct an exact and practical geometrical definition of the 
topological charge for non-Abelian theories, 
unless the gauge fields are very smooth
\cite{Luscher:1981zq}
(note that (\ref{eq:topgeo}) gives non-integers).
But even if we choose larger $\epsilon\gg 1/20$, it is quite 
possible that the barriers among the topological sectors are 
high enough to suppress topology changes for hundreds of
HMC trajectories, since any gauge action has a tendency to
prevent the topology transitions in the continuum limit. 

In Table.~\ref{tab:Qstab} we present our data 
of the topological charge stability, 
\begin{eqnarray}
  \mbox{Stab}_Q \equiv \frac{N_{\mathrm{trj}}}
  {\tau_{\mathrm{plaq}}\times \#Q},
\end{eqnarray}
where $\tau_{\mathrm{plaq}}$ is the plaquette 
autocorrelation time, which is measured according to
Appendix E of \cite{Luscher:2004rx}.
$N_{\mathrm{trj}}$ denotes the total length of the HMC
trajectories and 
$\#Q$ is the number of topology changes during the
trajectories.
This definition, $\mbox{Stab}_Q$, gives 
a mean number of uncorrelated gauge configurations
which can be sampled without changing the topology along
the simulations. 
But we should mention that 
it only gives an upper limit, because the topological charge
is measured only once per 10--20 trajectories and we may
underestimate the topology changes if $Q$ jumps and returns quickly
to the original value within this interval.
Therefore, $\mbox{Stab}_Q$ is not reliable 
when the topology change is very frequent.

\begin{table}[tbp]
\begin{center}
\begin{tabular}{cccccccc}
\hline\hline
Lattice &$1/\epsilon$ & $\beta$ & $r_0/a$ &
$N_{\mathrm{trj}}$ & $\tau_{\mathrm{plaq}}$ & $\#Q$ & Stab$_Q$ \\
\hline\\
$12^4$ & 1   & 1.0  & 3.257(30) & 18000 & 2.91(33) & 696&9\\
quenched & & 1.2  & 4.555(73) & 18000 & 1.59(15) & 265 &43\\
           & & 1.3  & 5.140(50) & 18000 & 1.091(70) & 69& 239\\
       & 2/3 & 2.25 & 3.498(24) & 18000 & 5.35(79) & 673 & 5\\
          &  & 2.4  & 4.386(53) & 18000 & 2.62(23) & 400 & 17\\
          &  & 2.55 & 5.433(72) & 18000 & 2.86(33) & 123 & 51\\
       & 0   & 5.8  & [3.668(12)] & 18205 & 30.2(6.6) & 728 & 1\\
          &  & 5.9  & [4.483(17)] & 27116 & 13.2(1.5) & 761 & 3\\
          &  & 6.0  & [5.368(22)] & 27188 & 15.7(3.0) & 304 & 6\\
\\
$16^4$ & 1   & 1.3  & 5.240(96) & 11600 & 3.2(6) & 78 & 46\\
quenched  & & 1.42 & 6.240(89) & 5000 & 2.6(4) & 2 & 961\\ 
       & 2/3 & 2.55 & 5.290(69) & 12000 & 6.4(5) & 107 & 18\\
           & & 2.7  & 6.559(76) & 14000 & 3.1(3) & 6 & 752\\
       & 0   & 6.0  & [5.368(22)] & 3500 & 11.7(3.9) & 14 & 21\\ 
          &  & 6.13 & [6.642(--)] & 5500 & 12.4(3.3) & 22 & 20\\
\\
$20^4$ & 1   & 1.3  & --- & 1240 & 2.6(5) & 14 & 34\\
quenched &  & 1.42 & --- & 7000 & 3.8(8) & 29 & 64 \\ 
       & 2/3 & 2.55 & --- & 1240 & 3.4(7) & 15 & 24\\
          &  & 2.7  & --- & 7800 & 3.5(6) & 20 & 111\\
       & 0   & 6.0  & --- & 1600 &14.4(7.8) & 37 & 3 \\
          &  & 6.13 & --- & 1298 & 9.3(2.8) &4  &35 \\
\\
$14^4$ 
& 1   & 0.75  &4.24(15) & 3500 & 5.05(82) & 0 & $>$693\\
with $\det H_W^2$ 
       & 2/3 & 1.8 & 4.94(19) & 5370 & 11.1(2.1) & 0 & $>$483\\
       & 0   & 5.0  & 4.904(90) & 3120 & 21.4(6.5) & 0 & $>$146 \\
\\
$16^4$ 
& 1   & 0.8  &4.81(17) & 627 & 0.69(10) & 0 & $>$908\\
with $\det H_W^2$ 
       & 2/3 & 1.75 & 4.71(19) & 580 & 1.64(40) & 0 & $>$353\\
       & 0   & 5.2  & 7.09(17) & 730 & 1.54(27) & 0 & $>$474 \\
\\
\hline
\end{tabular}
\end{center}
\caption{
  The stability of the topological charge, $\mbox{Stab}_Q$.
  The HMC trajectory length $N_{\mathrm{trj}}$,
  the plaquette autocorrelation time 
  $\tau_{\mathrm{plaq}}$,
  and the number of topology change $\#Q$ are also
  summarized. 
  }
  \label{tab:Qstab}
\end{table}

\begin{figure}[tbp]
  \centering
  \includegraphics[width=10cm]{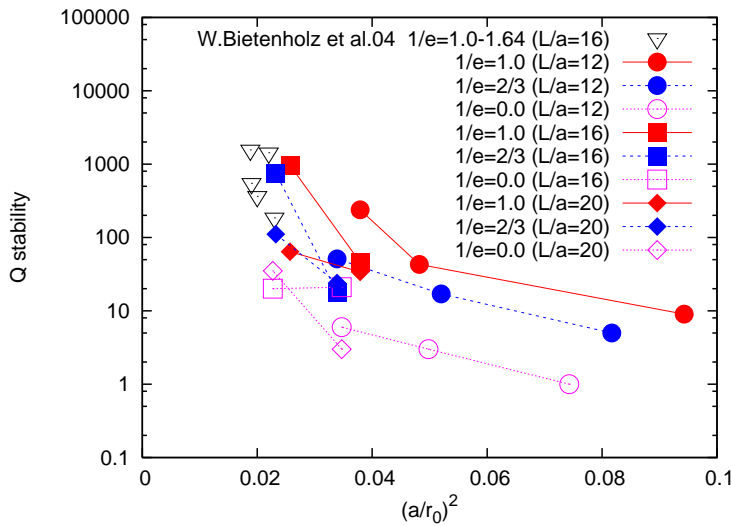}
  \caption{
    Stability of the topological charge in the quenched study.
    The results of different lattice size 
    are plotted with different symbols;  
    $L/a$ = 12 (circles), 16 (squares), 20 (upward
    triangles). 
    The value of $1/\epsilon$ is distinguished by the line
    type:
    $1/\epsilon$ = 0 (dotted), 2/3 (dashed), 1 (solid).
    Downward triangles are the data of
    \cite{Bietenholz:2004mq} measured on a 16$^4$ lattice. 
  }
  \label{fig:Qstab}
\end{figure}

In Fig.~\ref{fig:Qstab} the results are plotted as a function
of the lattice spacing squared.
Clearly, one can see that the stability goes better
as $1/\epsilon$ increases when the lattice spacing is the same.
Also, the stability gets worse as the lattice size is increased 
from $L/a=12$ to $L/a=20$.
This is no surprising, because the topology change occurs through
local dislocations of gauge field and its probability scales
as the volume.
An important notice here is that the topological charge can be
very stable in any case  in the continuum limit.
The stability quickly rises as $a\to 0$.

To study QCD in the $\epsilon$-regime 
in a fixed topological sector, the lattices
$(1/\epsilon, \beta, L)\sim (1,1.42,16)$ and 
$(2/3, 2.7, 16)$ would be appropriate.
Their physical size is around 
$L \sim$ 1.25~fm and the topological charge is stable for 
$(100-1000)\tau_{\mathrm{plaq}}$ trajectories.

\subsection{Negative mass Wilson fermion to fix topology}

In contrast to the case with topology conserving gauge action 
Eq.(\ref{eq:admiaction}), one expects that
the fermion determinant, $\det H_W^2$, can rigorously fix the
topology of gauge fields along the simulation.
In fact, as Table.~\ref{tab:Qstab} shows, the topology change has
never occurred in every run with different parameters,
which are chosen such that the lattice spacing is around
$a\sim 0.08$--0.1fm.
Our data show that 
$Q$ is unchanged for, at least,
100-1000 $\tau_{\mathrm{plaq}}$ trajectories,
even if one ignores the thermalization steps.

\newpage

\section{The effects on the overlap Dirac operator}\label{sec:overlap}

\subsection{Low-lying mode distribution of $H_W$}

As explained in Sec \ref{sec:setupov}, the order of
Chebyshev polynomial approximation, $N_{\mathrm{poly}}$
has to be proportional to the condition number,
$\kappa=\lambda_{\mathrm{max}}/\lambda_{\mathrm{min}}$, 
in order to keep a certain desired accuracy.
Since both topology conserving actions, Eq.(\ref{eq:admiaction})
and Eq.(\ref{eq:nwf}) would play a role to suppress the
occurrence of low-lying eigenvalues of 
$|aH_W|=|\gamma_5(aD_W-1-s)|$, they may be useful to
reduce the numerical cost of calculating 
the overlap Dirac operator.

\begin{figure}[btp]
  \centering
  \includegraphics[width=8cm]{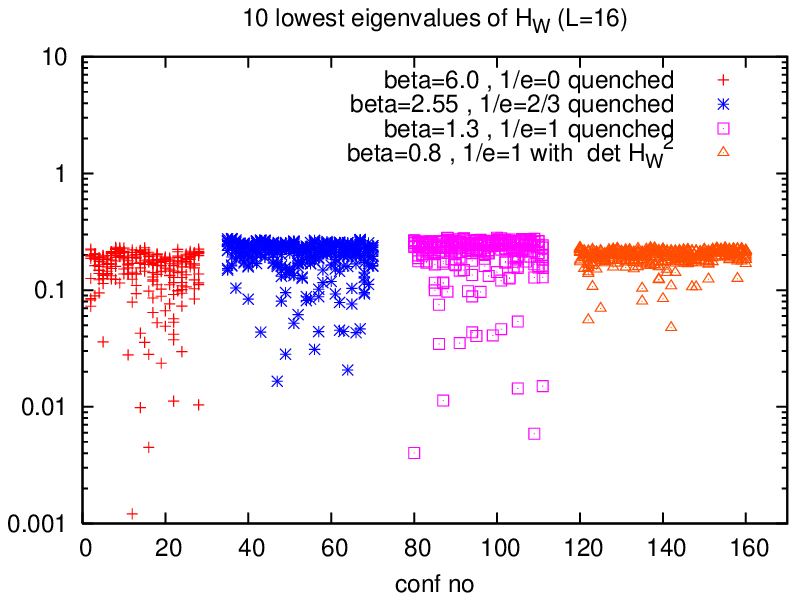}
  \includegraphics[width=8cm]{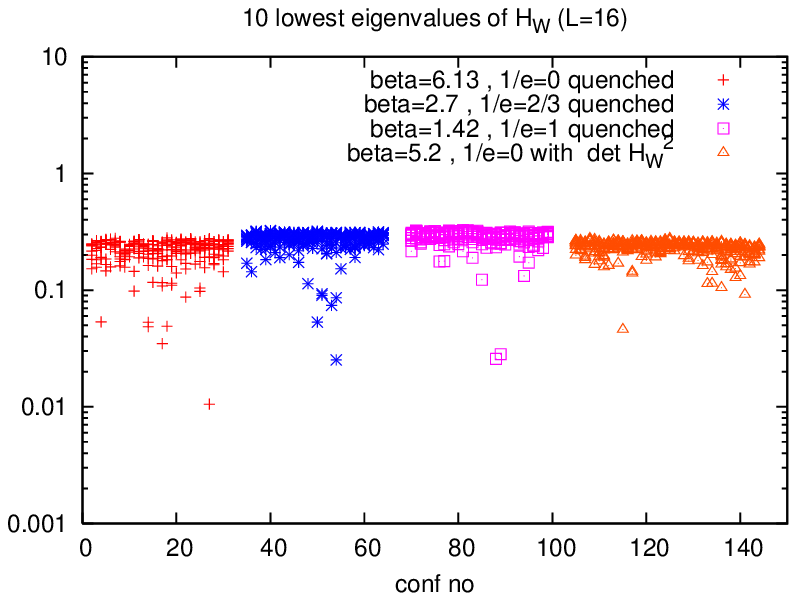}
  \caption{
    Ten lowest eigenvalues of $|aH_W|=|\gamma_5(aD_W-1.6)|$ 
    for gauge configurations with $r_0/a \simeq 5$ (left)
    and $r_0/a \simeq 6.5$-7 (right).
    Quenched data are shown for $1/\epsilon$ = 1 (pluses), 2/3
    (stars), 0 (squares) and triangles are the result with $\det H_W^2$.
    The lattice size is 16$^4$.
    }
  \label{fig:HwL20coarse}
\end{figure}

Fig.~\ref{fig:HwL20coarse} shows a typical comparison of
the eigenvalue distribution on a 16$^4$ lattice.
The values of $\beta$ and $\epsilon$ 
are chosen such that the Sommer scale
$r_0/a$ is roughly equal to 5 (left panel) 
and 6.5-7 (right panel) , which correspond to
$a\simeq 0.1$~fm and  $a\simeq 0.08$~fm respectively.
From the plot we observe that the density of the low-lying
modes is relatively small for larger values of $1/\epsilon$ in 
quenched case.
Let us discuss this more quantitatively.
In Table~\ref{tab:Hw} we list the probability,
$P_{\lambda_{\mathrm{min}}<0.1}$, 
that a configuration has the lowest eigenvalue lower than 0.1. 
For the above example (left panel), 
the probability is 74\% for
the standard Wilson gauge action 
($\beta=6.0$, $1/\epsilon$ = 0 quenched ), but it
decreases to 53\% (47\%) for $1/\epsilon$ = 2/3 (1)(quenched).
It is interesting to note that 
the data with $\det H_W^2$ show that 
the average of $\lambda_{\mathrm{min}}$ is not so large, but
the occurrence of very small eigenvalues $\ll$0.1 is strongly
suppressed. The lowest eigenvalue in these configurations
is 1.2e-03 in the quenched case at $\beta=6.0$ and $1/\epsilon=0$, 
while it is 0.042 in the case with $\det H_W^2$ at $\beta=0.8$ and 
$1/\epsilon=1$.

Also, for another lattice spacing ($r_0/a\simeq$~6.5) and lattice
size $20^4$, a similar trend can be seen.
In Table~\ref{tab:Hw} we also present the ensemble average
of the lowest eigenvalue $\lambda_{\mathrm{min}}$ and
the inverse of condition numbers
$\lambda_{\mathrm{max}}/\lambda_{\mathrm{min}}$ and
$\lambda_{\mathrm{max}}/\lambda_{\mathrm{10}}$, where
$\lambda_{\mathrm{10}}$ and $\lambda_{\mathrm{max}}$ denote
the 10th and the highest eigenvalues respectively.
We may conclude that the lowest eigenvalue is higher in
average when we use the topology conserving actions.
In the numerical implementation of the overlap-Dirac
operator, the low-lying eigen-modes of $aH_W$
are often subtracted and treated exactly.
Then the higher mode contributions are approximated 
by some polynomial or rational functions.
Here, we assume that 10 lowest eigen-modes are subtracted and
compare the relative numerical cost on the gauge
configurations with different values of $1/\epsilon$.
From Table~\ref{tab:Hw} we observe that the reduced
condition number is about a factor 1.2--1.4 smaller for
$1/\epsilon$ = 1 than that for the standard Wilson gauge
action.

\begin{table}[tbp]
\begin{center}
\begin{tabular}{cccccccc}
\hline\hline
lattice size & $1/\epsilon$ & $\beta$ & $r_0/a$ & 
$P_{\lambda_{\mathrm{min}}<0.1}$ &
$\lambda_{\mathrm{min}}$ &
$\lambda_{\mathrm{min}}/\lambda_{\mathrm{max}}$&  
$\lambda_{\mathrm{10}}/\lambda_{\mathrm{max}}$  \\
\hline
\\
$20^4$ & 1 & 1.3  & 5.240(96)  &  0.64 & 0.0882(84) & 0.0148(14) & 0.03970(29)\\
quenched    &  2/3 & 2.55 & 5.290(69)  &  0.75 & 0.0604(53) & 0.0101(08) & 0.03651(27)\\
    &    0 & 6.0  & [5.368(22)]&  0.97 & 0.0315(57) & 0.0059(34) & 0.02766(46)\\
    &    1 & 1.42 & 6.240(89)  &  0.22 & 0.168(13)  & 0.0282(21) & 0.04765(32)\\
    &  2/3 & 2.7  & 6.559(76)  &  0.19 & 0.151(11)  & 0.0251(19) & 0.04646(37)\\
    &    0 & 6.13 & [6.642(--)]&  0.45 & 0.0861(83) & 0.0126(15) & 0.03775(50)\\
\\
$16^4$ & 1 & 1.3  & 5.240(96)  &  0.47 & 0.111(12)  & 0.0187(21) & 0.04455(31)\\
quenched    &  2/3 & 2.55 & 5.290(69)  &  0.53 & 0.1038(98) & 0.0174(16) & 0.04239(36)\\
    &    0 & 6.0  & [5.368(22)]&  0.74 & 0.0692(90) & 0.0116(15) & 0.03451(62)\\
    &    1 & 1.42 & 6.240(89)  &  0.07  & 0.219(13)  & 0.0367(21) & 0.05233(26)\\
    &  2/3 & 2.7  & 6.559(76)  &  0.13  & 0.191(12)  & 0.0320(19) & 0.05117(29)\\
    &    0 & 6.13 & [6.642(--)]&  0.27 & 0.139(10)  & 0.0232(17) &
 0.04384(38)\\
\\
$16^4$ & 1 & 0.8  & 4.81(17)  &  0.12 & 0.1502(87)  & 0.0255(11) & \\
with $\det H_W^2$    &  2/3 & 1.75 & 4.71(19)  &  0.55 & 0.0999(44) & 0.0170(10) & \\
    &    0 & 5.2  & 7.09(17)&  0.05 & 0.1771(69) & 0.0297(12) &\\
\\
\hline
\end{tabular}
\end{center}
\caption{
  The probability, $P_{\lambda_{\mathrm{min}}<0.1}$, 
  to sample a configuration such that $\lambda_{\mathrm{min}}<0.1$
where $\lambda_{\mathrm{min}}$ is the lowest eigenvalue 
 of the hermitian Wilson-Dirac operator
  $|aH_W|=|\gamma_5(aD_W-1.6)|$.
  Averages of the lowest eigenvalue and the
  inverse of condition numbers are also listed.
  The Sommer scale $r_0/a$ is the results from $L=16$ lattice simulations.
  The values with [$\cdots$] are from \cite{Necco:2001xg} with an
  interpolation in $\beta$.
}
\label{tab:Hw}
\end{table}

\begin{figure}[tbp]
  \centering
  \includegraphics[width=8cm]{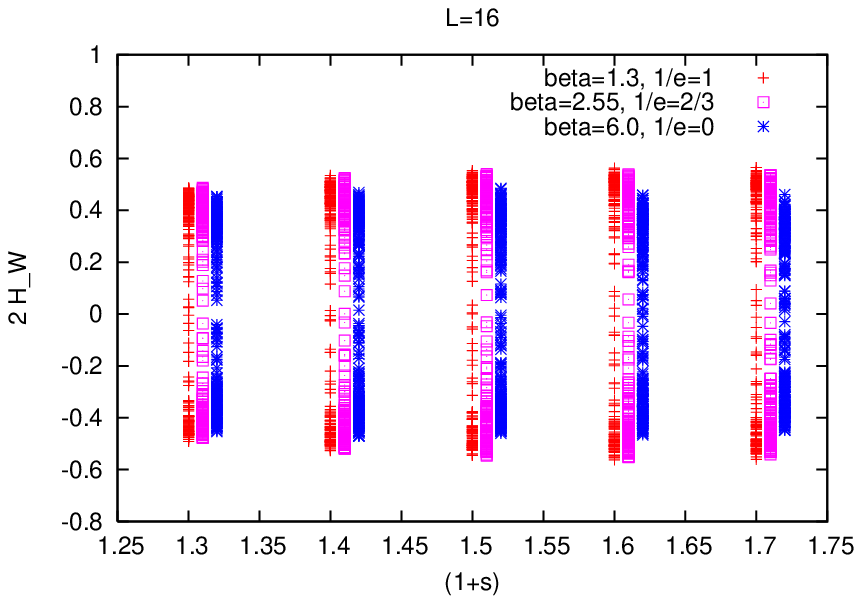}
  \includegraphics[width=8cm]{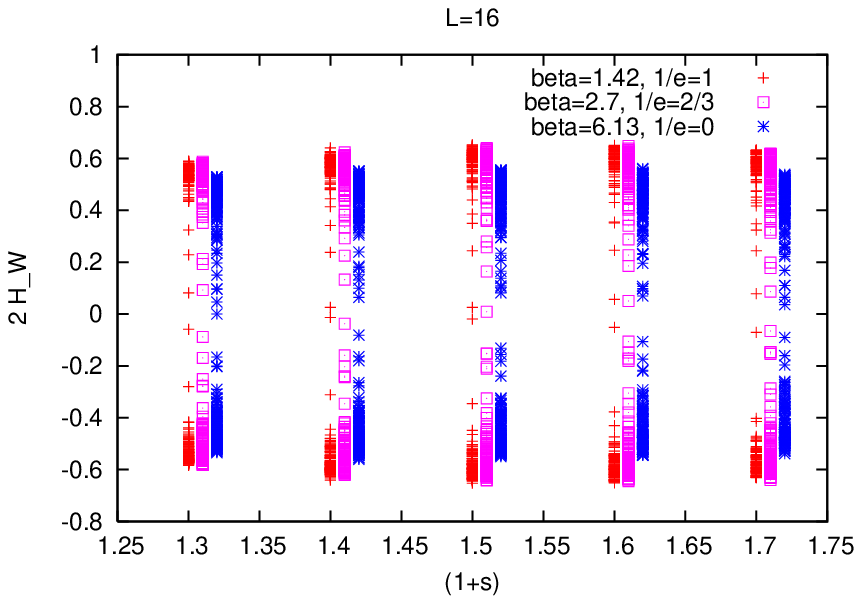}
  \caption{
    Ten near-zero eigenvalues on $16^4$ lattices with
    $r_0/a\sim 5.3$ (left) and $r_0/a\sim 6.5$ (right).
    Results are plotted as a function of $1+s$.
    Plots for $1/\epsilon$ = 2/3 and 0 are slightly shifted
    for clarity.
}
\label{fig:flow}
\end{figure}

We also check that the above observation does not change when
one shifts the value of $s$ in a reasonable range.
In Fig.~\ref{fig:flow}, a typical distributions of the
low-lying eigen-modes for $s$ = 0.2--0.7 are plotted.
We find that the advantage of the topology conserving
actions does not change.
Also, from these plots we can see that
$s\sim 0.6$ is nearly optimal for all the cases.

\subsection{Locality}
\begin{figure}[tbp]
  \centering
  \includegraphics[width=8cm]{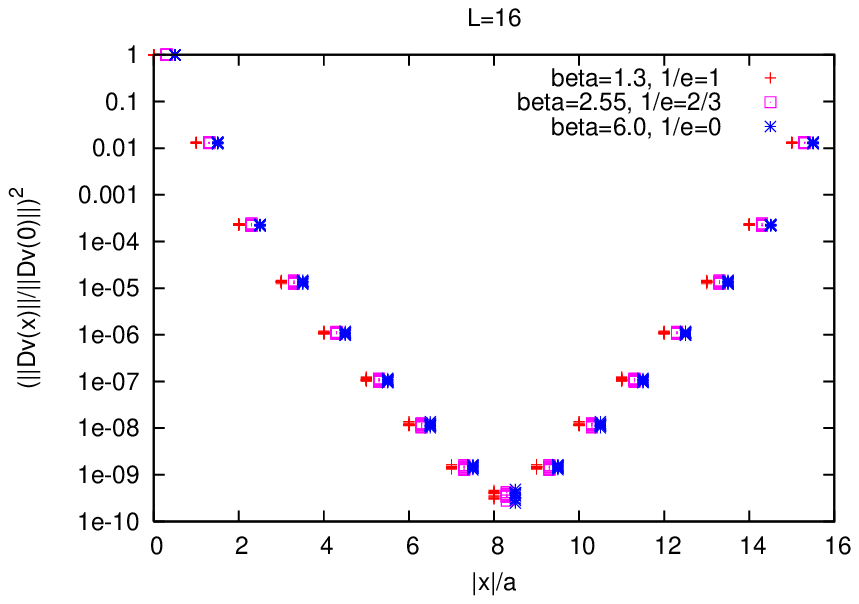}
  \includegraphics[width=8cm]{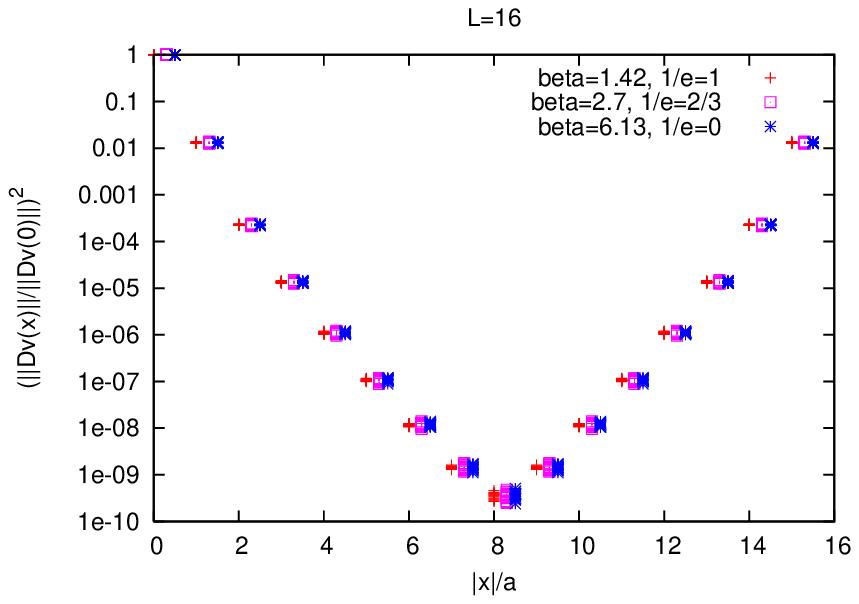}
  \caption{
    $(||D(x,y)v(y)||/||D(0,y)v(y)||)^2$ with $x_0=0$
    measured on 10 gauge 
    configurations for different values of $1/\epsilon$ (quenched).
    The lattice scale is
    $r_0/a\simeq 5.3$ (left) and 6.5 (right).
  }
  \label{fig:locality}
\end{figure}

When we talk about the locality of the overlap-Dirac operator,
it means that the norm 
$||D(x,y)v(y)||$ with a point source vector $v$ at $x_0$
should decay exponentially as a function of $|x-x_0|$
\cite{Hernandez:1998et}  
\begin{equation}
  ||D(x,y)v(y)|| \sim C \exp (-D|x-x_0|),
\end{equation}
where $C$ and $D$ are constants.
Actually we observe this property, as seen in
Fig.~\ref{fig:locality}.
The plot shows the results for different values of $1/\epsilon$ at
the lattice scales $r_0/a\simeq$ 5.3 (left) and 6.5 (right).
We find no remarkable difference on the locality when we
change $1/\epsilon$.

Recently, it has been indicated that the mobility edge is
more crucial quantity which governs the locality of the
overlap-Dirac operator 
\cite{Svetitsky:2005qa,Golterman:2005xa,Golterman:2005fe,Golterman:2004cy,
Golterman:2003qe}.
It would be interesting and important to see the dependence of the
mobility edge on the parameters in the topology conserving
actions, which should be done in the future works.

\newpage
\section{Lattice QCD in the $\epsilon$-regime with fixed $Q$}
\label{sec:eregime}

 In previous sections we have discussed 
the topology conserving actions 
which might be helpful to the full QCD 
simulation with the overlap fermion determinant.
Here we would like to show great uses of the overlap 
Dirac operator itself to measure physical observables.
Once a set of configurations of $N_f=2$ or 3 full QCD
is generated, it would be much easier and more reliable 
to extract physical quantities
such as pion mass, decay constants, Kaon bag parameters, and so on,
using the overlap Dirac operator which respects 
the ``exact'' chiral symmetry.
One of attractive numerical applications would be the so-called 
$\epsilon$-regime, where both of chiral symmetry and topology 
have significant effects on the observables.

In this section we demonstrate the analysis of quenched lattice QCD
in the $\epsilon$-regime to determine the low energy constants of
the quenched chiral perturbation theory (qChPT).
Of course, both quenched QCD and qChPT are unphysical
and the data of pion decay constant or the chiral condensate
have little relation to the true value of them in nature.
However, this quenched study is still interesting 
to see how these low-energy constants can be extracted
when the full QCD simulation is established in the future.

\subsection{Meson correlators in the $\epsilon$-regime}

It is believed that the low energy limit of QCD is described by 
the pion effective theory, or (q)ChPT.
To determine the fundamental parameters of (q)ChPT is one
of relevant issues of lattice QCD.
In the so-called $\epsilon$-regime 
\cite{Gasser:1983yg, Gasser:1987ah,Gasser:1987zq,Hansen:1990un,Hansen:1990yg, Leutwyler:1992yt},
where the size of the space-time box is smaller
than the pion Compton wave length; $L \ll 1/m_{\pi}$
(but larger than the QCD scale 
$1/\Lambda_{\mathrm{QCD}}\ll L$,
which should be guaranteed so that 
the pion can be treated as a point particle 
and other heavier hadrons are decoupled.), 
(q)ChPT with a expansion parameter 
$\epsilon^2\sim m_\pi/\Lambda\sim p_\pi^2/\Lambda^2$, 
is still applicable.
Here $\Lambda$ is a cutoff scale of the chiral 
Lagrangian roughly around 1~GeV.
An important notice is that the low energy constants in
(q)ChPT are defined at the cutoff scale and 
it does not depend on whether the system is in the 
$\epsilon$-regime or the large volume regime.
Therefore, once $F_{\pi}$ or $\Sigma$ are determined 
in the $\epsilon$-regime, one can use them 
in the standard (q)ChPT in a larger volume.
Analytic calculations of meson correlation functions in the
$\epsilon$-regime have been widely studied for both ChPT and qChPT
\cite{Damgaard:2001js,Damgaard:2002qe,Hernandez:2002ds}.

In the quenched case, the fundamental parameters of 
the effective theory are the pion decay constant,
$F_{\pi}$, the chiral condensate, $\Sigma$, the singlet mass,
$m_0$, and the coefficient of an additional singlet 
kinetic term, $\alpha$. 
Here we just present some results of 1-loop calculation, 
which are relevant to our numerical study. 
The details of calculation and the definition of the functions
and coefficients are summarized in Appendix \ref{sec:ChPT}.
The scalar condensate at 1-loop level in 
$Q$ topological sector is given by 
\begin{eqnarray}
  \label{eq:scond} 
  -\langle \bar{\psi}\psi\rangle =
\Sigma_Q(\mu^{\prime})=
\Sigma_{\mathrm{eff}} \mu^{\prime}
  (I_{|Q|}(\mu^{\prime})K_{|Q|}(\mu^{\prime})+I_{|Q|+1}
  (\mu^{\prime})K_{|Q|-1}(\mu^{\prime}))
  +\Sigma_{\mathrm{eff}}\frac{|Q|}{\mu^{\prime}}.
\end{eqnarray}
The triplet axial vector, scalar and pseudo-scalar correlators 
are given by
\begin{eqnarray}
\label{eq:tripletmesonAA}
  \langle A^a_0(x)A^a_0(0)\rangle_Q
  &=&
  -\frac{F_{\pi}}{V}-2m\Sigma_Q(\mu)\bar{\Delta}(x),
\\
\label{eq:tripletmesonSS}
  \langle S^a(x)S^a(0)\rangle_Q &=& C_S^a
  +\frac{\Sigma^2}{2F_{\pi}^2}\left[
    \frac{c_-}{N_c}(m_0^2\bar{G}(x)+\alpha\bar{\Delta}(x))
    -\bar{\Delta}(x)b_-
  \right],
  \\
\label{eq:tripletmesonPP}
  \langle P^a(x)P^a(0)\rangle_Q &=& C_P^a
  -\frac{\Sigma^2}{2F_{\pi}^2}\left[
    \frac{c_+}{N_c}(m_0^2\bar{G}(x)+\alpha\bar{\Delta}(x))
    -\bar{\Delta}(x)b+
  \right].
\end{eqnarray}
The singlet scalar and pseudo-scalar correlators are 
\begin{eqnarray}
  \langle S^0(x)S^0(0)\rangle_Q
  &=&
  C_S^0
  +\frac{\Sigma^2}{2F_{\pi}^2}\left[
    \frac{a_-}{N_c}(m_0^2\bar{G}(x)+\alpha\bar{\Delta}(x))
    -\bar{\Delta}(x)\frac{a_++a_--4}{2}
  \right],
  \\
  \langle P^0(x)P^0(0)\rangle_Q
  &=&
  C_P^0
  -\frac{\Sigma^2}{2F_{\pi}^2}\left[
    \frac{a_+}{N_c}(m_0^2\bar{G}(x)+\alpha\bar{\Delta}(x))
    -\bar{\Delta}(x)\frac{a_++a_-+4}{2}
  \right].
\end{eqnarray}
As mentioned in Appendix \ref{sec:ChPT}, the results are
valid only when $|Q|$ is small, which is a particular restriction
of quenching the fermion determinant.

As these equations of qChPT show, the meson correlators in the
$\epsilon$-regime should be quite sensitive to the topological
charge and the fermion mass. 
These prominent $m$ and $Q$ dependences are used to 
evaluate the low energy constants,
or namely, we fit our lattice data  with the above 
equations, in which 
$F_\pi$, $\Sigma$, $m_0$ and
$\alpha$ are treated as free parameters.
The parameter $\Sigma$ always appears associated with the
quark mass $m$ as $m\Sigma$ is renormalization scale 
and scheme independent.
The value of $\Sigma$ in the following
analysis should be understood as a bare quantity 
in the lattice regularization at a scale $1/a$.
To relate them with the conventional scheme such as the
$\overline{\mbox{MS}}$ scheme requires perturbative or
non-perturbative matching, 
which is beyond the scope of this work. 

\subsection{Lattice observables with the exact chiral symmetry}

We generate gauge link variables at $\beta=5.85$ and $1/\epsilon=0$
in the quenched approximation on a $L^3T=10^3\times 20$ lattice.
The spacial length $L$ of the box is about 1.23~fm.
We employ the massive Dirac operator for the valence quark,
\begin{equation}
D_m  =  \left(1-\frac{\bar{a}m}{2}\right)D + m,
\end{equation}
where the bare quark mass are chosen to be very small;
$am=$0.0016, 0.0032, 0.0048, 0.0064, 0.008, 
which corresponds 2.6--13MeV.
The number of configurations for each topological
sector is given in Table~\ref{tab:confnum}.
We analyze the gauge configurations in $|Q|\leq 3$ sectors. 

When the exact inversion of the overlap-Dirac operator is
needed, we use the techniques described in
Ref.~\cite{Giusti:2002sm}; 
for a given source vector $\eta$, we solve the equation 
\begin{equation}
  D_m\psi=\eta,
\end{equation}
by separating the left and right handed components as
$\psi = P_- \psi + P_+ \psi$ and solving two equations
\begin{eqnarray}
  P_-\psi &=& (P_-D^{\dagger}_mD_mP_-)^{-1}P_-D^{\dagger}_m\eta,
  \\
  P_+\psi &=& (P_+D_mP_+)^{-1}(P_+\eta-P_+D_mP_-\psi),
\end{eqnarray}
consecutively.
(The above equations apply to positive $Q$ cases and the same
procedure applies with a replacement $P_+\leftrightarrow
P_-$ to negative $Q$ cases.)
The conjugate gradient (CG) algorithm is used to invert
the chirally projected matrices with the low-mode
preconditioning in which 20 lowest eigen-modes are subtracted.

\begin{figure}[bthp]
\begin{center}
\includegraphics[width=10cm]{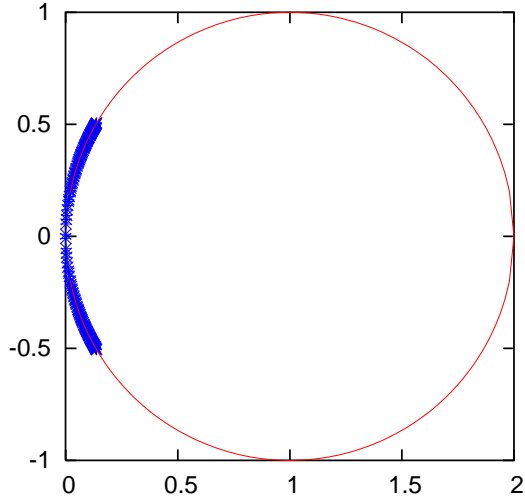}
\caption{
  Lowest 202 eigenvalues of the overlap-Dirac operator $\bar{a}D$
  at beta=5.85 on a $10^320$ lattice with
  topological charge $Q=-2$.
  The eigenvalues cover a $\pi/3$ arc of the circle. 
}\label{fig:eigenV}
\end{center}
\end{figure}

\begin{figure}[tbp]
\begin{center}
\includegraphics[width=12cm]{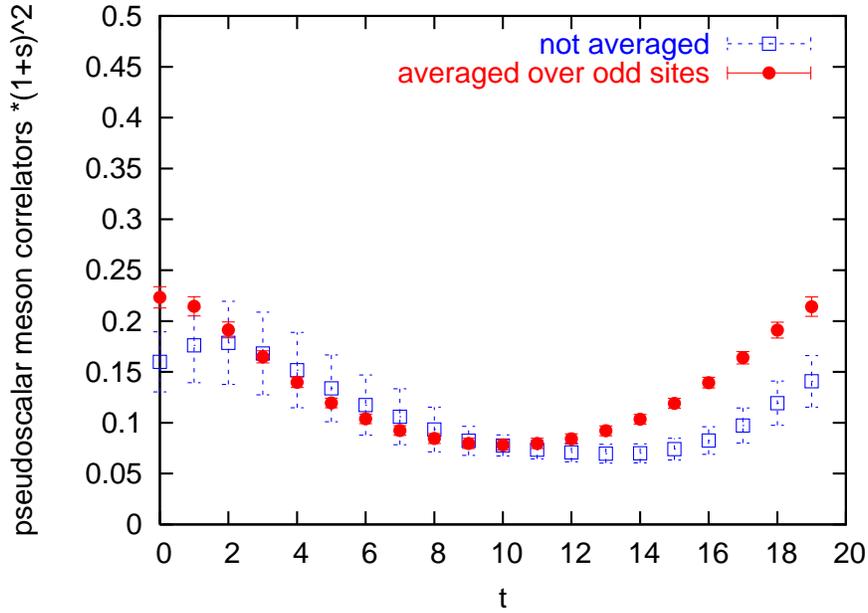}
\caption{
  The pseudo-scalar correlator at $am=0.008$ and $|Q|=1$.
  Filled symbols denote the data with the low-mode
  averaging, where we use $(L/2)^3\times T/2$ source points,
  while open symbols are not averaged.
}\label{fig:LMA}
\end{center}
\end{figure}

As one of special features of finite volume regime,
a remarkable low-mode dominance in the chiral limit 
is expected
\cite{DeGrand:2000gq, Neff:2001zr, DeGrand:2003sf}
.
Using ARPACK \cite{ARPACK} again, we calculate 200 + $|Q|$ 
lowest mode eigenvalues and their eigenfunctions
of the overlap Dirac operator. The index $Q$ is calculated at the same time.
Note that these eigenvalues cover more than 15\% 
of the circle in the complex space of the eigenvalues of $\bar{a}D$
as Fig.~\ref{fig:eigenV} shows. 
Then, the inverse of the overlap operator is decomposed as 
\begin{equation}
  \label{eq:eigen_decomp}
  D^{-1}_m(x,y) =
  \sum_{i=1}^{N_{\mathrm{low}}}
  \frac{1}{(1-\bar{a}m/2) \lambda_i+m} v_i(x)v_i^{\dagger}(y)
  +\Delta D_m^{-1}(x,y),
\end{equation}
where  $\lambda_i$'s are eigenvalues of $D$ and $v_i(x)$'s are
their eigenvectors. We set $N_{\mathrm{low}}=200+|Q|$.
In the following analysis, 
we often use the low-mode approximation
of the expectation value (We will denote 
$\langle \cdots \rangle_{\mathrm{low}}$ instead of a simple bracket.)
by inserting the low-mode part only,
\begin{equation}
  \label{eq:eigen_decomp2}
  D^{-1}_m(x,y)_{\mathrm{low}} =
  \sum_{i=1}^{N_{\mathrm{low}}}
  \frac{1}{(1-\bar{a}m/2) \lambda_i+m} v_i(x)v_i^{\dagger}(y),
\end{equation}
to the propagators.
This would be valid only when it is proved that
the higher mode's contributions are
sufficiently small or can be canceled in the combination of 
different operators.
An great advantage of using $D^{-1}_m(x,y)_{\mathrm{low}}$ instead
of exact $D^{-1}_m(x,y)$ is that 
the inversion at any $x$ and $y$ is obtained
without performing the CG algorithm, so that one can easily
average the correlators over the space-time arguments, like
\begin{equation}
  \left\langle
    \sum_{\vec{x}} \bar{\psi}\Gamma_i\psi(\vec{x},t) 
    \bar{\psi}\Gamma_i\psi(\vec{0},0)
  \right\rangle^Q_{\mathrm{low}}
  \to \frac{1}{T L^3}
  \sum_{\vec{x}_0,t_0}
  \left\langle 
    \sum_{\vec{x}}\bar{\psi}\Gamma_i\psi(\vec{x},t+t_0) 
    \bar{\psi}\Gamma_i\psi(\vec{x}_0,t_0)
  \right\rangle^Q_{\mathrm{low}},
\end{equation}
where $\Gamma_i$ denotes some gamma matrix.
This so-called low-mode averaging dramatically reduces 
the fluctuation of the correlators as shown in
Fig.~\ref{fig:LMA}, which is also reported in
Refs.~\cite{
DeGrand:2004qw,
Giusti:2004yp
}.

\begin{table}[hbtp]
\begin{center}
\caption{
  Number of configurations in each topological sector.
}\begin{tabular}{ccccc}
\hline\hline
$|Q|$ & 0 & 1 & 2 & 3 \\
\hline
\# of confs. & 20 & 45 & 44 & 24 \\
\hline
\end{tabular}\label{tab:confnum}
\end{center}
\end{table}


\newpage
\subsection{Numerical results}

\subsubsection{$F_{\pi}$ from the axial-vector
  correlator}

\begin{figure}[tbp]
  \begin{center}
    \includegraphics[width=8cm]{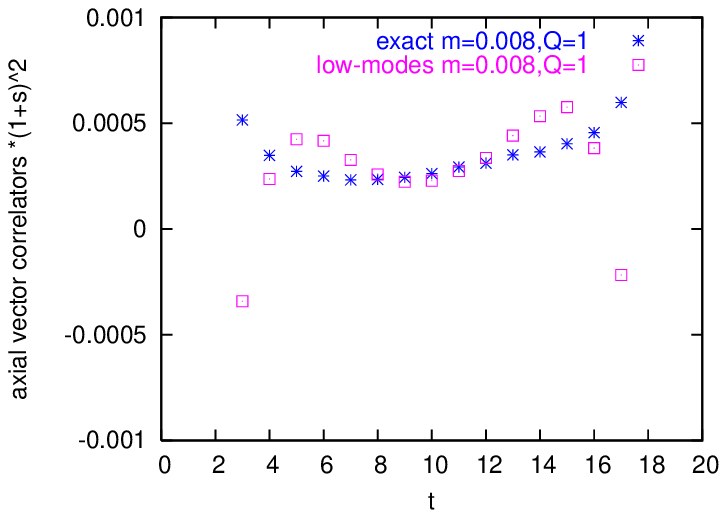}
    \includegraphics[width=8cm]{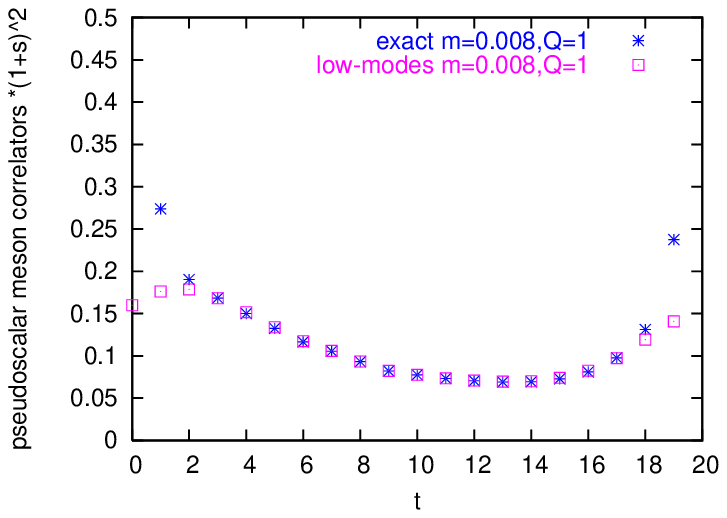}
    \caption{
      Triplet axial-vector (left) and pseudo-scalar (right)
      correlators at $am$ = 0.008 and $|Q|=1$.
      The low-mode-approximated correlator 
      is compared with the corresponding exact one.
      The 20+$|Q|$ low-modes are not sufficient to evaluate the
   axial correlators but they are enough for the pseudo-scalar correlators
   at long distances.
    }\label{fig:accuracy}
  \end{center}
\end{figure}

First let us consider 
the axial-vector current correlator 
(\ref{eq:tripletmesonAA}), which is most
sensitive to $F_\pi$ and not contaminated by the parameters
$m_0$ and $\alpha$.
Here we do not use the low-mode approximation because
200+$|Q|$ modes are not sufficient to estimate the total
propagator, as the left panel of Fig.\ref{fig:accuracy} shows.

A naive definition of the axial vector current is
$A_{\mu}^a(x)=\bar{\psi}(x)\gamma_5\gamma_{\mu}(\tau^a/2)\psi(x)$,
constructed from the overlap fermion field $\psi(x)$.
Note that it is not the conserved current associated with the
lattice chiral symmetry, and (finite) renormalization is needed
to relate it to the continuum axial-vector current.
We follow the method applied in Refs.~\cite{Giusti:2001yw,Giusti:2001pk}
to calculate the $Z_A$ factor non-perturbatively.
Namely, we calculate
\begin{eqnarray}
  \label{eq:R_rho}
  a\rho_m(t) \equiv 
  \frac{
    a\sum_{\vec{x}}
    \langle \bar{\nabla}_0 A^a_0(\vec{x},t)P^a(0,0)\rangle
  }{
    \sum_{\vec{x}}\langle P^a(\vec{x},t)P^a(0,0)\rangle
  },
\end{eqnarray}
where $\bar{\nabla}_0$ denotes a symmetric lattice
derivative and $P^a(x)=\bar{\psi}(\tau^a/2)\gamma_5\psi$
\footnote{
We should use the
chirally improved operator,
$P_{\mathrm{imp}}^a(x)=\bar{\psi}(\tau^a/2)\gamma_5(1-\frac{\bar{a}}{2}D)\psi$,
but for the on-shell matrix elements such as the one considered
here, one can use the equation of motion to replace the
$\bar{a}D$ by $-\bar{a}m/(1-\bar{a}m/2)$, which is negligible
for our quark masses.
We therefore use the local operator for $P^a(x)$.
}. 

\begin{figure}[tbp]
\begin{center}
\includegraphics[width=8cm]{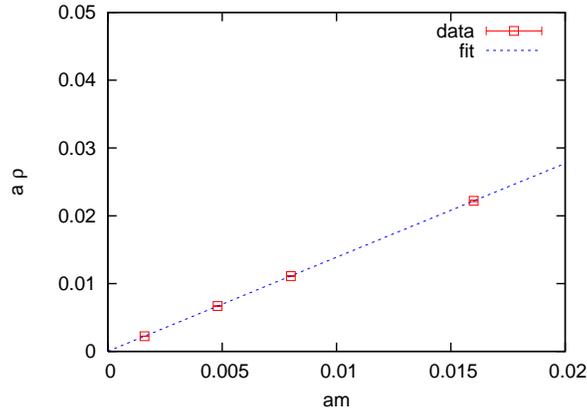}
\caption{
  $a\rho$ as a function of the bare quark mass. 
  Dashed line is a result of the quadratic fit:
  $a\rho=0.00001+1.390(ma)-0.19(ma)^2$. 
}\label{fig:ZAfit}
\end{center}
\end{figure}

Fitting the average of $\rho_m(t)$ over all topological
sectors with a constant in the range $7\leq t \leq 13$,
we obtain 
\begin{equation}
  a\rho(ma)
  \equiv
  \frac{a
    \langle \bar{\nabla_{\mu}}A^a_{\mu}(x)P^a(0)\rangle
  }{
    \langle P^a(x)P^a(0)\rangle
  }
  =
  \frac{2ma}{Z_A}+O(a^2),
\end{equation}
at four quark masses $am$ = 0.0016, 0.0048, 0.008, 0.016,
as plotted in Fig.~\ref{fig:ZAfit}, which shows a very
good chiral behavior.
With a quadratic fit we obtain 
\begin{equation}
  a\rho(ma)=0.00001(2)+1.390(14)(ma)-0.19(74)(ma)^2.
\end{equation}
The constant term is perfectly consistent with zero and we
extract an accurate value, $Z_A$ = 1.439(15)
, which is consistent with the value $Z_A$ = 1.448(4) 
reported in Ref.~\cite{Bietenholz:2004wv} which was done
with the same $\beta$ and $s$.

\begin{figure}[tbp]
\begin{center}
\includegraphics[width=8cm]{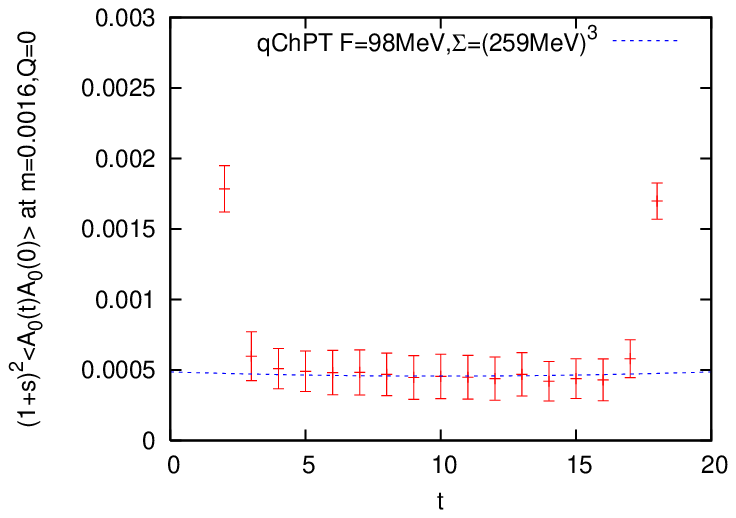}
\includegraphics[width=8cm]{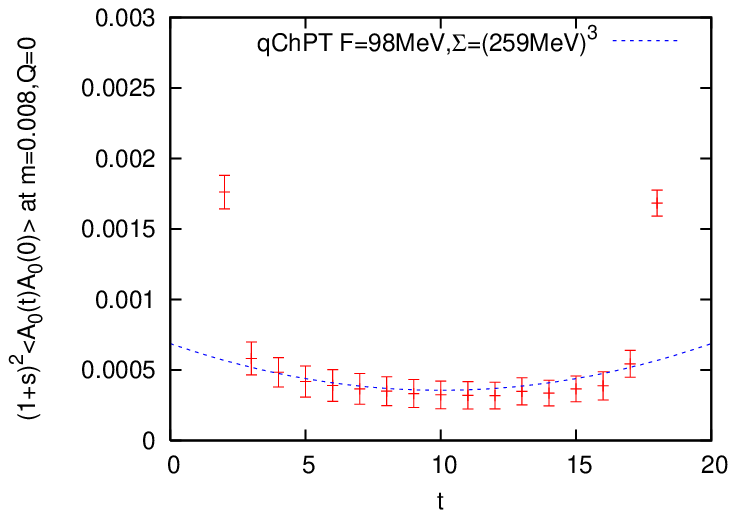}
\includegraphics[width=8cm]{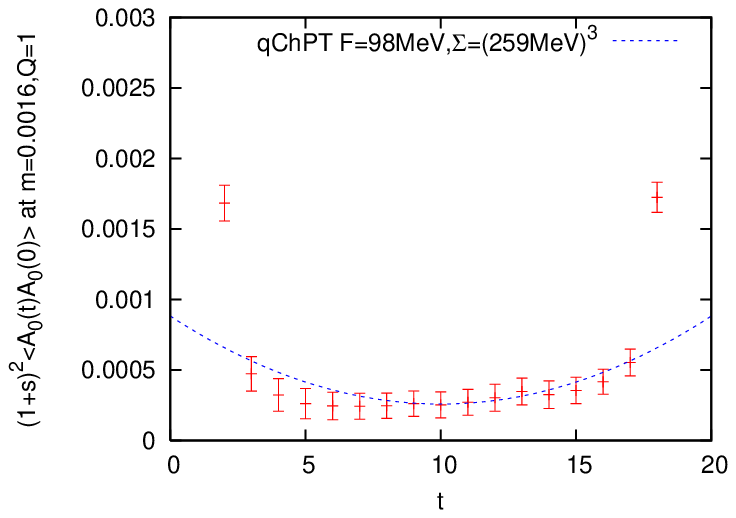}
\includegraphics[width=8cm]{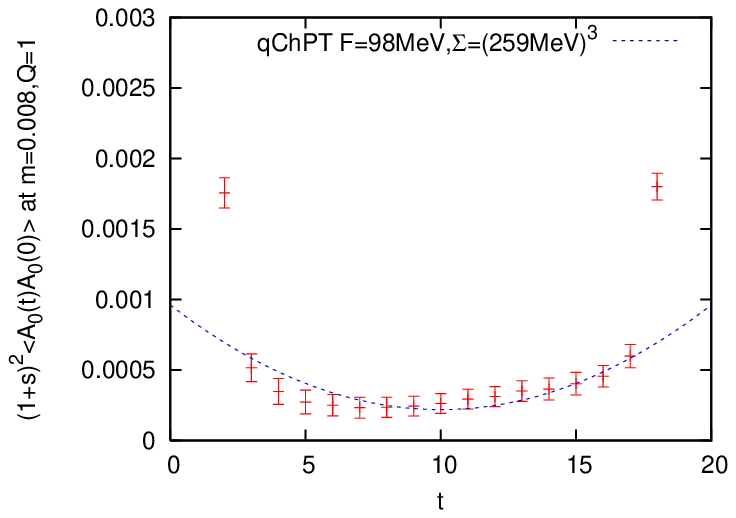}
\caption{
  Axial-vector current correlators at $am$ = 0.0016 (left)
  and 0.008 (right) for $|Q|$ = 0 (top) and 1 (bottom).
  The dashed lines are the result of simultaneous fitting of
  the data for $Q$ = 0 and 1 at $am$ = 0.0016, 0.0048, 0.008
  in the region $7\leq t\leq 13$.
}\label{fig:axialcorr01}
\end{center}
\end{figure}

Now one can compare the renormalized axial-vector correlation 
function with the qChPT result
\begin{equation}
  \label{eq:AA_QChPT}
  2Z_A^2\sum_{\vec{x}}
  \langle A_0(\vec{x},t)A_0(0,0)\rangle^Q 
  = 2\left(\frac{F_{\pi}^2}{T}+2m\Sigma_{|Q|}(\mu)Th_1(|t/T|)
  \right).
\end{equation}
to extract $F_{\pi}$ and $\Sigma$.
In Ref.~\cite{Bietenholz:2003bj} it is reported that the
correlators suffer from large statistical fluctuation when
$|Q|=0$ and the data in the 
other topological sectors are insensitive to
$\Sigma$,  but
it turned out that 
two-parameter ($F_{\pi}$ and $\Sigma$) fitting does 
work well when we treat the data of different topology and
fermion masses simultaneously. 
As shown in Fig.~\ref{fig:axialcorr01}, 
our data at $am$ = 0.0016, 0.0048, 0.008 in the 
$|Q|\leq 1$ sectors are well described by the qChPT formula
(\ref{eq:AA_QChPT}).
A simultaneous fit in the range $7\leq t\leq 13$ yields
$F_\pi$ = 98.3(8.3)~MeV and $\Sigma^{1/3}$ = 259(50)~MeV 
with $\chi^2/\mbox{dof}$ = 0.19.
The result for $F_\pi$ is in a good agreement with that of the
previous work \cite{Giusti:2004yp}, 102(4)~MeV.

\begin{figure}[tbp]
\begin{center}
\includegraphics[width=8cm]{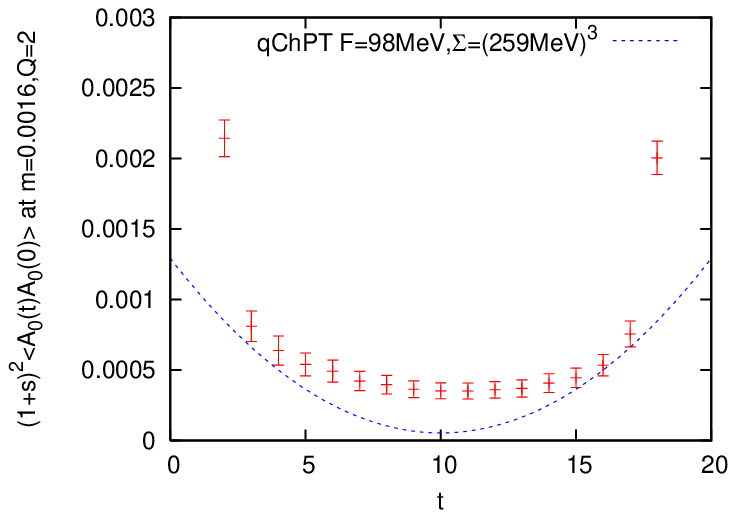}
\includegraphics[width=8cm]{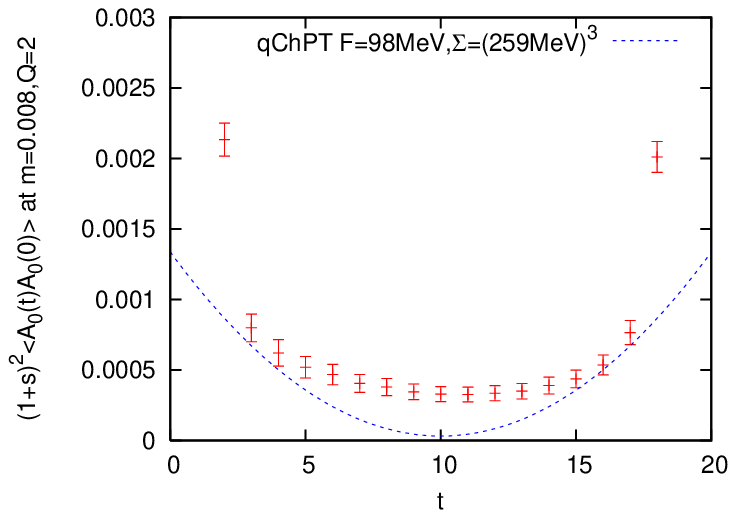}
\caption{
  Axial-vector current correlators at $am$ = 0.0016 (left)
  and 0.008 (right) for $|Q|$ = 2.
  The dashed lines are the qChPT prediction with parameters
  determined through $|Q|$ = 0 and 1 sectors.
}\label{fig:axialcorr2}
\end{center}
\end{figure}

On the other hand, 
the correlators at $|Q|$ = 2 do not agree with the
above fit parameters as shown in
Fig.~\ref{fig:axialcorr2}.
As discussed before, it may indicate that
the topological sector $|Q|$ = 2 is already too large to
apply the qChPT in the $\epsilon$-regime.

\subsubsection{$\Sigma$, $\Sigma_{\mathrm{eff}}$ and $\alpha$
from connected S and PS correlators}

We find that the scalar and pseudo-scalar triplet 
correlators are approximated 
precisely with the lowest 200$+|Q|$ eigen-modes 
at small quark masses ($m$ = 2.6--13~MeV) 
as the right panel of Fig.\ref{fig:accuracy} shows.
In the range $7\leq t\leq 13$, 
the systematic error of this low-mode approximation 
is estimated to be only $1\%$ for
the scalar correlators in $|Q|\geq 1$
sectors and the pseudo-scalar correlators in 
all the topological sectors. 
In this way we measure 
\begin{eqnarray}
  \label{eq:connectedcorr_SS}
  \langle \mathcal{S}(t)\rangle^Q
  \equiv
  -2 \sum_{\vec{x}} (1+s)^2
  \langle S^3(x+x_0)S^3(x_0)\rangle^Q_{\mathrm{low}},
  \\
  \label{eq:connectedcorr_PP}
  \langle \mathcal{P}(t)\rangle^Q
  \equiv
  2 \sum_{\vec{x}} (1+s)^2 
  \langle P^3(x+x_0)P^3(x_0)\rangle^Q_{\mathrm{low}},
\end{eqnarray}
at $am$ = 0.0016, 0.0032, 0.0048, 0.0064, and 0.008.
We take an average of the source point $x_0$ over 
$(L/2)^3\times (T/2)$ lattice sites.
In qChPT formulas
(\ref{eq:tripletmesonSS}) and
(\ref{eq:tripletmesonPP}), 
we have five parameters to be determined:
$F_\pi$, $\Sigma$, $\Sigma_{\mathrm{eff}}$, $m_0^2$ and $\alpha$.
Since these correlators are weakly depending on $F_\pi$,
we use the jackknife samples of $F_\pi$ obtained from
the axial-vector current correlator,
$F_\pi$ = 98.3(8.3)~MeV.
Unfortunately, there still remain too many parameters
to fit with qChPT expressions.
Therefore,
we use the relation (\ref{eq:topsus})
and an input $r_0^4\chi\equiv r_0^4\langle Q^2\rangle/V=0.059(3)$ 
from a recent work \cite{DelDebbio:2004ns}, which
gives $m_0$ = 940(80)(23)~MeV, where the second error 
reflects the error of $r_0^4\chi$.

\begin{figure}[tbhp]
\begin{center}
\includegraphics[width=8cm]{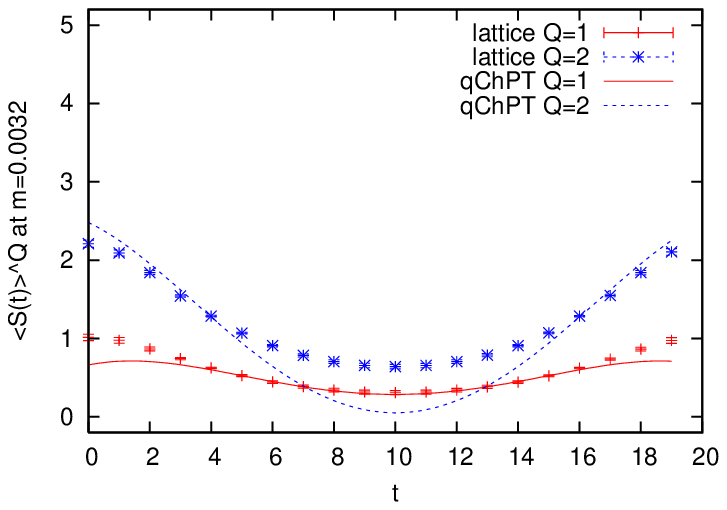}
\includegraphics[width=8cm]{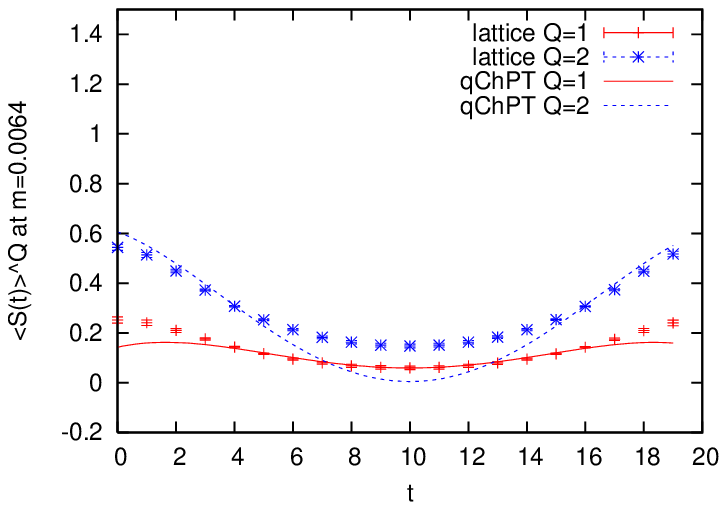}
\includegraphics[width=8cm]{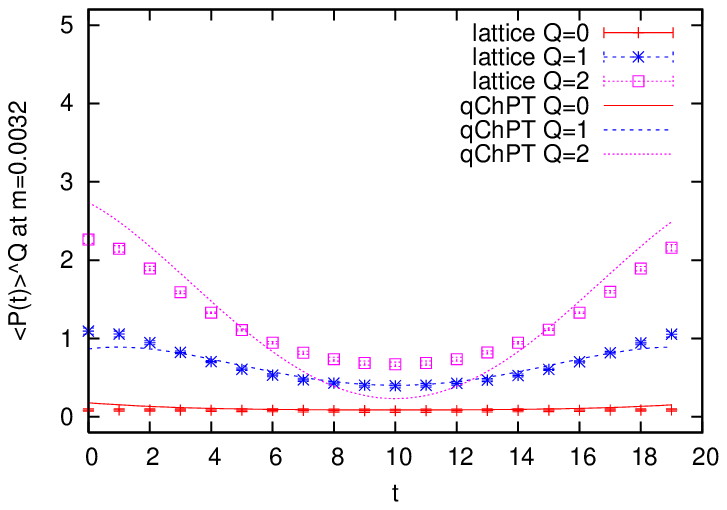}
\includegraphics[width=8cm]{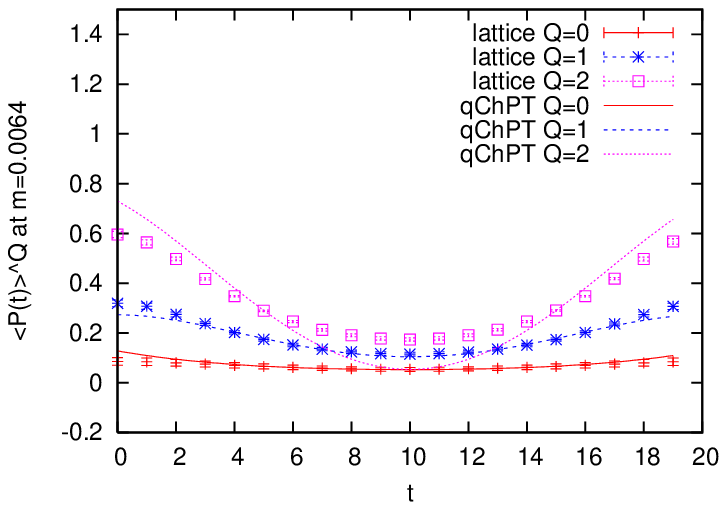}
\caption{
  Scalar (top) and pseudo-scalar (bottom) correlators
  at $am$ = 0.0032 (left) and 0.0064 (right).
  The dotted lines are the fit results with all available
  mass parameters $am$ = 0.0016, 0.0032, 0.0064 and 0.008
  for topological charges 0 and 1.
}\label{fig:connectprop}
\end{center}
\end{figure}

With these input values, we fit the correlators (\ref{eq:connectedcorr_SS}) and
(\ref{eq:connectedcorr_PP}) in the range 
$7\leq t\leq 13$ at different $Q$ and $m$ simultaneously.
Fig.~\ref {fig:connectprop} shows the correlators with fitting curves.
For $|Q|\leq 1$ sectors, the data at all available quark masses
$am$ = 0.0016, 0.0032, 0.0064 and 0.008 are fitted well, and
we actually get
$\chi^2/\mathrm{dof}$ $\sim$ 0.7. (Note that
the correlations between different $t$'s, $m$'s and channels
(PS and S) are not taken into account.)
This fit yields
$\Sigma^{1/3}$ = 257 $\pm$ 14 $\pm$ 00~MeV,
which is consistent with Ref.~\cite{Bietenholz:2003mi},
$\Sigma_{\mathrm{eff}}^{1/3}$ = 271 $\pm$ 12 $\pm$ 00~MeV, and
$\alpha$ = $-$4.5 $\pm$ 1.2 $\pm$ 0.2, where the first error
is the statistical error and the second one is from uncertainty 
of $\langle Q^2\rangle$. 

Here are some remarks. The ratio 
\begin{equation}
  \frac{\Sigma_{\mathrm{eff}}}{\Sigma} = 
  1+\frac{1}{N_cF_\pi} \left(
    m_0^2 \bar{G}(0) + \alpha \bar{\Delta}(0)
  \right),
\end{equation}
should indicate the size of the NLO correction 
in the $\epsilon$ expansion. 
Our data, 1.163(59), implies that the
$\epsilon$ expansion is actually converging.
$\alpha$ has large negative value,
which is also reported in Refs.~\cite{Bietenholz:2005kq} and
\cite{Shcheredin:2005ma}. These results contradict
with a previous precise calculation
\cite{Bardeen:2003qz}, which obtained a smaller value 
$\alpha$ = 0.03(3).
If we instead assume $\alpha$ = 0, and fit the data with 
$F_\pi$ as a free parameter, we obtain $F_\pi$ = 136.9(5.3)~MeV and
$\Sigma$ and $\Sigma_{\mathrm{eff}}$ are almost unchanged.
(Detailed numbers are summarized in
Table~\ref{tab:fitresults}.) 
A possible cause is that 
$|Q|=1$ is not small enough to derive
the partition function Eq.(\ref{eq:Z_Q}) 
(See Appendix \ref{sec:ChPT}).
Eq.(\ref{eq:topsus}) may also have 
a systematic error due to finite $V$ as well as finite $a$. 
Look at the data with $|Q|$ = 2, 
which are plotted in Fig.~\ref {fig:connectprop}.
They do not agree with expectations from the qChPT
shown by dashed curves in the plots, which is
also seen in the case with the axial-vector correlators.
A simultaneous fit with all the data including $|Q|$ = 0, 1
and 2 gives a bad $\chi^2/\mathrm{dof}$ ($\simeq$ 12).
Thus our numerical data might be posing a problem
whether the partition function of qChPT $Z_Q$ 
is reliable or not, when $|Q|/\langle Q^2\rangle \ll 1$ 
is not satisfied very much; 
$\langle Q^2\rangle\sim 4.34(22)$ on our lattice. 

\newpage
\subsubsection{Chiral condensates}

\begin{figure}[tbp]
\begin{center}
\includegraphics[width=8cm]{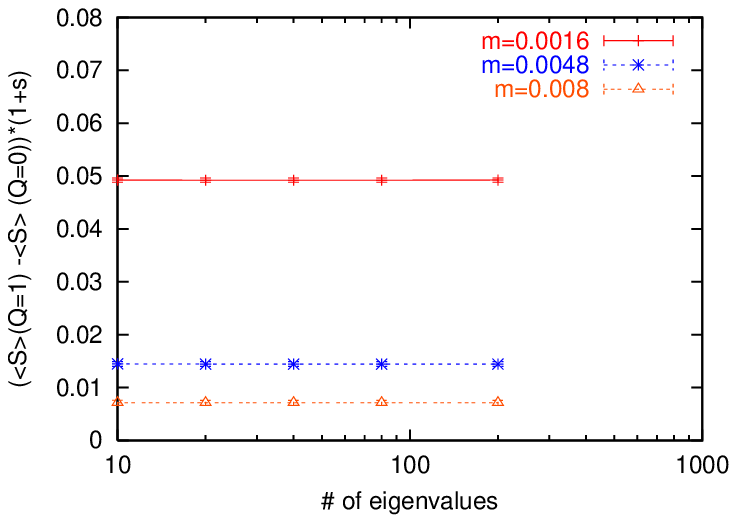}
\includegraphics[width=8cm]{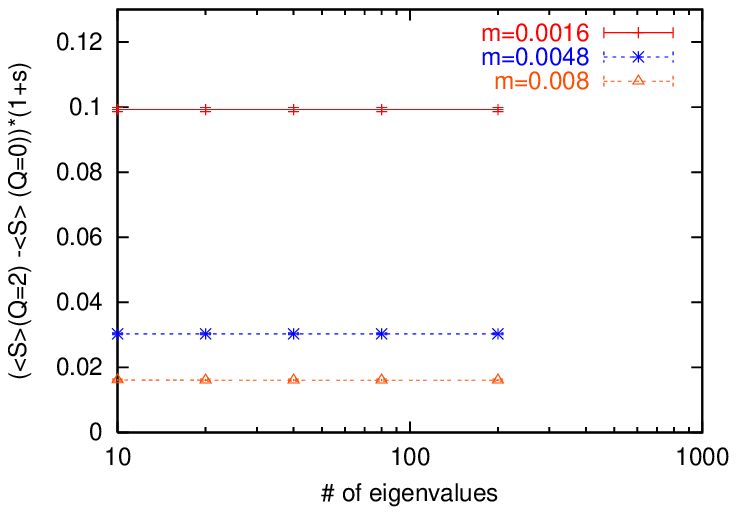}
\caption{
  Low-mode dominance of 
  $ -(\langle \bar{\psi}\psi\rangle_{\mathrm{low}}^Q - 
\langle \bar{\psi}\psi\rangle_{\mathrm{low}}^0)$
as a function of $N_{\mathrm{low}}$.
The data are averaged over 
20 configurations at  $Q=1$ (left) and $Q=2$ (right).
}\label{fig:saturationconden2}
\end{center}
\end{figure}

\begin{figure}[htbp]
\begin{center}
\includegraphics[width=8cm]{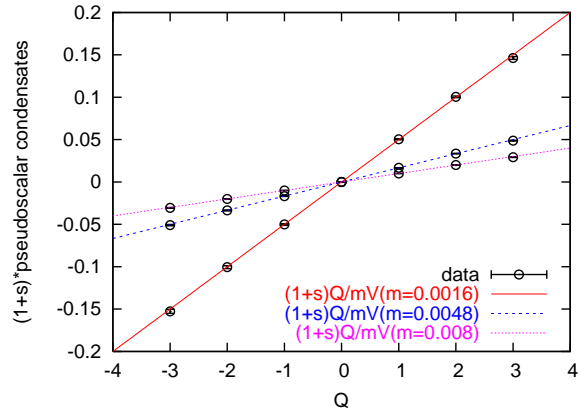}
\caption{
  Pseudo-scalar condensates approximated with 
  $N_{\mathrm{low}}=200+|Q|$
  low-modes.
  The lines represent the expectation (\ref{eq:pscond}).
}\label{fig:psconden}
\end{center}
\end{figure}

Consider the low-mode contribution to the scalar and
pseudo-scalar condensates with a fixed topological charge $Q$, 
\begin{eqnarray}
  \langle \bar{\psi}\psi(x)\rangle^Q 
  &=&
  - \langle \mbox{tr}D_m^{-1}(x,x)\rangle^Q
  \nonumber\\
  &=& 
  -\left\langle \mbox{tr}\left(
      \sum_{i=1}^{N_{\mathrm{low}}}
      \frac{1}{(1-\bar{a}m/2)\lambda_i+m}v_i(x)v_i^{\dagger}(x)
      +\Delta D_m^{-1}(x,x)
    \right)\right\rangle^Q,
  \\
  \langle \bar{\psi}\gamma_5\psi(x)\rangle^Q 
  &=&
  - \langle\mbox{tr}\gamma_5D_m^{-1}(x,x)\rangle^Q
  \nonumber\\
  &=&
  -\left\langle\mbox{tr}\left(
      \sum_{i=1}^{N_{\mathrm{low}}}\frac{1}
      {(1-\bar{a}m/2)\lambda_i+m}\gamma_5v_i(x)v_i^{\dagger}(x)
      +\gamma_5\Delta D_m^{-1}(x,x)
    \right)\right\rangle^Q.\nonumber\\
\end{eqnarray}
If $N_{\mathrm{low}}$ is large enough, 
the higher mode contribution, $\Delta D_m^{-1}(x,x)$, should
be insensitive to the link variables $U_{\mu}(y)$ 
if $|y-x|\gg 0$ and thus to the global structure of
the gauge field configuration, such as the topological charge. 
We, therefore, expect that the
difference of the scalar condensates with different topology
can be approximated with low-modes,
\begin{equation}
\label{eq:scdiff}
  -(\langle \bar{\psi}\psi(x)\rangle^Q 
  - \langle \bar{\psi}\psi(x)\rangle^0)
\sim -(\langle \bar{\psi}\psi(x)\rangle_{\mathrm{low}}^Q 
  - \langle \bar{\psi}\psi(x)\rangle_{\mathrm{low}}^0).
\end{equation}
In fact, Fig.\ref{fig:saturationconden2} shows that 
the convergence up to $N_{\mathrm{low}}=200+|Q|$ 
is really good.
For the pseudo-scalar condensate, 
the situation is much easier, since the condensate is
determined by the zero modes only;
\begin{equation}
\label{eq:pscond}
  - \langle \bar{\psi}\gamma_5\psi\rangle_Q = \frac{Q}{mV}.
\end{equation}
Note that the contributions from
non-zero eigen-modes cancel 
because of the orthogonality among different eigenvectors. 
As shown in Fig.~\ref{fig:psconden}, our data with
$N_{\mathrm{low}}=200+|Q|$ low-modes perfectly agree
with this theoretical expectation.

\begin{figure}[hbtp]
\begin{center}
\includegraphics[width=8cm]{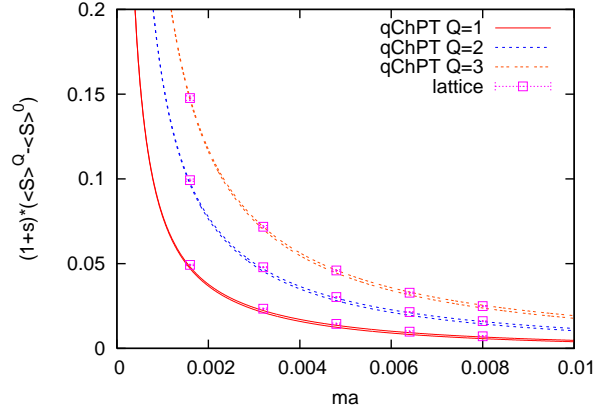}
\caption{
  $-(\langle \bar{\psi}\psi\rangle^Q
  -\langle\bar{\psi}\psi\rangle^0)$
  as a function of quark mass.
  Data points are at $am$ = 0.0016, 0.0032, 0.0048, 0.0064,
  0.008.
  The lines are qChPT predictions with
  $\Sigma_{\mathrm{eff}}^{1/3}$ = 271(12)~MeV.
}\label{fig:scalarconden}
\end{center}
\end{figure}

The free parameter in the scalar condensate is
$\Sigma_{\mathrm{eff}}$ as seen in (\ref{eq:scond}) and
(\ref{eq:muprime}).
We compare our numerical result of Eq.(\ref{eq:scdiff})
with the qChPT result $\Sigma_Q(\mu')-\Sigma_{Q=0}(\mu')$.
We use the low-mode averaging over 
$(L/2)^3\times (T/2)$ lattice sites.
Fig.~\ref{fig:scalarconden} shows the results as a
function of quark mass for $|Q|$ = 1, 2 and 3.
The lattice data agree remarkably well with
the qChPT expectation with $\Sigma_{\mathrm{eff}}$ =
271(12)~MeV as determined from the (pseudo-)scalar connected
correlators as presented in the previous section. 
In fact, if we fit these data with
$\Sigma_{\mathrm{eff}}$ as a free parameter, we obtain
256(14)~MeV which is consistent with the result above.

\subsubsection{Disconnected PS correlators}
\begin{figure}[tbp]
\begin{center}
\includegraphics[width=8cm]{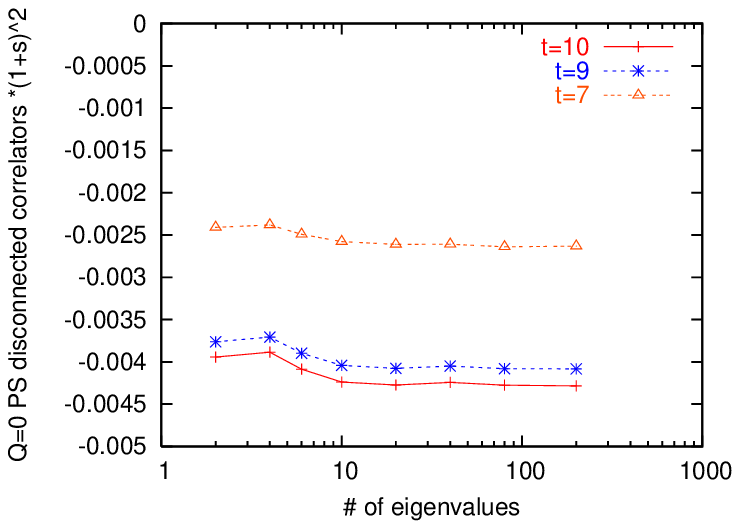}
\includegraphics[width=8cm]{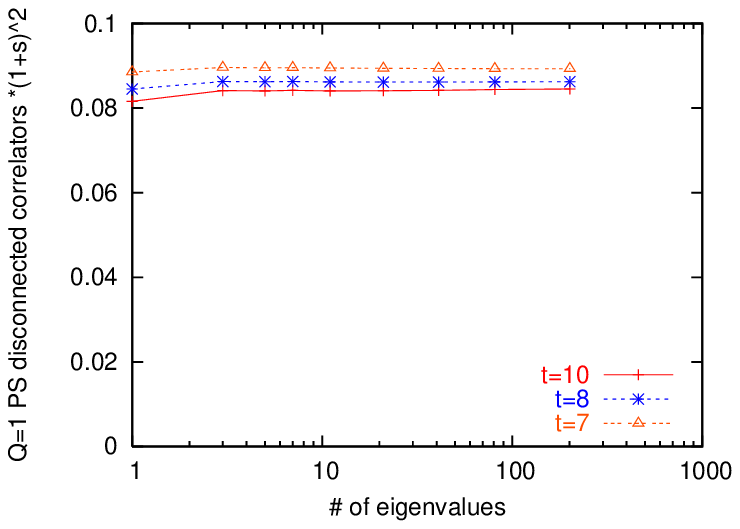}
\caption{
  Convergence of the pseudo-scalar disconnected correlators
  for one sample configuration
  at $am=0.008$ for $Q=0$ (left) and $Q=1$ (right).
}\label{fig:saturationdis}
\end{center}
\end{figure}

Let us rewrite the disconnected pseudo-scalar correlator in a 
fixed topological sector,
\begin{eqnarray}
  \langle P(t)P(0)\rangle_{\mathrm{disc}}^Q 
  & \equiv & 
  \left\langle
    \sum_{\vec{x}} \mbox{tr}\left(
      \gamma_5 D_m^{-1}(\vec{x},t; \vec{x},t)
    \right)
    \mbox{tr}\left(
      \gamma_5 D_m^{-1}(\vec{0},0; \vec{0},0)
    \right)
  \right\rangle^Q,
  \nonumber\\
  & \stackrel{t\gg0}{\longrightarrow}& 
  \left\langle
    \sum_{\vec{x}}\mbox{tr}\left(
      \gamma_5 
      \sum_{i=1}^{200+|Q|}
      \frac{1}{(1-\bar{a}m/2)\lambda_i+m}
      v_i(x)v_i^{\dagger}(x)
    \right)\right.
  \nonumber\\
  &&
  \left.
    \times\mbox{tr}\left(
      \gamma_5 \sum_{i=1}^{200+|Q|}
      \frac{1}{(1-\bar{a}m/2)\lambda_i+m}
      v_i(0)v_i^{\dagger}(0)
    \right)
  \right\rangle^Q
  \nonumber\\
  & + &
  2 \left\langle\mbox{tr}
    \left(\gamma_5\Delta D_m^{-1}(x,x)\right)
  \right\rangle^{\prime}
  \nonumber\\
  &&
  \times \left\langle \mbox{tr}
    \left(\gamma_5
      \sum_{i=1}^{200+|Q|}
      \frac{1}{(1-\bar{a}m/2)\lambda_i+m}v_i(0)v_i^{\dagger}(0)
    \right)
  \right\rangle^Q
  \nonumber\\
  & + &
  \left[
    \left\langle \mbox{tr}
      \left(\gamma_5\Delta D_m^{-1}(x,x)\right)
    \right\rangle^{\prime}
  \right]^2,
\end{eqnarray}
where $x=(\vec{x},t)$.
Here we assume that higher mode's contribution does not have
correlation with any local operator $O(y)$ separated enough
from $x$, {\it i.e.}
\begin{eqnarray}
  \langle \Delta D_m^{-1}(x,x) O(y) \rangle^Q
  \stackrel{|x-y|\gg0}{\longrightarrow}  
  \langle \Delta D_m^{-1}(x,x) \rangle'
  \times \langle O(y) \rangle^Q,
\end{eqnarray}
where the expectation value $\langle\cdots\rangle'$
represents insensitivity to the topological charge.
We also use the translational invariance
$\langle O(x)\rangle=\langle O(0)\rangle$.
Since we know 
$\langle \mbox{tr}\left(\gamma_5\Delta D_m^{-1}(x,x)\right)
\rangle^{\prime}=0$, the above equation 
should be well approximated with low-modes.
In fact, Fig.~\ref{fig:saturationdis} shows
a very good convergence with the lowest 200+$|Q|$ eigen-modes. 
Similar results were also obtained previously
in the study of the $\eta'$ propagator with the Wilson fermion
\cite{Neff:2001zr} and with the overlap fermion \cite{DeGrand:2002gm}.

In the qChPT the correlator is written as
\begin{eqnarray}
  \label{eq:disconnected}
  \langle \mathcal{P}^d(t)\rangle^Q 
  &\equiv&
  \int\! d^3x\, (1+s)^2 
  \langle 2P^3(x)P^3(0)-P^0(x)P^0(0)\rangle^Q
  \nonumber\\
  & = &
  \int\! d^3x\, (1+s)^2
  \left[
    C_P^d - \frac{\Sigma^2}{2F_{\pi}^2}\left(
      \frac{d_+}{N_c}(m_0^2\bar{G}(x)+\alpha
      \bar{\Delta}(x))
      -e_+\bar{\Delta}(x)
    \right)
  \right],
\end{eqnarray}
where
\begin{eqnarray}
  C_P^d  =  \frac{Q^2}{m^2V^2},
  &
  d_+  =  -4\left(1+\frac{Q^2}{\mu^2}\right),
  &
  e_+  = 
  -2\left(\left(\frac{\Sigma_Q(\mu)}{\Sigma}\right)^{\prime}
    -\frac{\Sigma_Q(\mu)}{\mu\Sigma}\right).
\end{eqnarray}

\begin{figure}[tbp]
\begin{center}
\includegraphics[width=8cm]{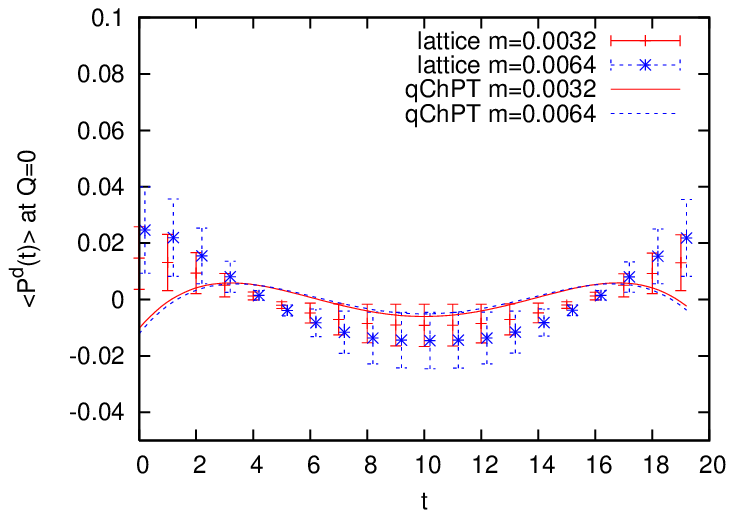}
\includegraphics[width=8cm]{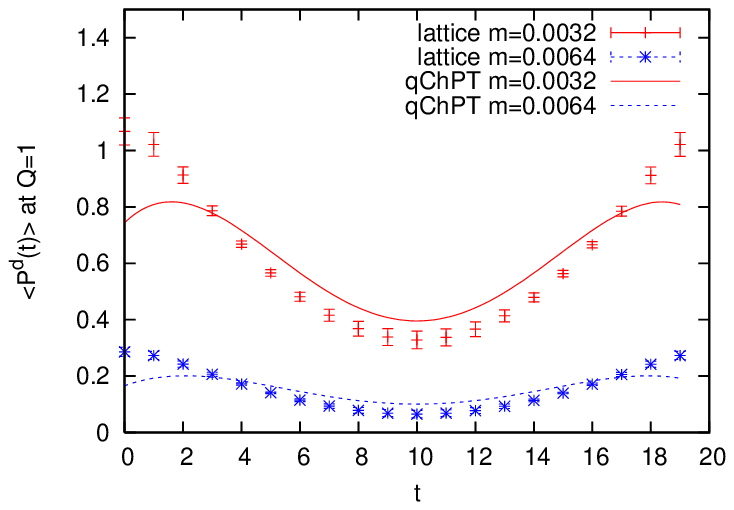}
\includegraphics[width=8cm]{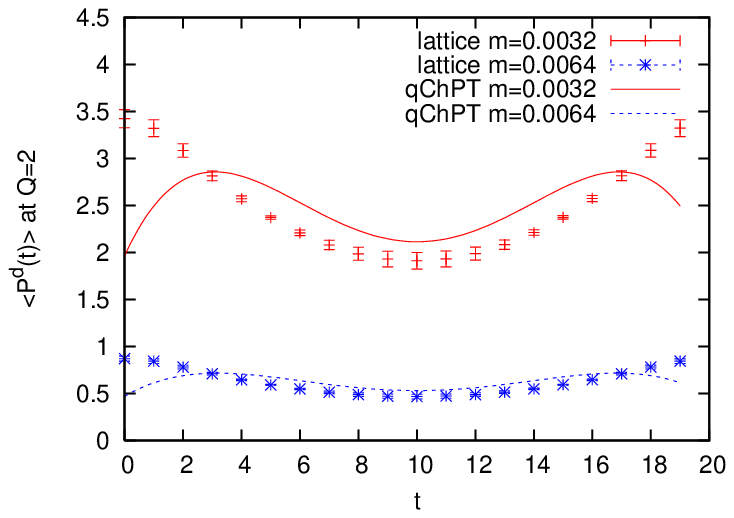}
\includegraphics[width=8cm]{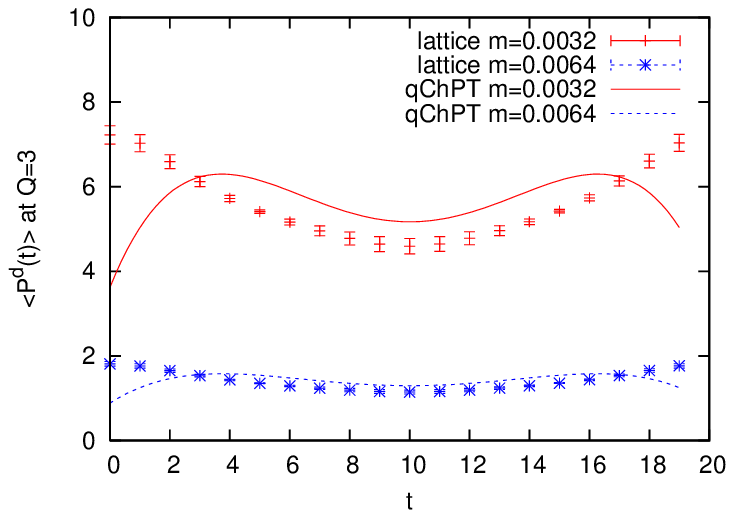}
\caption{
  Disconnected pseudo-scalar correlators in the $0\leq
  |Q|\leq 3$ sectors 
  at $am$ = 0.0032 and 0.0064.
  The curves represent the results of qChPT
  with $\Sigma^{1/3}$ = 257~MeV,
  $F_{\pi}$ = 98.3~MeV, $m_0$ = 940~MeV and 
  $\alpha$ = $-$4.5. 
}\label{fig:disconnectprop}
\end{center}
\end{figure}

In  Fig.~\ref{fig:disconnectprop}, 
we present the data in $|Q|$ = 0--3
topological sectors at two
representative quark masses $am$ = 0.0032 and 0.0064.
The qChPT predictions are plotted with the parameters
determined from the axial-vector and (pseudo-)scalar
connected correlators:
$\Sigma^{1/3}$ = 257~MeV,
$F_{\pi}$ = 98.3~MeV, $m_0$ = 940~MeV, and 
$\alpha$ = $-$4.5. 
One finds that the agreement is marginal, though the
correlator's magnitude and shape are qualitatively well
described.
Instead, if we fit the disconnected correlator with
$\Sigma$ and $\alpha$ as free parameters while fixing
$F_{\pi}$ and $m_0$ to the same value, we obtain
$\Sigma^{1/3}$ = 227(32)~MeV and $\alpha$ = $-$3.5(1.2),
which are statistically consistent with those input
numbers. 
We conclude that not only the connected correlators,
but also disconnected correlators can be consistently 
expressed by the qChPT in the $\epsilon$-regime
when $|Q|$ is small.
Details of the fit results are listed in
Table~\ref{tab:fitresults}.

\begin{table}[btp]
\rotatebox[origin=c]{90}{
\begin{minipage}{\textheight}
\caption{
  Summary of the fitting results in the $\epsilon$-regime.
  The first column denotes the topological sectors used in
  the fit.
  The values in $[\cdots]$ are input parameters.
  The first error is statistical.
  The second and third errors reflect the uncertainty in the
  input parameters,
  $\langle Q^2 \rangle$ and $F_{\pi}$, respectively.
}
\begin{center}
\let\tabularsize\footnotesize
\begin{tabular}{ccccccc}
\hline
\hline
correlators
& $F_{\pi}$(MeV) & $\Sigma^{1/3}$ (MeV) &
$\Sigma_{\mathrm{eff}}^{1/3}$ (MeV)
& $\alpha$ & $m_0$ (MeV) & $\chi^2/\mathrm{dof}$ \\
\hline
\multicolumn{7}{l}{axial vector}\\
{\footnotesize $|Q|=0$} & 98(17) & 279(65) &&&&0.02\\
$0\leq |Q|\leq 1$ & 98.3(8.3) & 259(50) & & & & 0.19\\
$0\leq |Q|\leq 2$ & 117.9(4.3) & 335(16) & & & & 2.8 \\
\hline
\multicolumn{7}{l}{connected PS+S}\\
$0\leq |Q|\leq 1$ & [98.3(8.3)] & 257(14)(00)
&271(12)(00) &$-$4.5(1.2)(0.2) &[940(80)(23)] & 0.7\\
$0\leq |Q|\leq 1$ & 136.9(5.3)(0.9) & 250(13)(00)
&258(11)(00) & [0] &[674(26)(16)] & 0.3\\
$0\leq |Q|\leq 2$ & [98.3(8.3)] & 258(12)(00)
&264(11)(00) &$-$3.8(0.5)(0.2) &[940(80)(23)] & 11.8\\
\hline
\multicolumn{7}{l}{disconnected PS}\\
$0\leq |Q|\leq 1$ & [98.3(8.3)] & 227(32)(00)
& &$-$3.5(1.2)(0.3) &[940(80)(23)] & 1.0\\
$0\leq |Q|\leq 1$ & 125.7(5.6)(0.9) & 223(29)(00)
& & [0] &[734(33)(14)] & 0.7\\
$0\leq |Q|\leq 2$ & [98.3(8.3)] & 229(33)(00)(03)
& &$-$3.6(0.2)(0.3)(1.0) &[940(80)(23)] & 1.0\\
$0\leq |Q|\leq 2$ & 135.0(4.9)(1.4) & 237(32)(00)
& & [0] &[684(25)(13)] & 1.7\\
$0\leq |Q|\leq 3$ & [98.3(8.3)] & 229(33)(01)(05)
& &$-$3.6(0.1)(0.2)(0.8) &[940(80)(23)] & 1.0\\
$0\leq |Q|\leq 3$ & 139.3(4.1)(1.4) & 244(32)(00)
& & [0] &[663(19)(12)] & 1.9\\
\hline
\multicolumn{7}{l}{scalar condensate 
($\langle \bar{\psi}\psi\rangle^0-\langle \bar{\psi}\psi\rangle^Q$)}\\
$1\leq |Q|\leq 3$ & & & 256(14) & & &1.2\\
\hline
\end{tabular}
\end{center}
\label{tab:fitresults}
\end{minipage}
}
\end{table}

\newpage
\section{Conclusions and discussions}\label{sec:conclusion}

The exact chiral symmetry is established on the lattice,
with the overlap Dirac operator which satisfies the 
Ginsparg-Wilson relation.
However, the overlap Dirac operator 
becomes ill-defined 
at certain points where the zero-mode of $H_W$ appears.
Also, practically, these points make the numerical study 
very difficult.
One has to carefully evaluate the discontinuity of the 
overlap fermion determinant (reflection/refraction)
and the polynomial or rational expression 
of the overlap Dirac operator itself goes worse near
this discontinuity.
In fact, these dangerous points are topology boundaries;
it is known that crossing $H_W=0$ changes the
index of the overlap Dirac operator, or topological charge. 

In the continuum limit, $H_W=0$ is automatically excluded
but it would be better if we can construct the lattice 
gauge theory which does not allow $H_W=0$ from the beginning
with a finite lattice spacing. 
There have been proposed two promising strategies.
One is the gauge action which satisfies 
L\"uscher's ``admissibility'' condition and the other
is an additional fermion determinant, $\det H_W^2$.
Both of them are designed for the use of the hybrid Monte Carlo 
algorithm, in which global and small (smooth) 
updates are performed.
Therefore, we regard them as ``topology conserving actions''. 
In this thesis, 
we have investigated the possibility of lattice QCD in a fixed
topological sector.
We studied the properties of the topology conserving
actions with no light quarks (namely, pure $SU(3)$ theory.). 
Although the admissibility condition with 
small $\epsilon$ ($<$ 1/20) can strictly prohibit the 
topology changes, our interest is the case with 
$\epsilon\sim O(1)$ for practical purposes.
In the (quenched) Hybrid Monte Carlo updates, 
we found that the topology change is strongly suppressed for
$1/\epsilon$ = 2/3 and 1, compared to the standard Wilson
plaquette gauge action.
The topological charge seems more stable for finer
lattices, and it is possible to preserve the topological
charge for O(100)-O(1,000) HMC trajectories at $a\simeq$ 0.08~fm
and $L\simeq$~1.3~fm, which might be applicable to the
$\epsilon$-regime.
The action Eq.(\ref{eq:admiaction}) 
has, thus, been proved to be useful 
to accumulate gauge configurations 
in a fixed topological sector.
While the gauge action Eq.(\ref{eq:admiaction}) 
with $1/\epsilon = 1$ and 2/3 allows the topology changes,
at times, the topology conservation with the
fermion determinant $\det H_W^2$ seems perfect.
The numerical cost for this determinant is, of course,
much more expensive than the quenched case, but it would be
negligible compared to the cost of the 
dynamical overlap fermion.
The full QCD overlap fermion simulation with
the additional determinant $\det H_W^2$ may be very 
efficient, if we can omit the reflection/refraction procedures.

To test their practical feasibility, 
we measured the heavy quark potential with these actions.
The lattice spacing is determined from the Sommer scale $r_0$.
With these measurements we also investigated the scaling
violation at short and intermediate distances.
The probe in the short range is the violation of the 
rotational symmetry, and a ratio $r_c/r_0$ of two different
scale can be used for the intermediate distance.
For both of these we found that the size of the scaling
violation is comparable to the case with 
the Wilson plaquette gauge action,
which is consistent with the expectation that the term with
$1/\epsilon$ introduces a difference at most $O(a^4)$ or
the additional large negative mass Wilson fermions are decoupled.
These actions show no disadvantage as
far as Wilson loops are concerned.
We also found that the perturbative expansion of 
the coupling (after mean-field improvement)
shows very good convergence even if 
$1/\epsilon$ term is introduced.
The coupling constant in a certain scheme at a given scale
is consistent among different values of $1/\epsilon$.

As another advantage of the (approximate) topology conservation, 
the low-lying eigenvalues of the Wilson-Dirac operator in the
negative mass regime is suppressed.
This reduces the cost of the numerical implementation of the 
overlap-Dirac operator.
We observed that the topology conserving actions have
a gain about a factor 2--3 at the same
lattice spacing compared to the standard Wilson gauge action.
Comparison with the other improved gauge actions, such as the
L\"uscher-Weisz, Iwasaki and DBW2 would be interesting.\\

In a fixed topological sector,
one of the very interesting applications would be
the QCD in the $\epsilon$-regime.
In the chiral limit, 
the (q)ChPT analysis shows that the meson correlators
are largely affected by the fermion zero-modes, and thus by
the topology of the background gauge field.
In order to study lattice QCD in such a regime,
the chiral symmetric Dirac operator is essential, otherwise
the large lattice artifacts would contaminate the fundamental
points of the analysis, such as the definition
of the topological charge, or what is the zero-mode, etc.
To demonstrate how much effective the overlap Dirac operator
is, we studied the quenched QCD with very small quark masses
in the range $2.6$--13MeV.
In this chiral regime, we observed that the chiral behavior is
really nice, as seen in, for example, 
accurate determination of $Z_A$, or the pseudo-scalar condensates 
$\bar{\psi}\gamma_5\psi$.
We also found the importance of the low 
lying modes and used them to extract the low-energy constants.
(Note that the low lying modes are not affected by
the $a\to0$ limit, since it would add 
higher modes only, which are 
irrelevant to the low energy dynamics.)
From triplet meson correlators with $Q=0$ and 1, 
we extracted 
$F_\pi$ = 98.3(8.3)~MeV, 
$\Sigma^{1/3}$ = 257(14)(00)~MeV
($\Sigma_{\mathrm{eff}}^{1/3}$ = 271(12)(00)~MeV),
$m_0$ = 940(80)(23)~MeV, and $\alpha$ = $-$4.5(1.2)(0.2).
In these numerical results the second error reflects the
error of $\langle Q^2 \rangle$.
We also obtained consistent results from disconnected
pseudo-scalar correlator and the chiral condensates.
Although we observed minor inconsistencies and problems 
due to finite $Q$ correction, which is a special restriction 
of the quenched ChPT, these remarkable successes
would encourage us to go further to $N_f\neq 0$ full QCD
in the $\epsilon$-regime.\\

The exact chiral symmetry with the overlap 
Dirac operator is surely a most remarkable progress in
the lattice gauge theory and it would be more and more
important in both of the numerical works and theoretical 
works, in the future.
In order to perform the path integrals, 
one should, however, note the fact that there is 
several points where $H_W$ has zero modes, which makes 
the overlap Dirac operator ill-defined, 
its locality is doubtful, and the numerical cost is 
suddenly enhanced.
Our observation shows that the lattice QCD with fixed topology 
would be one of interesting and promising solutions to this.
A number of applications such as $\theta$-vacuum, 
finite temperatures, and so on, might be possible as well.\\

For the future works, we would like to give a few remarks.

We have observed a large $\beta$-shift when
the negative mass Wilson fermion Eq.(\ref{eq:nwf}) 
is added, which may also 
cause a unwanted large scaling violation.
We thus propose another ``topology stabilizer''
which would have effects on $H_W\sim 0$ modes only;
\begin{equation}
\label{eq:nwf2}
\det \frac{a^2H_W^2}{a^2H_W^2+a^2m_t^2},
\end{equation}
where the denominator is corresponding to
the twisted mass ghost 
\cite{Frezzotti:1999vv,Frezzotti:2000nk,Frezzotti:2001du}

with a large negative mass
$-(1+s)/a$ and a small twisted mass $m_t$. 
The effect from both determinant
should be canceled unless $H_W$ has small eigenvalues.
Note that $\det H_W^2/(H_W^2+a^2m_t^2)\to 1$ with $m_t$ fixed.

As a final remark, 
let us consider the contribution from different topological sectors.
In each topological sector, one measures the expectation value of
operator $O$ with a fixed topological charge,
\begin{eqnarray}
\langle O \rangle^Q &\equiv &
\frac{\int dU^Q  O \det (D+m)^{N_f} e^{-S_G}}
{\int dU^Q  \det (D+m)^{N_f} e^{-S_G}}
\equiv
\frac{\int dU^Q  O \det (D+m)^{N_f} e^{-S_G}}{Z^Q},
\end{eqnarray}
where $dU^Q$ denotes the integral over the gauge fields with 
topological charge $Q$ and $\det (D+m)^{N_f}$ is the determinant
of $N_f$ flavor overlap fermion with quark mass $m$. 
In order to calculate the total expectation value 
(in $\theta$-vacuum);
\begin{eqnarray}
\langle O \rangle^{\theta} &=&
\frac{\sum_Q e^{i\theta Q} \int dU^Q  O \det (D+m)^{N_f} e^{-S_G}}
{\sum_Q e^{i\theta Q} \int dU^Q  \det (D+m)^{N_f} e^{-S_G}}
= \frac{\sum_Q e^{i\theta Q}\langle O \rangle^Q Z_Q/Z_0}
{\sum_Q e^{i\theta Q} Z_Q/Z_0},
\end{eqnarray}
the ratio $Z_Q/Z_0$ has to be evaluated.
In fact, there are several proposals to calculate this ratio of the
partition function,
including the tempering method
\cite{Marinari:1992qd, Hukushima,Marinari:1996dh,Joo:1998ib,
Ilgenfritz:2001jp}.
Here, we would like to propose an easier method.
With an assumption that the ratio should be Gaussian;
\begin{equation}
Z_Q/Z_0 \propto 
\exp\left(-\frac{Q^2}{2\langle Q^2 \rangle^{\theta=0}}
\right), 
\end{equation}
in a large volume $V$, one only has to calculate 
the topological susceptibility
\cite{Giusti:2004qd, Luscher:2004fu},
\begin{equation}
\chi = \frac{\langle Q^2\rangle^{\theta=0}}{V}, 
\end{equation}
to evaluate the ratio $Z_Q/Z_0$ with any value of $Q$.
In principle, the topological susceptibility can be
evaluated with the local topological charge 
density operator $q(x)=-\mbox{Tr}\gamma_5 \bar{a}D(x,x)/2$
in a semi-local region (assuming the cluster decomposition principle);
\begin{eqnarray}
\chi = \int d^4x \langle q(x)q(0) \rangle^{\theta=0}
= \int_{V^{\prime}} d^4x \langle q(x)q(0) \rangle^{\theta=0},
\end{eqnarray}
where $\int_{V^{\prime}}d^4 x$ is performed over the 
small volume $V^{\prime}$ around the origin,
where $V^{\prime}\gg 1/\Lambda_{QCD}$.
Thus, if the topological susceptibility in a fixed 
topological sector,
\begin{eqnarray}
\chi^Q \equiv 
\int_{V^{\prime}} d^4x \langle q(x)q(0) \rangle^Q,
\end{eqnarray}
shows sufficiently small $Q$ dependence and the finite volume effects
due to $V$ and $V^{\prime}$ are both negligible,
then $\chi^Q$ should be a good approximation of $\chi$
\footnote{In this argument we omit the 
renormalization or mixing of $q(x)$ for simplicity.}.
What we would like to emphasize here is 
that summing up all the topology with the weight $Z_Q/Z_0 $
(in a $\theta$-vacuum) is not so difficult and maybe not so important
unless $\theta$ term is considered\footnote{In Ref.\cite{Brower:2003yx},
it is shown that the observables in a fixed topology are equivalent to
those in the $\theta=0$ vacuum up to correction terms 
proportional to $V^{-1}$.},
since $Z_Q/Z_0 \to_{V\to \infty} 1$, although the numerical studies
has to be done in the future works to verify this quite optimistic argument.
It seems an appropriate way to preserve the topology along the
simulations, since any lattice gauge action would eventually create
large barriers between the topological sectors in the continuum limit.

\section*{ACKNOWLEDGMENTS}\label{sec:acknowlegments}

I would like to thank Prof. Tetsuya Onogi, 
my supervisor, for his great advices, encouragements 
and many nice lectures.
I also thank S.Hashimoto, K.Ogawa, T.Hirohashi and
H.Matsufuru for the collaborations, on which
I really enjoyed and discussed many interesting topics.
This work is greatly owing to M.L\"uscher who gave me
a lot of crucial advices and helped me staying in Geneva
very much when I visited CERN.
Also I would like to thank L.Del Debbio, L.Giusti, 
C.Pena, S.Vascotto and all the members of CERN
for fruitful discussions and 
their very warm hospitality during my stay
which was really impressive experience for me.
I acknowledge W.Bietenholz, K.Jansen and S.Shcheredin
for giving me many meaningful advices.
Many instructions on computational works are given by 
T.Umeda, and I would like to express special thanks to him.
I thank all the members of YITP and the Department of 
Physics of Kyoto Univ. for happy everyday life. 

Simulations are done on 
NEC SX-5 at RCNP, Alpha workstations at YITP
, and SR8000 and Itanium2 workstations at KEK.\\

Finally I would like to thank my parents, my brother,
and my grandmothers, 
for continuous encouragement and supports.

\appendix
\section{Notations}\label{sec:notations}

Lorentz indices $\mu, \nu$ run from 0 to 3.
The lattice spacing and size are denoted by $a$ and $L$ respectively.
The gauge field $U_{\mu}(x)\in SU(3)$ is located on the link from
$x$ to $x+\hat{\mu} a$, where $\hat{\mu}$ is the unit vector in
direction $\mu$. 
The plaquette is denoted by $P_{\mu\nu}(x)=U_{\mu}(x)U_{\nu}(x+a\hat{\mu})
U^{\dagger}_{\mu}(x+a\hat{\nu})U^{\dagger}_{\nu}(x)$. 
The fermion field $\psi(x)$ is located on the site $x$.
The forward and backward covariant difference of the 
fermion field $\psi(x)$ are defined by
\begin{eqnarray}
\label{eq:diffoperators}
\nabla_{\mu}\psi(x) &=& \frac{U_{\mu}(x)
\psi(x+\hat{\mu}a)-\psi(x)}{a},\nonumber\\
\nabla_{\mu}^*\psi(x) &=& 
\frac{\psi(x)-U_{\mu}^{\dagger}(x)\psi(x-\hat{\mu}a)}{a}.
\end{eqnarray}
Euclidean Dirac matrix $\gamma_{\mu}$ satisfies 
\begin{eqnarray}
\gamma_{\mu}^{\dagger}=\gamma_{\mu},\;\;\;
\{\gamma_{\mu},\gamma_{\nu}\}=2\delta_{\mu,\nu},\;\;\;
\gamma_5=\gamma_0\gamma_1\gamma_2\gamma_3.
\end{eqnarray}
The expectation values of an operator $O$ 
with a fixed topology is denoted by $\langle O \rangle^Q$.

\section{The hybrid Monte Carlo algorithm}\label{sec:HMC}

We explain how to generate the configurations in
our numerical simulations.
The hybrid Monte Carlo algorithm (HMC) is one of the most efficient
\cite{Duane:1987de}
algorithms and widely used in the lattice QCD studies.

Consider updating fields, $q_i$'s, which have an action $S(q)$.
The HMC algorithm consists of two parts;
\begin{enumerate}
\item generate a candidate of updating field $q_i^c$.
\item judge whether $q_i^c$ is accepted or rejected.
\end{enumerate}
The former is called molecular dynamics steps,
since it is very similar to the trajectory of 
a classical particle, randomly walking
in configuration space, of which path is determined by the
equation of motion;
\begin{eqnarray}
\frac{d q_i(\tau)}{d \tau}=\pi_i(\tau),
&&
\frac{d \pi_i(\tau)}{d \tau}=-\frac{\partial S(q)}{\partial q_i}(\tau),
\end{eqnarray}
where the initial momentum $\pi_i(\tau=0)$ 
is randomly given with probability density
\begin{equation}
P(\{\pi_i(0)\})=\left(\prod_i \frac{1}{\sqrt{2\pi}}
\exp (-\sum_i \pi_i^2(0)/2)\right).
\end{equation}
The final point, $q_i(\tau=\tau_f)$, obtained in this way 
(after a fixed 'time' $\tau_f$), is chosen to be the candidate $q_i^c$.
This step is known to satisfy ``the detailed balance'' 
which is a sufficient condition for an algorithm 
to sample the distribution
with the correct Boltzmann weight $\exp (-S(q))$.
However, in the numerical studies,
the evolution is done by the leap-frog updates with a
finite step-size $\Delta \tau$,
\begin{eqnarray}
q_i(\tau+\Delta \tau)-q_i(\tau)&=&\Delta \tau*\pi_i(\tau+\Delta \tau/2),
\nonumber\\
\pi_i(\tau+\Delta \tau/2)-\pi_i(\tau-\Delta \tau/2)&=&
-\Delta \tau*\frac{\partial S(q)}{\partial q_i}(\tau).
\label{eq:hmcev}
\end{eqnarray}
Therefore, one needs the latter procedure, or the Metropolis test,
to judge that the new configuration is accepted or not;
$q_i^c$ is accepted with probability
\begin{equation}
P = \mbox{min} 
\left\{1, \frac{e^{-\pi_i^{c 2}/2-S(q^c)}}{e^{-\pi_i^2/2-S(q)}}
\right\},
\end{equation}
which realizes the exact detailed-balance.
The above steps ($N_{mds}$ molecular dynamics steps with the step-size
$\Delta \tau$ and the Metropolis test) are done in every 'trajectory'.
In this way, the configurations are generated performing the hundreds or
thousands of trajectories, $N_{\mathrm{trj}}$.\\

Let us consider the path integral
\begin{eqnarray}
Z &=& \int dU_{\mu} \exp (-S_G) \det (D+m)^2\nonumber\\ 
  &=& \int dU_{\mu} \exp (-S_G) \int d\phi d\phi^* 
\exp \left(- \int d^4x \phi(x)^* ((D+m)^2)^{-1}\phi(x)\right),
\end{eqnarray}
where the pseudo-fermion field $\phi$ is introduced to 
evaluate the fermion determinant $\det (D+m)^2$, 
which can not be directly 
calculated when the lattice size is large.

Since the pseudo-fermion part of the action is a simple 
bilinear, $\phi$ can be updated with a conventional heat-bath 
algorithm.
The HMC algorithm is applied to
updating the link fields $U_{\mu}$'s
with fixed $\phi=\phi_h$ which is generated by the heat-bath method;
\begin{eqnarray}
Z_{\phi_h} 
  &=& \int dU_{\mu} \exp (-S_G)
\exp \left(- \int d^4x \phi_h(x)^* (H_W^2)^{-1}\phi_h(x)\right)
\nonumber\\
&\equiv& \int dU_{\mu} \exp(-S^{\phi_h}_{\mathrm{eff}}(U_{\mu})).
\end{eqnarray}
Thus, our lattice QCD simulation is performed as follows,
\begin{enumerate}
\item Choose a starting link variable configuration.
\item Choose $\eta \equiv (D+m)^{-1}\phi_h$ to be a field of Gaussian
      noise (heat-bath).
\item Choose the momentum of link variables, $\pi_{\mu}(x)$ from a
      Gaussian noise.
\item $N_{mds}$ molecular dynamics with the step-size $\Delta \tau$ are
      done.
\item Accept or reject the new configuration (Metropolis test).
\item Store the (new or old) configuration.
\item Return to step 2. 
\end{enumerate}

Note that both of
\begin{equation} 
S^{\phi_h}_{eff},\;\;\; \mbox{and}\;\;\; \frac{\partial S^{\phi_h}_{eff}}{\partial U_{\mu}}
\end{equation}
involves the calculation of $(D+m)^{-1}$, which requires a very 
expensive computational cost in the chiral limit, $m\to 0$,
since a lot of low eigenvalues appear.
Also, it is notable that 
when $(D+m)$ is not smooth with respect to the gauge link
field $U_{\mu}$, sudden jumps of momentum evolution in 
Eq.(\ref{eq:hmcev}) should be monitored, otherwise over(under)
estimation of the momentum would break the algorithm.
Finally, we should refer to the fact 
that each step of HMC is small variation of
the fields, and therefore, tends to keep the initial 
topological properties of the link variables.

\section{Quenched chiral perturbation theory in the 
$\epsilon$-regime}\label{sec:ChPT}

This appendix is devoted to a brief review of
quenched chiral perturbation theory (qChPT) in the $\epsilon$-regime 
\cite{Damgaard:2001js}
and we
summarize the relevant formulas for our analysis of
meson correlation functions.

The partition function of qChPT with $N_v$ valence quarks is defined 
\begin{eqnarray}
  \label{eq:fullpart}
  Z(\theta, M) 
  &=& 
  \int dU \exp \left(
    -\int d^4x \mathcal{L}^{\theta}_M(x)\right),
\end{eqnarray}
where the Lagrangian $\mathcal{L}^{\theta}_M$ is given by
\begin{eqnarray}
\label{eq:chptlag}
  \mathcal{L}^{\theta}_M(x) 
  &=&
  \frac{F_{\pi}^2}{4}\mbox{Str} (\partial_{\mu}U(x)^{-1}\partial_{\mu}U(x))
  -\frac{m\Sigma}{2} \mbox{Str}(U_{\theta}U(x)+U(x)^{-1}U_{\theta}^{-1})
  \nonumber\\
  &&
  +\frac{m_0^2}{2N_c}\Phi(x)^2
  +\frac{\alpha}{2N_c}\partial_{\mu}\Phi(x)\partial_{\mu}\Phi(x),
\end{eqnarray}
at the leading order of pion mass, $m_\pi^2$, and pion momentum,
$p_\pi^2$, expansion,
where $N_c$ denotes the number of color, $F_{\pi}$ is the pion decay
constant and $\Sigma$ is the chiral condensate.
$U(x)$ is integrated over a sub-manifold
of the super-group $Gl(N_v|N_v)$, called the maximally symmetric
Riemannian sub-manifold which is characterized by a matrix of form
\begin{equation}
  U = \left(
    \begin{array}{cc}
      A & B \\
      C & D 
    \end{array}
  \right),
  \;\;\;\;\; A \in U(N_v), 
  \;\;\;\;\; D \in Gl(N_v)/U(N_v),
\end{equation}
and Grassmannian $N_v\times N_v$ matrices $B$ and $C$.
Str denotes the super-trace.
The quark mass is involved in the mass matrix 
$M=(mI_v+m\tilde{I_v})$ where $I_v$ and $\tilde{I_v}$ are 
the identity matrix in the fermion-fermion and boson-boson
blocks respectively.
The $\theta$ term which violates CP symmetry, enters through
$U_\theta\equiv \exp(i\theta/N_v)I_{N_v}+\tilde{I}_{N_v}$.
Note that in the quenched approximation the singlet field
$\Phi(x)\equiv \frac{F_{\pi}}{\sqrt{2}}\mbox{Str}[-i\ln U(x)]$
is not decoupled, and its mass $m_0^2$ and coupling $\alpha$
are introduced \cite{Sharpe:1992ft}.

The $\epsilon$-regime
\cite{Gasser:1983yg, Gasser:1987ah,Gasser:1987zq,Hansen:1990un,Hansen:1990yg, Leutwyler:1992yt} 
is a special regime where the quark mass is
so small that the pion Compton wavelength $\sim 1/m_\pi$
is larger than the linear extent of the space-time $L$.
In this regime, the zero mode of pion gives
important contribution and one must explicitly integrate
out the constant mode of $U(x)$.
The other mode's contributions are perturbatively evaluated
with the expansion parameter,
$\epsilon^2\sim m_\pi/4\pi F_\pi\sim 1/(LF_\pi)^2$.
Namely, separating the zero mode and the other modes;
\begin{equation}
  U(x) = U_0 \exp i\frac{\sqrt{2}\xi(x)}{F_\pi},
\end{equation}
the Lagrangian Eq.(\ref{eq:chptlag}) can be rewritten.
Together with $\epsilon$-expansion,
one obtains the partition function with a fixed topological
charge $Q$ by Fourier transforming (\ref{eq:fullpart}) as follows,
\begin{eqnarray}
  \label{eq:ZQinduce}
  Z_{Q}(M) &\equiv& 
  \frac{1}{2\pi}\int^{\pi}_{-\pi}\! d\theta e^{i\theta Q}Z(\theta,M)
  \nonumber\\
  &=&
  \frac{1}{2\pi}\int^{\pi}_{-\pi}\! d\theta 
  \int dU^{\prime}_0d\xi (\mbox{Sdet}U_0^{\prime})^Q
  \exp \left[-\int d^4x\left(
      \mathcal{L}^{\theta}_M(x)+i\frac{\sqrt{2}Q}{F_{\pi}V}\Phi_0\right)\right]
  \nonumber\\
  &=&
  \frac{1}{2\pi}\int^{\pi}_{-\pi}\! d\theta 
  \int dU^{\prime}_0d\xi (\mbox{Sdet}U_0^{\prime})^Q
  \exp\left[
    -\frac{Vm_0^2}{2N_c}\left(\Phi_0^{\prime}
      -\frac{F_{\pi}\theta}{\sqrt{2}}\right)^2
    -\frac{\sqrt{2}iQ}{F_{\pi}}
    \left(\Phi_0^{\prime}-\frac{F_{\pi}\theta}{\sqrt{2}}\right)
  \right]
  \nonumber\\
  & &
  \times
  \exp \left[
    \frac{m\Sigma  V}{2}\mbox{Str}(U_0^{\prime}+U_0^{\prime -1})
  \right.
  \nonumber\\
  & &
  \left.
    \;\;+\int d^4x\left(
      -\frac{1}{2}\mbox{Str}(\partial_{\mu}\xi\partial_{\mu}\xi)
      -\frac{m_0^2}{2N_c}(\mbox{Str}\xi)^2
      -\frac{\alpha}{2N_c}(\partial_{\mu}\mbox{Str}\xi)^2
    \right)
  \right]
  \nonumber\\
  & = &
  \frac{1}{\sqrt{2\pi\langle Q^2\rangle}}e^{-Q^2/2\langle Q^2\rangle}
  \int dU^{\prime}_0d\xi (\mbox{Sdet}U^{\prime}_0)^Q \exp\left[
    \frac{m\Sigma V}{2}\mbox{Str}(U^{\prime}_0+U^{\prime -1}_0)\right.
  \nonumber\\
  & & 
  +\left.\int d^4x \left(
      -\frac{1}{2}\mbox{Str}(\partial_{\mu}\xi\partial_{\mu}\xi)
      -\frac{m_0^2}{2N_c}(\mbox{Str}\xi)^2
      -\frac{\alpha}{2N_c}(\partial_{\mu}\mbox{Str}\xi)^2
    \right)+ O(\epsilon^4)\right],
\end{eqnarray}
where we use
\begin{eqnarray}
  U & = & U_0e^{i\sqrt{2}\xi/F_{\pi}},
  \\
  \Phi_0 & \equiv &
  \frac{F_{\pi}}{\sqrt{2}}\mbox{Str}(-i\ln U_0),
  \\
  U_0^{\prime} & = & U_{\theta}U_0,
  \\
  e^{iQ\theta} & = &
  (\mbox{Sdet}U_0^{\prime})^Q
  \exp\left(-\int d^4x
    \frac{\sqrt{2}iQ}{F_{\pi}V}\Phi_0\right),
  \\
  \Phi_0^{\prime} & \equiv &
  \frac{F_{\pi}}{\sqrt{2}}\mbox{Str}(-i\ln U_0^{\prime})
  =\Phi_0+\frac{F_{\pi}\theta}{\sqrt{2}}.
\end{eqnarray}
In the last line of (\ref{eq:ZQinduce}), we perform
$\theta$ integral as a Gaussian;
\begin{eqnarray}\label{eq:Qgauss}
  \lefteqn{
  \frac{1}{2\pi}\int^{\pi}_{-\pi} d\theta 
  \exp\left[
    -\frac{Vm_0^2F_{\pi}^2}{4N_c}
    \left(\theta-\frac{\sqrt{2}}{F_{\pi}}\Phi_0^{\prime}\right)^2
    +iQ\left(\theta-\frac{\sqrt{2}}{F_{\pi}}\Phi_0^{\prime}\right)
  \right]
  }
  \nonumber\\
  & = &
  \exp\left(-\frac{Q^2}{2\langle Q^2\rangle}\right)
  \frac{1}{2\pi}\int^{\pi}_{-\pi} d\theta^{\prime} 
  \exp\left[-\frac{\langle Q^2\rangle}{2}\left(
      \theta^{\prime}-\frac{iQ}{\langle Q^2\rangle}
  \right)^2\right]
  \nonumber\\
  & \sim &
  \frac{1}{\sqrt{2\pi\langle Q^2\rangle}}
  \exp\left(-\frac{Q^2}{2\langle Q^2\rangle}\right),
\end{eqnarray}
where $\langle Q^2\rangle=Vm_0^2F_{\pi}^2/2N_c$ and 
$\theta^{\prime}=\theta-\sqrt{2}\Phi^{\prime}_0/F_{\pi}$.
We need a condition, $|Q|/\langle Q^2\rangle \ll 1$,
in order to justify this Gaussian integral,
otherwise the integral
Eq.(\ref{eq:Qgauss}) should depend on $\Phi_0^{\prime}$, which
means that $\Phi_0^{\prime}$ and $\theta$ can not be treated
independently and the last line of the 
partition function Eq.(\ref{eq:ZQinduce}) is invalid.
Therefore, all the results shown below are reliable only for
small $|Q|$.

With redefinition $U_0^{\prime}=U_0$, the partition function in
a fixed topological sector is obtained
\begin{eqnarray}
  \label{eq:Z_Q}
  Z_{Q}(M) &\equiv& 
  \frac{1}{2\pi}\int_{-\pi}^{+\pi}\!
  d\theta e^{i\theta Q}Z(\theta,M)
  \nonumber\\
  &=&
  \frac{1}{\sqrt{2\pi\langle Q^2\rangle}}e^{-Q^2/2\langle Q^2\rangle}
  \int dU_0d\xi\, (\mathrm{S}\det U_0)^Q \exp\left[
    \frac{m\Sigma V}{2}\mbox{Str}(U_0+U_0^{-1})
  \right.
  \nonumber\\
  &&
  +\left.
    \int\! d^4x \left(
      -\frac{1}{2}\mbox{Str}(\partial_{\mu}\xi\partial_{\mu}\xi)
      -\frac{m_0^2}{2N_c}(\mbox{Str}\xi)^2
      -\frac{\alpha}{2N_c}(\partial_{\mu}\mbox{Str}\xi)^2
    \right)+ O(\epsilon^4)
  \right],\nonumber\\
\end{eqnarray}
where $dU_0$ denotes the Haar measure of the maximally Riemannian
sub-manifold of \\ $Gl(N_v | N_v)$.
Note that the topological charge distributes as a Gaussian with variance
\begin{equation}
  \label{eq:topsus}
  \frac{\langle Q^2 \rangle}{V} = \frac{F_{\pi}^2m_0^2}{2N_c},
\end{equation}
which is in a good contrast with the full theory, 
for which $\langle Q^2 \rangle=m\Sigma V/N_f$ 
is expected with $N_f$ flavors. 


In the following we consider $N_v$ = 1 and 2 cases only, 
as we are interested in the system with two light quarks.
Any correlators or condensates are obtained by 
the perturbation of $\xi$ fields and 
the exact integration over zero mode, $U_0$,
which can be written in terms of the Bessel functions.

First let us calculate the condensates.
At the tree-level the scalar condensate is given  by
\begin{eqnarray}
  -\langle \bar{\psi}\psi\rangle_Q \equiv 
  \Sigma_Q(\mu) = \Sigma\mu
  (I_{|Q|}(\mu)K_{|Q|}(\mu)+I_{|Q|+1}
  (\mu)K_{|Q|-1}(\mu))
  +\Sigma\frac{|Q|}{\mu},
\end{eqnarray}
with $\mu\equiv m\Sigma V$. 
$I_{|Q|}(\mu)$ and $K_{|Q|}(\mu)$ denote the modified Bessel
functions.
One-loop contribution does not change its functional form 
\cite{Damgaard:2001xr}
\begin{eqnarray}
  \label{eq:sigmaeff} 
  \Sigma^{\mathrm{1-loop}}_Q(\mu) 
  = \Sigma_{\mathrm{eff}} \mu^{\prime}
  (I_{|Q|}(\mu^{\prime})K_{|Q|}(\mu^{\prime})+I_{|Q|+1}
  (\mu^{\prime})K_{|Q|-1}(\mu^{\prime}))
  +\Sigma_{\mathrm{eff}}\frac{|Q|}{\mu^{\prime}} 
  =\Sigma_Q(\mu^{\prime}),
\end{eqnarray}
but the parameters $\mu$ and $\Sigma$ are shifted to $\mu'$
and $\Sigma_{\mathrm{eff}}$:
\begin{eqnarray}
  \label{eq:muprime}
  \mu^{\prime}  \equiv  m\Sigma_{\mathrm{eff}}V,
  &&
  \Sigma_{\mathrm{eff}}  \equiv  \Sigma
  \left(
    1+\frac{m_0^2\bar{G}(0)+\alpha\bar{\Delta}(0)}{N_c F_\pi^2}
  \right).
\end{eqnarray}
Here, parameters $\bar{G}(0)$ and $\bar{\Delta}(0)$ are
ultraviolet divergent tadpole integrals,
\begin{eqnarray}
  \bar{G}(x) \equiv  
  \frac{1}{V}\sum_{p\neq 0}\frac{e^{ipx}}{p^4},
  &&
  \bar{\Delta}(x) \equiv 
  \frac{1}{V}\sum_{p\neq 0}\frac{e^{ipx}}{p^2},
\end{eqnarray}
which need renormalization.


Next, we consider the flavor-singlet meson operators
\begin{eqnarray}
  S^0(x)  \equiv  \bar{\psi}(x)I_{N_V}\psi(x),
  &&
  P^0(x)  \equiv  \bar{\psi}(x)i\gamma_5I_{N_V}\psi(x).
\end{eqnarray}
To add these operators to the QCD Lagrangian as source terms; 
\begin{equation}
  \mathcal{L}\to\mathcal{L}+s(x)S^0(x)+p(x)P^0(x),
\end{equation}
corresponds to the substitution
\begin{equation}
  M \to M+s(x)I_{N_v}+ip(x)I_{N_v}
\end{equation}
in qChPT. 
The two-point correlation functions are obtained by
differentiating the generating functional by $s(x)$ and $p(x)$. 
To $O(\epsilon^2)$ the results are
\begin{eqnarray}
  \langle S^0(x)S^0(0)\rangle_Q
  &=&
  C_S^0
  +\frac{\Sigma^2}{2F_{\pi}^2}\left[
    \frac{a_-}{N_c}(m_0^2\bar{G}(x)+\alpha\bar{\Delta}(x))
    -\bar{\Delta}(x)\frac{a_++a_--4}{2}
  \right],
  \\
  \langle P^0(x)P^0(0)\rangle_Q
  &=&
  C_P^0
  -\frac{\Sigma^2}{2F_{\pi}^2}\left[
    \frac{a_+}{N_c}(m_0^2\bar{G}(x)+\alpha\bar{\Delta}(x))
    -\bar{\Delta}(x)\frac{a_++a_-+4}{2}
  \right],
\end{eqnarray}
where
\begin{eqnarray}
  a_+  = 
  4\left[\left(\frac{\Sigma_Q(\mu)}{\Sigma}\right)^{\prime}
    +1+\frac{Q^2}{\mu^2}\right],
  &&
  a_-  =  
  4\left[-\frac{1}{\mu}\frac{\Sigma_Q(\mu)}
    {\Sigma}+1+\frac{Q^2}{\mu^2}\right],
\end{eqnarray}
and the constant terms are given by 
\begin{eqnarray}
  C_S^0 & = &
  \frac{\Sigma^2_{\mathrm{eff}}}{4} a_+^{\mathrm{1-loop}}
  =\Sigma^2_{\mathrm{eff}}\left[
    \left(\frac{\Sigma_Q(\mu')}{\Sigma_{\mathrm{eff}}}\right)^{\prime}
    +1+\frac{Q^2}{\mu'^2}
  \right],
  \\
  C_P^0 & = &
  -\frac{\Sigma^2_{\mathrm{eff}}}{4} a_-^{\mathrm{1-loop}}
  =\Sigma^2_{\mathrm{eff}}\left[
    \frac{1}{\mu'}\frac{\Sigma_Q(\mu')}{\Sigma_{\mathrm{eff}}}
    -\frac{Q^2}{\mu'^2}
  \right].
\end{eqnarray}
Note that the prime denotes the derivative with respect to $\mu$,
\begin{eqnarray}
  \left(\frac{\Sigma_Q(\mu)}{\Sigma}\right)^{\prime}
  =I_{|Q|}(\mu)K_{|Q|}(\mu)-I_{|Q|+1}(\mu)K_{|Q|-1}(\mu)
  -\frac{|Q|}{\mu^2}.
\end{eqnarray}

For flavor non-singlet case, we need a $N_v=2$ super-group
integral, which is also described with the Bessel functions.
The non-singlet operators are given by
\begin{eqnarray}
  S^a(x) & \equiv & \bar{\psi}(x) (\tau^a/2) I_{N_V}\psi(x),
  \\
  P^a(x) & \equiv & \bar{\psi}(x) (\tau^a/2) i\gamma_5I_{N_V}\psi(x),
\end{eqnarray}
with the Pauli matrices $\tau^a$.
To $O(\epsilon^2)$ the two-point functions are given by
\begin{eqnarray}
  \label{eq:SS_QChPTconnected}
  \langle S^a(x)S^a(0)\rangle_Q &=& C_S^a
  +\frac{\Sigma^2}{2F_{\pi}^2}\left[
    \frac{c_-}{N_c}(m_0^2\bar{G}(x)+\alpha\bar{\Delta}(x))
    -\bar{\Delta}(x)b_-
  \right],
  \\
  \label{eq:PP_QChPTconnected}
  \langle P^a(x)P^a(0)\rangle_Q &=& C_P^a
  -\frac{\Sigma^2}{2F_{\pi}^2}\left[
    \frac{c_+}{N_c}(m_0^2\bar{G}(x)+\alpha\bar{\Delta}(x))
    -\bar{\Delta}(x)b+
  \right],
\end{eqnarray}
where
\begin{eqnarray}
  b_+  =  2\left(1+\frac{Q^2}{\mu^2}\right),
  \;\;\;
  b_-  =  2\frac{Q^2}{\mu^2},
  \;\;\;
  c_+  =  2\left(\frac{\Sigma_Q(\mu)}{\Sigma}\right)',
  \;\;\;
  c_-  =- 2\frac{1}{\mu}\frac{\Sigma_Q(\mu)}{\Sigma},
\end{eqnarray}
and 
\begin{eqnarray}
  C_S^a  =  
  \frac{\Sigma^2_{\mathrm{eff}}}{2}\left(\frac{\Sigma_Q(\mu^{\prime})}
    {\Sigma_{\mathrm{eff}}}\right)^{\prime},
  &&
  C_P^a  =  
  \frac{\Sigma^2_{\mathrm{eff}}}{2}\left(\frac{\Sigma_Q(\mu^{\prime})}
    {\mu'\Sigma_{\mathrm{eff}}}\right).
\end{eqnarray}
For the flavor non-singlet axial-vector current
\begin{equation}
  A_\mu^a(x) = \bar{\psi}(x) (\tau^a/2) i\gamma_\mu\gamma_5 \psi(x),
\end{equation}
the correlator is obtained \cite{Damgaard:2002qe}
\begin{eqnarray}\label{eq:axial}
  \langle A^a_0(x)A^a_0(0)\rangle_Q
  &=&
  -\frac{F_{\pi}}{V}-2m\Sigma_Q(\mu)\bar{\Delta}(x).
\end{eqnarray}
An important observation here is that the axial-current
correlator does not depend on the parameters related to the
quenched artifact, {\it i.e.} $m_0^2$ and $\alpha$.
Also note that the constant term is proportional to
$F_\pi$ rather than $\Sigma$.
Therefore, this channel is suitable for an extraction of
$F_\pi$, whereas for the pseudo-scalar and scalar
correlators $F_\pi$ appears only in the coefficients of
$\bar{\Delta}(x)$ and $\bar{G}(x)$ terms.

For zero spatial momentum projection,
it is convenient to define
\begin{eqnarray}
  h_1(|t/T|)
  & = &
  \frac{1}{T}\int\! d^3x\, \bar{\Delta}(x)= 
  \frac{1}{2}
  \left[\left(\frac{|t|}{T}-\frac{1}{2}\right)^2
    -\frac{1}{12}\right],
  \\
  h_2(|t/T|)
  & = &
  -\frac{1}{T^3} \int\! d^3x\, \bar{G}(x)
  =\frac{1}{24}\left[\frac{t^2}{T^2}\left(\frac{|t|}{T}-1\right)^2
    -\frac{1}{30}\right].
\end{eqnarray}
As a Remarkable feature in the $\epsilon$-regime, 
the correlators do not behave as
the usual exponential fall-off $\exp(-Mt)$ with the mass gap
$M$.

\end{document}